\documentclass[11pt]{article}
\pdfoutput=1
\usepackage{jcappub}

\usepackage{bm}
\usepackage{aas_macros}
\usepackage[mathscr]{eucal}

\newcommand{\rhocrit}{\rho_{\rm crit}}
\newcommand{\rvir}{r_{\rm \Delta_c}}
\newcommand{\rsph}{r_{\rm sph}}
\newcommand{\mvir}{m_{\rm vir}}

\newcommand{\nparthalo}{N_{\rm halo-part.}}
\newcommand{\seed}{\xi^{\rm seed}}
\newcommand{\om}{\Omega_{m}} 

\newcommand{\xv}{\mathbf{x}}
\newcommand{\rv}{\mathbf{r}}

\newcommand{\rconv}{r_{\rm conv}}

\newcommand{\rperp}{\rv_{\perp}}
\newcommand{\rpara}{r_{\parallel}}

\newcommand{\excor}{\mathscr{C}}

\newcommand{\axis}{\mathbf{a}} 
\newcommand{\orient}{\mathbf{\varepsilon}}

\newcommand{\sscale}{\tilde{s}} 
\newcommand{\baxis}{b_{\rm shape}}

\title{The shapes and alignments of \\dark matter halos}
\author[a,b]{Michael D. Schneider,}
\author[b]{Carlos S. Frenk,}
\author[b]{and Shaun Cole}

\affiliation[a]{Lawrence Livermore National Laboratory, \\P.O. Box 808 L-210, Livermore, CA 94551-0808, USA.}
\affiliation[b]{Institute for Computational Cosmology, Department of Physics, Durham University, \\South Road, Durham, DH1 3LE, UK.}

\emailAdd{schneider42@llnl.gov}
\emailAdd{c.s.frenk@durham.ac.uk}
\emailAdd{shaun.cole@durham.ac.uk}

\subheader{LLNL-JRNL-513132}

\abstract{  
We present
measurements of the triaxial dark matter halo shapes and alignment correlation functions in the
Millennium and Millennium-2 dark matter $N$-body simulations. 
These two simulations allow us to measure the distributions of 
halo shapes down to 10\% of the virial radius over a halo mass range of 
$6\times10^{9}$ -- $2\times10^{14}\,h^{-1}M_{\odot}$.   
We largely confirm previous results on the distributions of halo axis ratios as
a function of  halo mass, but we find that the median angle between halo major axes at different halo radii
can vary by a factor of 2 between the  Millennium-1 and 2 simulations because of the different mass
resolution.  Thus, error in the shape  determinations from limited resolution is potentially 
degenerate with the misalignment of halo inner and outer shapes  used to constrain 
Brightest Cluster Galaxy alignments in previous works.
We also present simplifying parameterizations for the 3-D halo-mass alignment correlation functions that are
necessary ingredients for triaxial halo models of large-scale structure and models of galaxy
intrinsic alignments as contaminants for cosmic shear surveys. 
We measure strong alignments between halos of all masses and the surrounding
dark matter overdensities  out to several tens of $h^{-1}$~Mpc, in agreement with observed 
shear--galaxy and cluster shape correlations.   
We use these measurements to forecast the contribution to the weak lensing 
signal around galaxy clusters from correlated mass along the line-of-sight. 
For prolate clusters with major axes aligned with the line-of-sight the fraction of the weak lensing 
signal from mass external to the cluster can be twice that predicted if the excess 
halo alignment correlation is assumed to be zero.
}

\arxivnumber{1111.5616}

\begin{document}
\maketitle
\flushbottom

\section{Introduction}
\label{sec:introduction}
The cold dark matter (CDM) model is now the paradigm for explaining the statistics of large-scale structure as well as galaxy formation and evolution. By using $N$-body simulations it is possible to make detailed predictions of the cosmological distribution and evolution of CDM that can be compared with a wide array of observations. The large-scale statistics of CDM are a good fit to galaxy clustering and weak lensing observations while weak and strong lensing, galaxy velocity measurements, and galaxy cluster mass estimates are well-described by the predicted abundances and density profiles of virialized dark matter structures in $N$-body simulations on scales larger than $\sim100$~kpc.

While the spherically averaged density profiles of dark matter halos first described by~\citep{Navarro1997} (NFW) are good fits to both simulations and many observations, deviations from halo sphericity can be important for detailed comparisons of observations with the CDM model~\citep{1988ApJ...327..507F,Hayashi:2006p2600,2011MNRAS.413.1973W}. Measurements of halo mass density profiles for specific halos must account for asphericity to accurately compare with CDM models~\citep{Meneghetti:2007p2431}.  Halo triaxiality can also lead to biases in ensemble statistics such as the halo mass function and $n$-point correlations when used to constrain cosmological parameters~\cite{2012MNRAS.421.1399B} as well as measurements of the Hubble constant from clusters~\cite{2006ApJ...643..630W}. Accounting for cluster triaxiality is further important for detailed interpretation of lensing and Sunyaev-Zeldovich measurements~\cite{2012MNRAS.421.1399B}.

The details of the distributions of halo shapes and alignments also carry information about the hierarchichal formation of halos and the filamentary structure in which they are embedded~\citep{2005ApJ...627..647B, 2007ApJ...655L...5A,VeraCiro:2011nb,Slosar:2008p2313, 2008MNRAS.389.1266L}. In the current understanding of hierarchical structure formation the small mass halos form first from the anisotropic collapse of ellipsoidal overdensities in the mass distrubution. Larger halos grow both through fairly steady mass accretion of the surrounding dark matter and through ``major mergers'' with other halos. The ``radial orbit instability''~\cite{2008ApJ...685..739B} can be important for determining the triaxial shapes of halos formed through steady mass accretion, but it remains to be determined whether mergers with larger halos or the anisotropic accretion of mass from filaments play larger roles in the determination of halo shapes. Other recent work has shown that the triaxial density profiles of halos may be determined by the initial conditions of the mass-density perturbations~\cite{2011MNRAS.414.3044V}. 

Halos become more spherical with time as their constituent particles complete many orbits and undergo ``violent relaxation''~\cite{1996grdy.conf..121W}. 
For a given halo mass and cosmological epoch, halos that formed more recently are also found to have smaller biases with respect to the dark matter distribution than halos that formed at earlier times~\cite{2005MNRAS.363L..66G, 2007ApJ...657..664J}.
Correlations between halo shapes and formation times have also been found in $N$-body simulations~\cite{2010ApJ...708..469F}. Together these results would seem to indicate that the halo formation time is a key parameter in determining the shapes and shape-dependent clustering of halos at given halo mass and epoch. However, the dependence of halo shape on formation time is nontrivial as many other halo properties, such as concentration, spin, and velocity and substructure statistics, have also been shown to be correlated with halo formation time and to influence the clustering of halos in ways that cannot be explained solely by the relationships between these parameters~\cite{2006ApJ...652...71W, 2007MNRAS.377L...5G, 2007ApJ...656..139W, 2007MNRAS.378...55M, 2010ApJ...708..469F}. This complicated dependence of halo properties and clustering is referred to as ``assembly bias''~\cite{2007MNRAS.377L...5G} and indicates that any model for the origins of halo shapes and orientation correlations should account for properties of the halo environment that have not yet been causally isolated.

There are many recent attempts to analytically model triaxial halo shapes~\citep{2005ApJ...632..706L,2010A&A...518A..38A,Desjacques:2008hq,2011ApJ...734..100L, 2012MNRAS.420.1693B} based on the gravitational collapse of initally ellipsoidal mass overdensities. These models show good qualitative agreement with both observations and simulations, but simulations are still required to make precise predictions of the triaxial shape distributions and correlations in CDM because of the highly nonlinear gravitational growth over many decades in length scales.

Predictions of large-scale clustering statistics using the halo model~\citep[e.g.][]{2002PhR...372....1C} typically assume the spherically-averaged NFW profile but can be altered in detectable ways by halo triaxiality~\citep{2005MNRAS.360..203S, 2006MNRAS.365..214S, 2011arXiv1110.4888V}. As shown by the pioneering work of~\cite{2005MNRAS.360..203S}, including triaxial halos in the halo model requires knowledge of the joint probability distributions of the halo axis lengths and orientations as functions of mass and redshift. Also, the ``two-halo'' term in the halo model requires a model for the correlations of halos as a function of the halo separation vector, and halo masses, axis lengths, and orientations. A major goal of the present paper is to provide these missing ingredients to the halo model.

Also, the intrinsic galaxy alignments that contaminate weak lensing measurements are expected to be related to the intrinsic alignments of dark matter halo shapes. Previous groups have attempted to quantify the predicted galaxy intrinsic alignment signal by measuring projected halo shape correlations in $N$-body simulations and then comparing with observed projected shape correlations to calibrate the amplitude of the correlations~\citep{2009RAA.....9...41F,2009ApJ...694..214O}. These groups further used the calibrated shape correlation function amplitudes as indications of the degree of misalignment between central galaxies and their parent dark matter halos. Because the distributions of projected triaxial shapes can be complicated to interpret~\citep{1983AJ.....88.1626F}, in this paper we focus on the measurements of 3D shapes and shape correlations in part so that the modeling of alignments between galaxies and dark matter can be constrained robustly from observations. In this paper we compare measurements of triaxial halo shapes in the Millennium and Millennium-II simulations to both produce statistically significant measurements over a range of halo masses and to assess the convergence of the halo shape statistics with mass resolution.

This paper is organized as follows.   We briefly describe the $N$-body simulations studied in section~\ref{sec:simulations}. Our definitions of halos and our methods for measuring shapes are described in section~\ref{sec:methods} and can be skipped by readers familiar with previous similar studies as well as those less interested in the details of our methods.
In section~\ref{sec:shapes} we update previously published halo shape distributions given our improved mass range and resolution and present new results on the orientations of halo shapes as a function of radius in section~\ref{sec:orientations}.
Section~\ref{sec:alignments} consists entirely of new measurements and analysis of the halo correlation functions binned in angles  between the halo major axes and the major axis of one halo and the separation vector to another halo. We also present some projected halo alignment correlations that can be directly compared with previous work. We present simple parameterizations for the halo alignment correlations in section~\ref{sec:alignmentmodels} and show how the measured alignment correlations can be applied to understand the bias in weak lensing cluster mass estimates from line-of-sight structures. We draw conclusions on the impact of halo alignments on several types of observations and describe future applications of our results in section~\ref{sec:conclusions}.

\section{Simulations}
\label{sec:simulations} 
The Millennium\footnote{\url{http://www.mpa-garching.mpg.de/galform/millennium/}}  
simulation~\citep{2005Natur.435..629S}  solved for the
positions and velocities of $10^{10}$ dark  matter tracer particles, each of mass
$8.6\times10^8$~$h^{-1}M_{\odot}$, in a cubic volume 500~$h^{-1}$Mpc on a side.   The
Millennium-2\footnote{\url{http://www.mpa-garching.mpg.de/galform/millennium-II/}}
simulation~\citep{2009MNRAS.398.1150B} used the same number of particles as the Millennium but in a
volume  100~$h^{-1}$Mpc on a side so that the particle masses were $6.89\times
10^6$~$h^{-1}M_{\odot}$.   Both simulations were run the following cosmological parameters:
$\Omega_m=0.25$, $\Omega_{\Lambda}=0.75$, $\Omega_b=0.045$, $n_s=1$, $\sigma_8=0.9$, $h=0.73$.
This cosmology was chosen to be consistent with the WMAP3~\cite{wmap3} constraints.  The improved 
cosmological constraints from WMAP7~\cite{wmap7} favor $\sigma_8\approx0.8$. While such a change 
in $\sigma_8$ will significantly alter the amplitudes of the halo correlation functions as 
well as the rate of nonlinear growth of structure, we do not expect large differences in the 
degree of halo triaxiality or alignments that cannot be understood with a simple time 
rescaling to match the linear growth functions as in~\cite{2010MNRAS.405..143A}.

\section{Halo definitions}
\label{sec:methods}

In this section we describe our algorithm for identifying halos and determining their masses and
triaxial shapes.

We use the ``Friends-of-Friends'' (FoF)~\cite{1985ApJ...292..371D} 
catalogue generated for each simulation to initially
identify the halos.   For each halo, we read in all the particles in a sphere centered on the FoF
center and with radius twice the FoF virial radius, $r_{\rm vir}^{\rm FoF}$,
(as given in the FoF catalogue~\cite{2009MNRAS.398.1150B}).  Next we redetermine the halo
center by iteratively computing the  center-of-mass of the particles within spheres whose radius is
reduced by 2.5\%~\citep{2003MNRAS.338...14P}  in each iteration.   The iteration is stopped when no
more than 500 particles remain in the sphere and the center-of-mass at this  stage is taken as the
new halo center.  (We discard all FoF halos that have fewer than 500 particles, which is consistent 
with the mass cuts we apply in Section~\ref{sec:shapes}.) 
We flag any halos whose new centers are more than 7\% of the  FoF virial radius
away from the most-bound halo particle~\citep{2007MNRAS.381.1450N}  
as determined by the SubFind algorithm~\citep{2001MNRAS.328..726S}. 
The offset between the most  bound particle and the center
of mass of a halo was previously used as a proxy for determining  ``unrelaxed'' halos
by~\cite{2007MNRAS.381.1450N,2007MNRAS.378...55M}, 
where \cite{2007MNRAS.378...55M} also found that halos with large center offsets tend to have
highly prolate shapes.  By selecting only halos with center offsets less than 7\%,  we therefore
intend to remove any bias in the shape distributions from unrelaxed halos.
However, note that~\citep{2007MNRAS.381.1450N} used the center-of-mass of all particles within 
the virial radius while we determine the center-of-mass only from the 500 particles remaining 
at the end of the iteration for determining the halo center. Our flag for unrelaxed halos based on the 
center offsets is therefore liable to miss some halos that would be flagged by the algorithm in
\citep{2007MNRAS.381.1450N}, but many of these are likely caught by our substructure flag described below.

We then define updated halo masses and radii by determining the radius $r$ of a sphere centered at the halo center where,
\begin{equation}
  \frac{M_x(<r_x)}{4\pi r_x^3/3} = \Delta_x(z)\, \rhocrit(z),
\end{equation}
where $M(<r)$ is the mass enclosed in the sphere of radius $r$ and $\rhocrit\equiv 3H^2(z)/8\pi G$.  We use two definitions for $\Delta_x$: $\Delta_{200}(z)\equiv 200$ 
and~\citep{1996MNRAS.282..263E, 1998ApJ...495...80B, 2010MNRAS.404.1137B}
\begin{equation}\label{eq:overdensity}
  \Delta_c(z) \equiv 18\pi^2 + 82\left(\om(z)-1\right)  -39\left(\om(z)-1\right)^2.
\end{equation}
At the three simulation snapshots we will consider at $z = $~0, 0.5, and 1, the overdensity 
$\Delta_c(z) = $~94, 131, and 152.
We define $\rvir(z)\equiv r_{\rm vir}(z)$ 
using $\Delta_x = \Delta_c(z)$ in equation~\ref{eq:overdensity} 
and use the notation $r_{200}$ to explicitly refer to radii 
computed at an overdensity of $\Delta_{200}$ (which is the overdensity often used to define 
the ``virial radius'').
We determine the radius that matches the target density by computing the enclosed density in expanding spheres; starting at $r_{\rm vir}^{\rm FoF}/2$ and increasing the radii 
in increments of $0.005\,r_{\rm vir}^{\rm FoF}$ until the density threshold is reached.

We label the triaxial halo axis lengths as $a(\rsph) < b(\rsph) < c(\rsph)$. The axis lengths are defined to 
be functions of the spherical radius $\rsph$ from the halo center to allow for different halo shapes 
at different radii. For a given $\rsph$, the axis lengths and orientations are defined to be 
the eigenvalues and eigenvectors, respectively, of the reduced inertia tensor of all the 
mass tracer particles contained within a sphere of radius $\rsph$ centered on the halo,
\begin{equation}\label{eq:redinertiatensor}
  I_{ij}(\rsph) \equiv \sum_{n=1}^{\nparthalo} \frac{x_{n,i} x_{n,j}}{R_n^2(\rsph)},
  \qquad i,j = 1,2,3,
\end{equation}
where $\nparthalo$ is the number of particles in the halo, 
$x_{n,i}$ is the $i$th coordinate of the $n$th particle in the halo measured with respect to a fixed 
cartesian coordinate system.
The normalization of the particle positions, $R_n(\rsph)$, 
is the elliptical radius of the $n$th particle
defined in terms of the axis lengths,
\begin{equation}\label{eq:ellipticalradius}
  R^2(\rsph) \equiv \frac{x^2}{a^2(\rsph)} + \frac{y^2}{b^2(\rsph)} + \frac{z^2}{c^2(\rsph)},
\end{equation}  
where $x,y,z$ are defined in the principal-axis frame of the halo.
Note that $R$ is dimensionless because the axis lengths have physical units.
The reduced inertia tensor gives less weight to particles at the outer edges of 
the halo than the standard inertia tensor (without the $1 / R^2(\rsph)$ weighting).
We finally define the two 3D axis ratios as, 
\begin{equation}
  s \equiv \frac{a}{c} \qquad
  q \equiv \frac{b}{c}.
\end{equation}

Following previous analyses~\citep{2006MNRAS.367.1781A, VeraCiro:2011nb, 2012MNRAS.420.3303B}, 
we compute the reduced inertia tensor using an iterative algorithm
that starts with all the particles within a sphere of radius $\rsph$ 
that is then deformed along the eigenvectors 
of the initial inertia tensor while keeping the volume within the (deformed) ellipsoid fixed.
Because the axis lengths $a(\rsph)$, $b(\rsph)$, and $c(\rsph)$ enter the definition 
of the inertia tensor explicity in equations~\ref{eq:redinertiatensor} and 
\ref{eq:ellipticalradius}, we rescaled the eigenvalues of the inertia tensor at 
each iteration by the quantity $\rsph / (a b c)^{1/3}$ to impose the fixed volume constraint.
Note that variations of the algorithm exist in the literature where the intermediate axis 
rather than the ellipsoidal volume is kept fixed between iterations. By keeping the enclosed 
volume fixed allows us to equate the shapes we measure at 
fixed spherical starting radius $\rsph$
with the shapes measured in~\cite{2002ApJ...574..538J} at fixed enclosed mean mass density.
Particles that are outside of the ellipsoid boundary are dropped and any new particles within 
the ellipsoid are added in the subsequent recalculation of the reduced inertia tensor.  
The algorithm is terminated when either the axis ratios converge to a given tolerance 
(${\rm max}\left((q'-q)^2/q^2, (s'-s)^2/s^2\right) < 5\times10^{-3}$) or the ellipsoid has 
fewer than 50 particles enclosed.
This means the enclosed volume of the ellipsoid with axis ratios $a,b,c$ is related to 
the spherical radius $\rsph$ from 
equation~(\ref{eq:ellipticalradius}) as $V=\frac{4}{3}\pi \rsph^3$.
Recently, \cite{2011ApJS..197...30Z} pointed out that our definition of the 
reduced inertia tensor may be more biased in recovering input triaxial shapes than other 
possible inertia tensor definitions. However, the definition of inertia tensor must 
ultimately be chosen with the final application in mind. In this paper we aim to 
measure triaxial halo statistics that can be compared with other CDM halo predictions and 
also input into a halo model for galaxies. Because our reduced inertia tensor definition 
measures a shape of all enclosed particles, it is potentially more useful than, e.g., 
the shape of all particles in an ellipsoidal shell. The latter was used effectively 
by \cite{2002ApJ...574..538J} to measure the triaxial density profile of halos, 
which we do not reconsider here.

We restrict the inertia tensor calculation to those particles identified by the SubFind algorithm 
as gravitationally bound to the parent halo (denoted ``SubFind halo-0'' or SF0)~\citep{2007MNRAS.381.1450N}.  
That is, we consider only the halos that might host isolated galaxies or groups and clusters and 
discard all gravitationally bound satellite halos or other substructures.  
In addition, any particles that are 
neither gravitationally bound to the parent halo or in substructures are also discarded. 
We compute the reduced inertia tensor of SubFind halo-0 at 32 radii for each halo linearly spaced from 
$0.1\,\rvir$ to $\rvir$.  
(Note these radii set the size of the sphere that is the starting point for the 
iterative inertia tensor calculation.) 

\begin{figure}[htpb]
  \centerline{
  \includegraphics[scale=0.6]{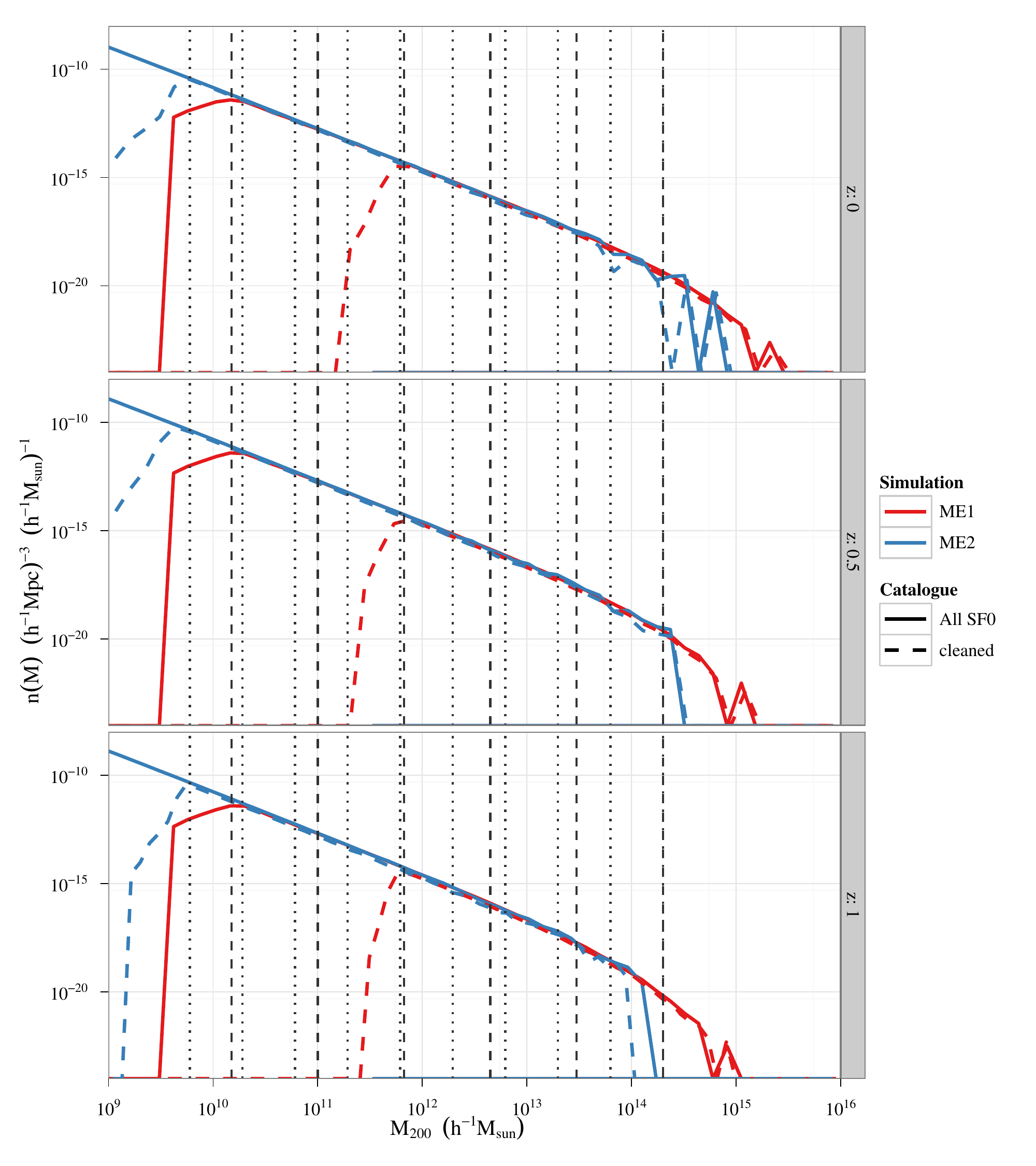}
  }
  \caption{\label{fg:massfunctions}Comparison of the mass functions for all isolated (``SF0'') halos in the 
  simulations and those halos whose shapes were successfully measured (``cleaned''). 
  The ``cleaned'' samples have cuts on the minimum number of particles in the halo (50), 
  the shift in the halo centers and Virial masses between the FoF catalogue and our algorithms, 
  the substructure mass fractions, and the sphericities (see text).
  The vertical dotted lines show the mass bins we use when describing the halo shape distributions 
  while the vertical dashed lines show the mass bins for the halo correlation function measurements.}
\end{figure}

We discard all halos in the FoF catalogue whose centers, virial masses, or shapes do not converge in our iterative algorithms, but this is only 3\% of halos in Millennium-1 and 1\% of halos in Millennium-2 at $z=0$ (see appendix~\ref{sec:halocounts} for more detail on the halo counts with various cuts).
We also discard all halos that have substructure mass fractions greater than 10\%, where the substructure mass is defined as the mass in all bound subhalos with masses greater than 1\% of the virial mass as 
determined by the SubFind algorithm. The substructure mass cut further reduces the halo catalogues at $z=0$ by 24\% and 15\%, respectively, for the Millennium-1 and 2 simulations.
The mass functions of the resulting halo catalogues are shown in figure~\ref{fg:massfunctions}.  
The vertical dotted and dashed lines in figure~\ref{fg:massfunctions} indicate the mass bins that we 
consider in the remainder of the paper.

Our decisions to consider only the shapes of relaxed halos with substructures removed 
is motivated by our primary interests in the CDM predictions for halo shape statistics and
in potential applications in modeling the shapes and orientations of central galaxies 
in the halo model. If we wanted to determine the elliptical halo shapes that 
could be detected in lensing measurements, we might instead measure the shapes 
of all FoF particles, perhaps in ellipsoidal shells rather than enclosed ellipsoids~\cite{2011ApJS..197...30Z}. 
Although, our halo shape definition based on the reduced inertia tensor of all enclosed particles 
may in fact be more relevant for comparing with observable 
halo lensing signals due to the uncertainties of the line-of-sight projection~\cite{2012MNRAS.420.3303B}.
In general, the choice of halo definitions and halo constituent 
particles will be application specific. We believe our choices 
are consistent with previous studies of CDM halo statistics and models for 
central galaxy alignments (which we explain further in Section~\ref{sec:projcorr}). 
In Section~\ref{sec:wlmassbias} we use the shapes of SF0 halos to estimate the bias 
in cluster weak lensing masses from correlated line-of-sight structures. 
Again, evaluating the validity of this choice will depend on the method used 
to select clusters and estimate their ellipticities.

\section{Halo shapes}
\label{sec:shapes}
In this section we update the fitting functions for the joint distribution of the axis ratios
first described in~\cite{2002ApJ...574..538J} to cover the larger mass and radius 
ranges available with the Millennium-2 simulation \citep[where we have particle masses 
of $6.89\times 10^6$~$h^{-1}M_{\odot}$ versus $6.2\times10^8$~$h^{-1}M_{\odot}$][]{2002ApJ...574..538J}. 
We also investigate the radial dependence of the axis ratios and the ``twisting'' of the 
halo axes with radius.  We expect these new measurements to be particularly useful for 
modeling the halo mass distributions for 
lensing studies~\citep[e.g.][]{2007MNRAS.380..149C} and for understanding the 
relative orientations of galaxies and their host halos.  
 
We do not update the triaxial density profile measurements of~\cite{2002ApJ...574..538J} 
because their high-resolution halo simulations already provided better resolution than 
in the large-volume simulations we investigate here.  However, we have confirmed that the 
triaxial density profile model of~\cite{2002ApJ...574..538J} is also an excellent 
fit to the halo shapes we measure.

\subsection{Axis ratio distributions}
\label{sec:axisratios}
\begin{figure*}[htpb]
  \centerline{
    \includegraphics[scale=0.68]{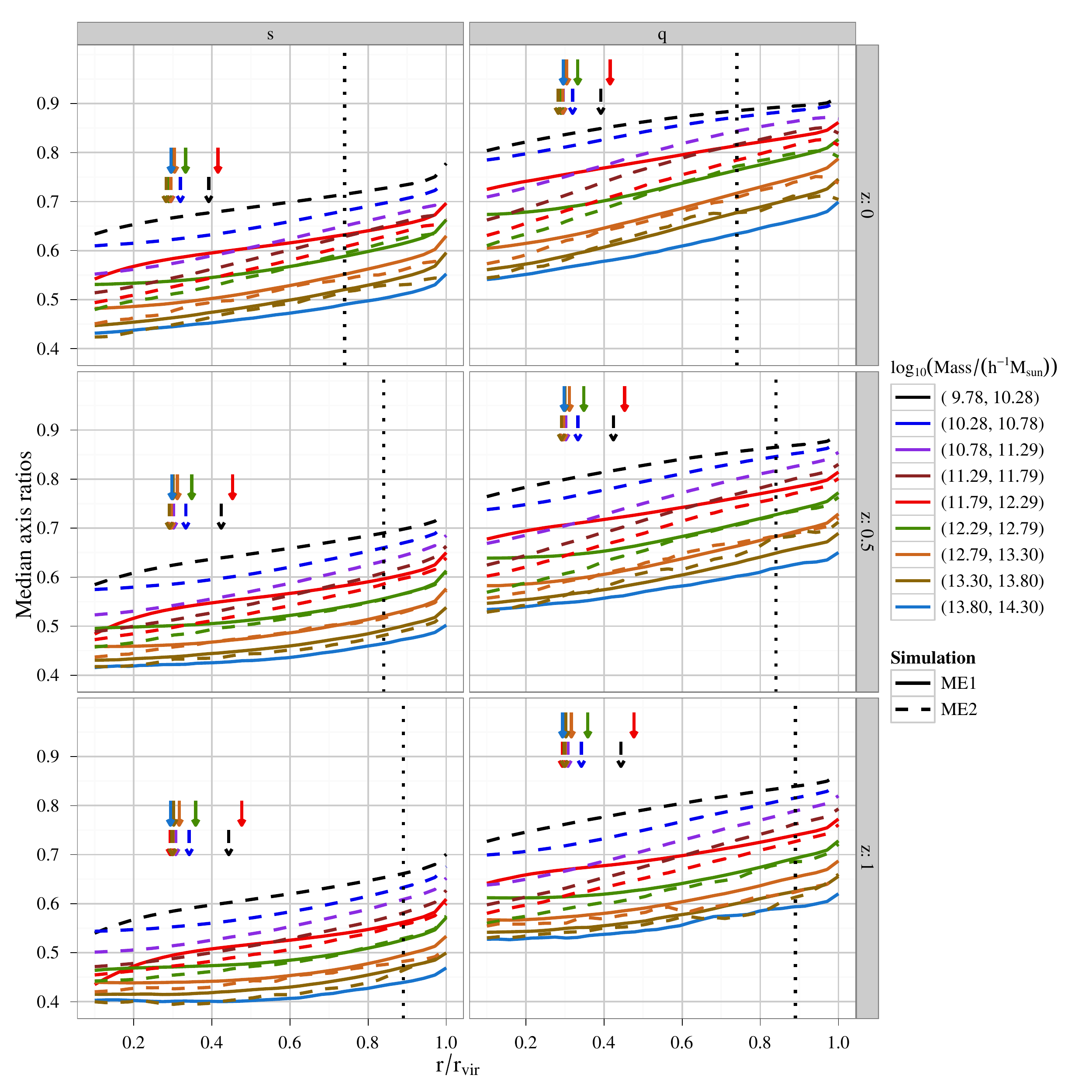}
  }
  \caption{\label{fg:medianaxisratios}Median axis ratios of the triaxial halo shapes as functions of elliptical radius from the center of the halo (normalized by $r_{\rm vir}\equiv \rvir$) 
  in bins in $\log_{10}(M_{200} / h^{-1}M_{\odot})$.  
  The left panels show the ratio of the minor to major axis, $s\equiv \frac{a}{c}$, while the 
  right panels show the ratio of the intermediate to major axis, $q\equiv \frac{b}{c}$. Note that 
  all halos are restricted to have $s \le 0.9$ to ensure reliable shape estimates.
  The axis ratios at redshifts $z=0,0.5,1$ are shown from top to bottom.
  The arrows denote the ``convergence radius'' in each mass bin for each simulation where 
  the ratio of the collisional relaxation time to the circular orbit timescale, $\kappa=7$.
  For a given simulation, lines are omitted in mass bins that do not have complete sampling 
  as indicated by the halo mass function.
  The vertical dotted lines denote the mean values of $r_{200}/\rvir$ for each redshift.}
\end{figure*}

The median axis ratios as functions of the elliptical halo radius are shown in 
figure~\ref{fg:medianaxisratios}. To avoid skewing the axis ratio distributions with halos 
that have poor shape determinations we have computed the medians after selecting only those 
halos with minor-to-major axis ratio $s \le 0.9$ (which removes only 0.2\% and 1\% of halos in the 
Millennium-1 and 2 simulations, respectively, at $z=0$).  
In all cases, the axis ratios increase with radius (meaning the halos become more spherical with increasing radius) and also increase with decreasing halo mass at fixed radius.
The arrows in figure~\ref{fg:medianaxisratios} show the ``convergence radius,'' $\rconv$, 
defined so that the ratio,
$\kappa \equiv t_{\rm relax}/t_{\rm circ}(r_{200}) = 7$, where 
$t_{\rm relax}$ is the ``collisional relaxation time'' and $t_{\rm circ}(r_{200})$ is the 
circular orbit timescale at $r_{200}$~\citep[as defined in][]{2003MNRAS.338...14P, 2010MNRAS.402...21N}.  
Ref.~\cite{VeraCiro:2011nb} found that $\kappa=7$ marked 
the convergence radius where the axis ratios of individual relaxed halos agreed in different 
resolution runs of a quiescent halo in the Aquarius simulations of Milky Way-mass dark matter halos.
For the intermediate mass bins where both simulations have good statistics, it is clear from 
figure~\ref{fg:medianaxisratios} that the median axis ratios are still slightly misestimated 
for $\rconv$ defined at $\kappa=7$.  This is likely due to the wide range of accretion histories 
of the halos in our catalogues that increase the scatter in the axis ratios at fixed halo radius.  
Because the Millennium-1 simulation has larger-mass particles, fewer substructures will be resolved 
and excised from the halos and will therefore contribute to the shape measurements.  
Also, smaller-mass halos will tend to have larger mass-fractions of unresolved 
substructures and should therefore deviate 
more from the Millennium-2 results than the highest mass halos. 
This is exactly the trend we see in the discrepancies between the solid 
and dashed lines in figure~\ref{fg:medianaxisratios}. Because we exclude all halos 
with fewer than 50 particles within a spherical radius of $0.1\rvir$, it is possible that 
our measurements of the axis ratio distributions are skewed at small radii in the lowest-mass 
bin for Millennium-1 because we are selecting only high-concentration halos 
(the Millennium-2 mass bins all have significantly more than 50 particles).
We have checked that increasing the minimum particle cut from 50 to 300 particles does not 
alter either the measurement of the convergence radius or the median 
axis ratios for radii above the convergence radius.
Finally, we note that figure~3 in~\cite{2012MNRAS.420.3303B} showed that the definition of the 
inertia tensor has a noticeable effect on the distributions and median axis 
ratios, with the weighted inertia tensor yielding slightly larger axis ratios at all halo masses 
than the unweighted inertia tensor. However, the differences are smaller than the trends with mass 
and radius that we show in figure~\ref{fg:medianaxisratios}.

Previous simulation studies have shown that more massive halos tend to have later 
formation times (as it takes longer to accrete mass and substructures for more 
massive halos)~\citep{2005MNRAS.363L..66G}. 
At a fixed time, massive halos then tend to be less relaxed than low-mass halos and are therefore
less spherical and have lower concentrations~\citep{2010MNRAS.407..581R,2011MNRAS.413.1973W}. 
The decreases in the axis ratios with both increasing mass and redshift in 
figure~\ref{fg:medianaxisratios} are consistent with this picture of the hierarchical 
formation of halos.

\begin{figure}[htpb]
  \centerline{
  \includegraphics[scale=0.6]{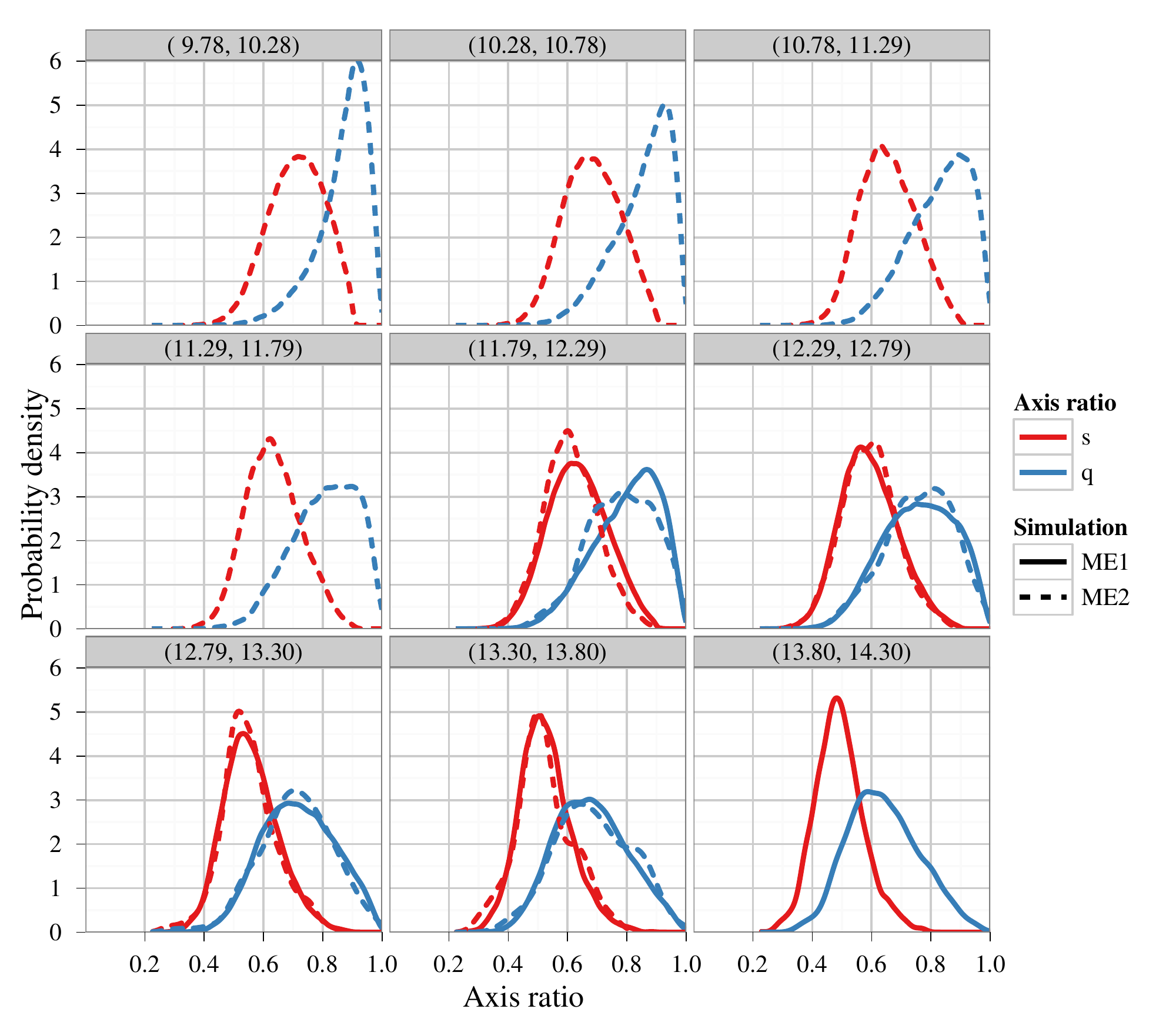}
  }
  \caption{\label{fg:histaxisratios}Marginal distributions of the axis ratios of the triaxial halo shapes 
  at $r_{200}$ and $z=0$.  $s$ is the ratio of the minor to major axes and $q$ is the ratio of the 
  intermediate to major axes.  The panels indicate bins in $\log_{10}(M_{200} / h^{-1}M_{\odot})$.}
\end{figure}

\begin{figure}[htpb]
  \centerline{
  \includegraphics[scale=0.6]{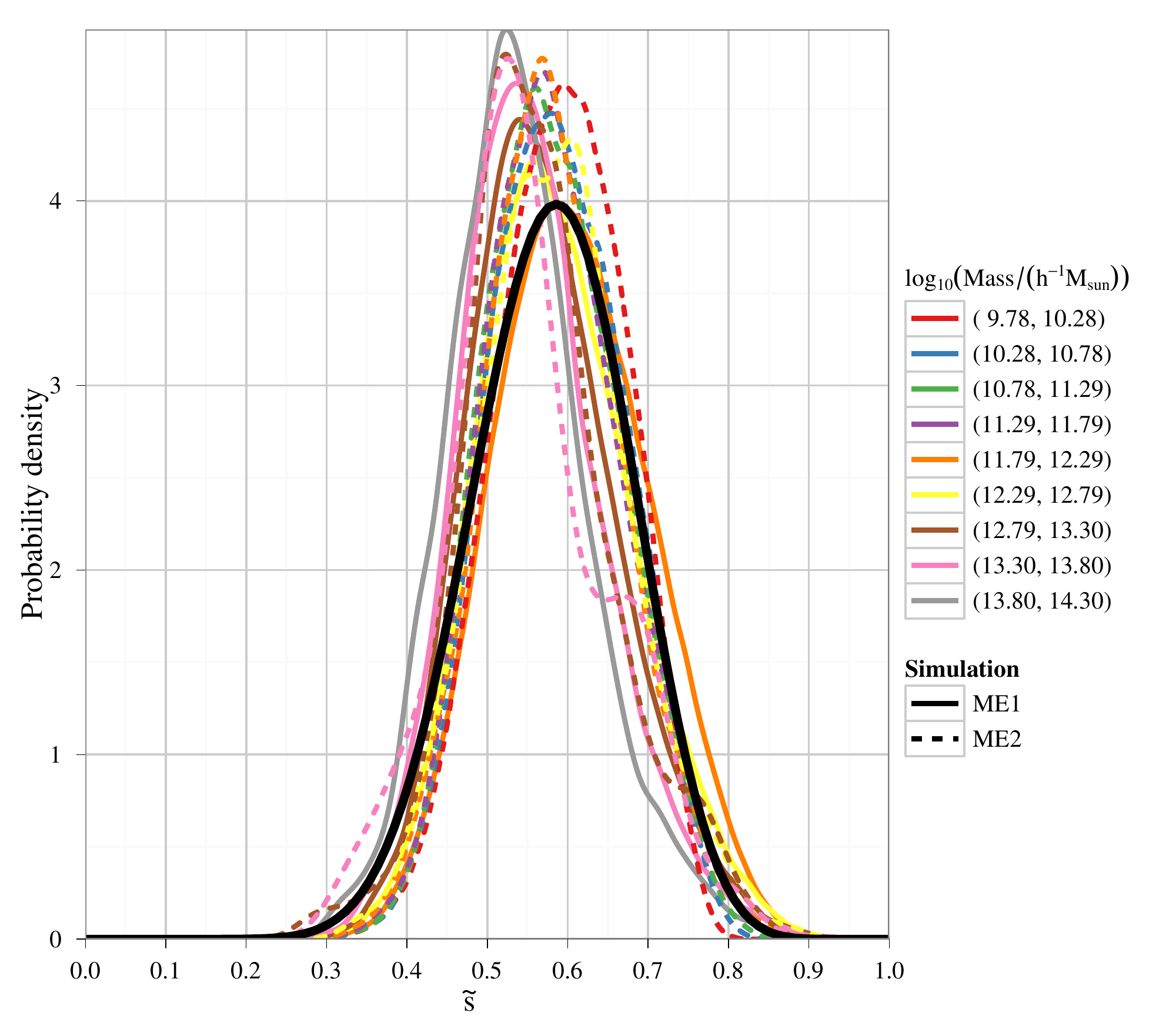}
  }
  \caption{\label{fg:sscalehist}Marginal distributions of the scaled minor-to-major axis 
  ratio $\sscale$ (defined in eq.~\ref{eq:sscale}) at $r_{200}$ and $z=0$.
  The thick black line indicates the fit to the distribution of the axis ratios for all masses 
  in the two simulations as described below equation~\ref{eq:scaledsdistfit}.}
\end{figure}

The marginal distribution of the axis ratios at fixed mass is shown in 
figure~\ref{fg:histaxisratios}.  
In figure~\ref{fg:sscalehist} we show the marginal distributions of the scaled axis ratio,
\begin{equation}\label{eq:sscale}
  \sscale \equiv s\, \left(\frac{\mvir}{M_{*}(z)}\right)^{0.0375[\Omega(z)]^{0.16}}
\end{equation}
in different mass bins, which are only weakly dependent on halo mass and redshift, 
where 
$M_{*}$ is the mass where the r.m.s. density in top-hat spheres $\sigma(M)$ is equal to the 
overdensity for spherical collapse, $\delta_{\rm sc}(z)$.  
The values of $M_{*}(z)$ for the three snapshots we consider are given in table~\ref{tab:mstar}.
\begin{table}[htpb]
\begin{center}
  \label{tab:mstar}
  \caption{Characteristic halo mass for our three snapshots.}
  \begin{tabular}{lc}
  \hline\hline
  Redshift & $M_{*} (h^{-1}M_{\odot})$ \\
  \hline
  0   & $6.2\times 10^{12}$ \\ 
  0.51 & $1.3\times 10^{12}$ \\
  0.98   & $3.0 \times 10^{11}$\\
  \hline
  \end{tabular}
\end{center}
\end{table} 
This scaling is motivated by~\cite{2002ApJ...574..538J} (their eq.~16), 
but we find we need a steeper mass dependence with index 
$0.0375[\Omega(z)]^{0.16}$ ($=0.030$ at $z=0$) 
rather than their index of $0.07[\Omega(z)]^{0.7}$ ($=0.027$ at $z=0$).
While we study simulations with only one choice of cosmological parameters, 
\cite{2006MNRAS.367.1781A} previously found that the dependence of the 
distributions of halo shapes on the amplitude of density perturbations, 
$\sigma_8$, was well described by the cosmology dependence of $M_{*}$ alone.

\begin{figure}[htpb]
  \centerline{
  \includegraphics[scale=0.6]{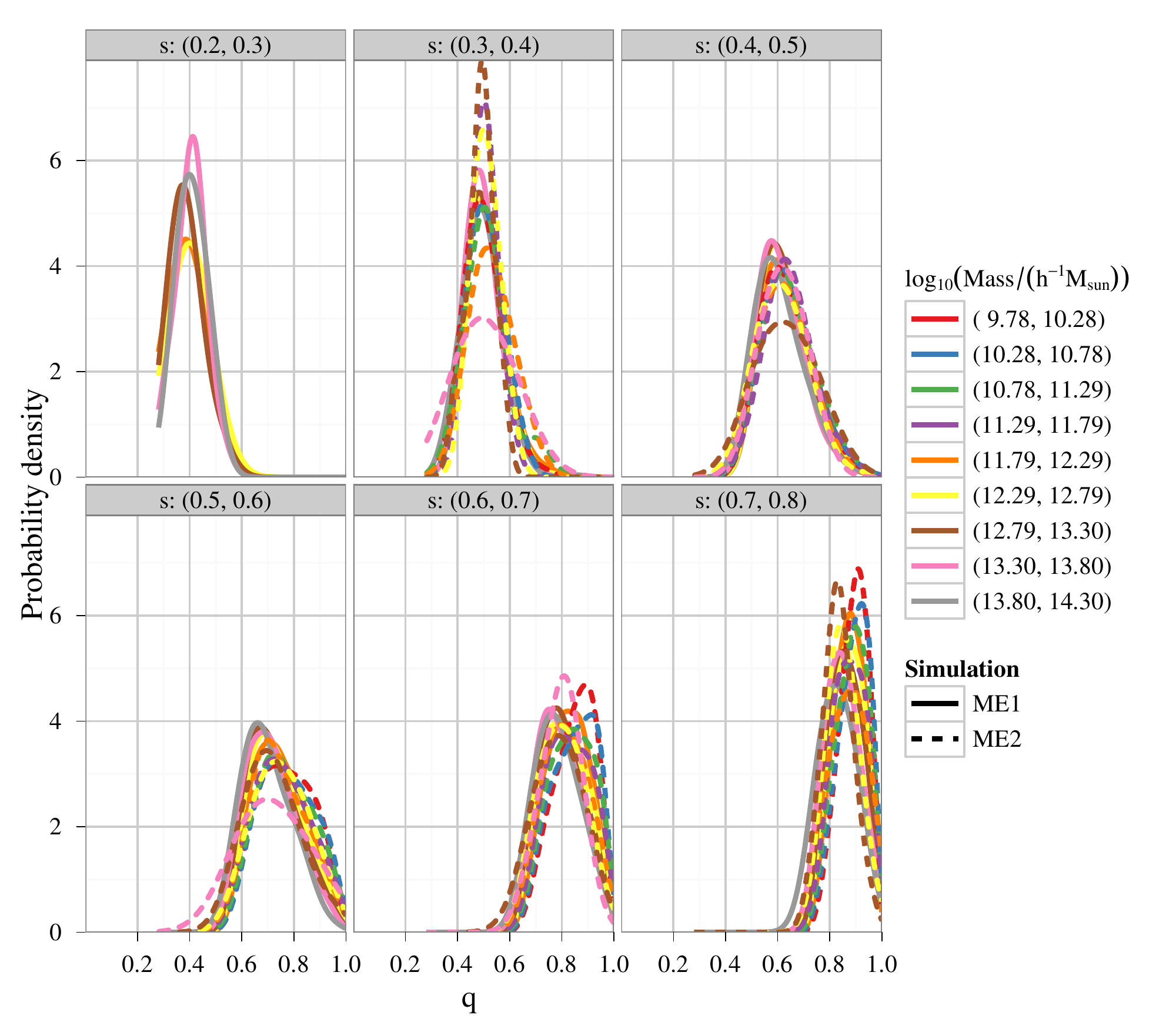}
  }
  \caption{\label{fg:histqconddist}Conditional distributions of intermediate-to-major 
  axis ratios of the triaxial halo shapes given the minor-to-major axis 
  ratio, $s$, at $r_{200}$ and $z=0$.
  The panels indicate bins in $\log_{10}(M_{200} / h^{-1}M_{\odot})$.}
\end{figure}

We fit the marginal distributions for $\sscale$ shown in figure~\ref{fg:sscalehist} with the same beta distribution (i.e. independent of mass), 
\begin{equation}\label{eq:scaledsdistfit}
  p(\sscale; \alpha_s, \beta_s) \propto \sscale^{\alpha_s-1}\,\left(1-\sscale\right)^{\beta_s-1},
\end{equation}
where we find $\alpha_s\approx 14.3 - 2.9 z$ and $\beta_s\approx10.4 - 1.8z$ as functions of redshift $z$.
To describe the joint probability distribution for $s$ and $q$ we again 
follow~\cite{2002ApJ...574..538J} and 
plot the conditional distribution $p(q|\sscale)$ 
in figure~\ref{fg:histqconddist}, which then gives us $p(\sscale,q) = p(\sscale)\,p(q|\sscale)$.  
The conditional distributions for $q$ naturally become narrower with increasing $s$ because of 
the definition $q \ge s$.
The distributions appear to be largely independent of halo mass as well as redshift (not shown).  
We again fit the conditional distributions of $q$ given $s$ with beta distributions as in 
equation~\ref{eq:scaledsdistfit}. But, we first transform $q$ onto the unit interval with the transformation $q'\equiv \frac{q-s}{1-s}$ to have the same support as the beta distribution. We find $s$-dependent, but redshift independent, fit parameters for conditional distributions of $q'$,  
$\alpha_{q'|s}(s) \approx 72 - 230s + 222s^2$ and
$\beta_{q'|s}(s) \approx 1.6 s^{-2.2}$.

\subsection{Halo morphologies and orientations}
\label{sec:orientations}
\begin{figure}[htpb]
  \centerline{
  	\includegraphics[scale=0.6]{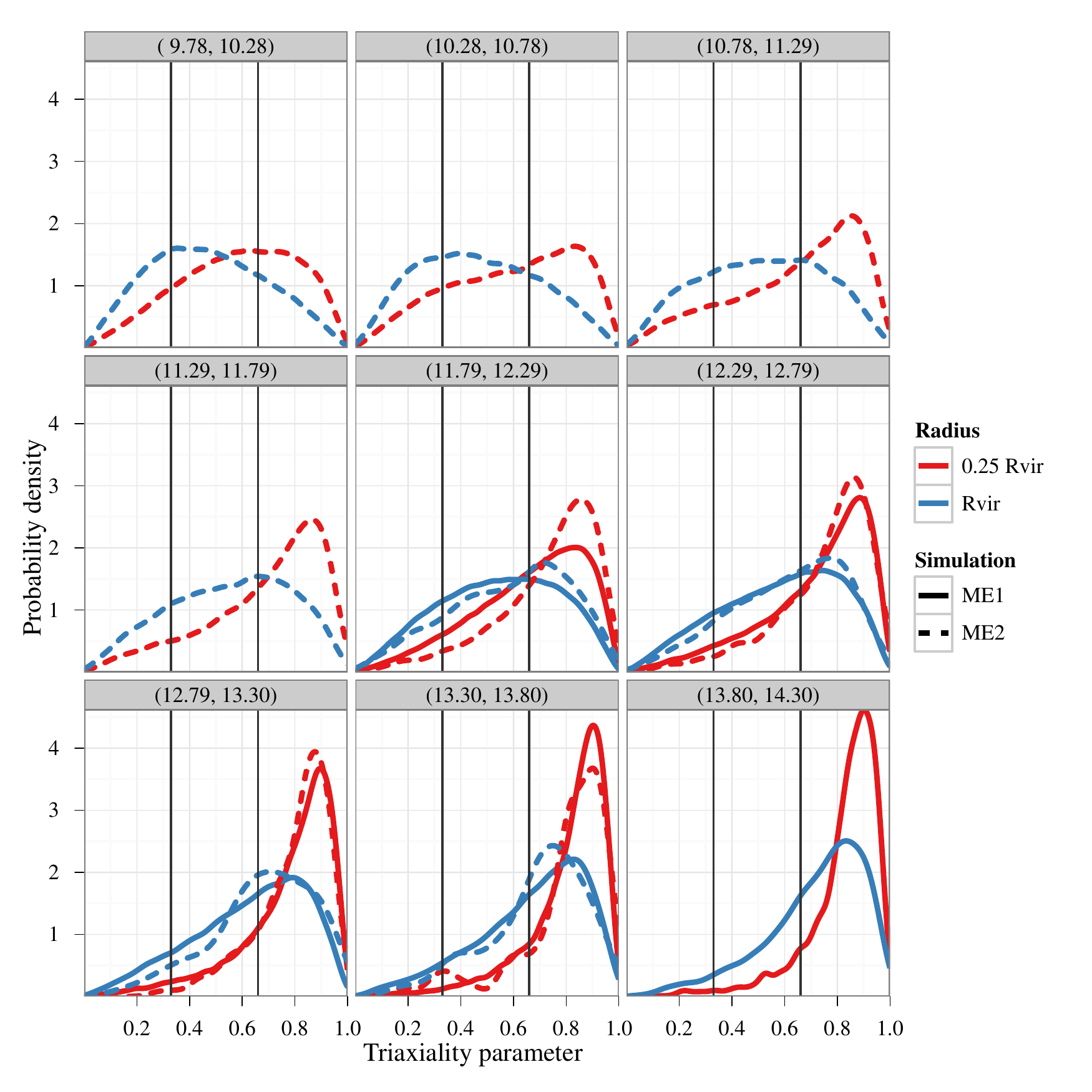}
  }
  \caption{\label{fg:triaxparameter}Triaxiality parameter, $T\equiv(a^2-b^2)/(a^2-c^2)$ at $z=0$.
  The panels indicate bins in $\log_{10}(M_{200} / h^{-1}M_{\odot})$.
  $T\lesssim 0.33$ indicates an oblate ellipsoid while $T\gtrsim0.66$ indicates a prolate ellipsoid.}
\end{figure}

The halo ``triaxiality parameter''~\citep{1991ApJ...383..112F, 2006MNRAS.367.1781A}, 
\begin{equation}
	T\equiv(a^2-b^2)/(a^2-c^2),
\end{equation}
is a convenient method of classifying the triaxial halo morphologies.  We show the distributions of the triaxiality parameter as a function of halo mass at $z=0$ in figure~\ref{fg:triaxparameter}.  
A $T \lesssim 0.33$ indicates an oblate ellipsoid, $T \gtrsim 0.66$ indicates a prolate ellipsoid, while triaxial ellipsoids have $0.33 \lesssim T \lesssim 0.66$.  
We confirm previous results that the halos become more prolate with increasing 
halo mass~\cite{2006MNRAS.366.1503P} and also see that the shapes of the 
halos at small radii ($0.25\rvir$) are more prolate than the 
shapes at $\rvir$.  The distribution of the triaxiality parameters at higher redshifts are 
qualitatively similar to figure~\ref{fg:triaxparameter}, although the distributions at all 
masses and radii become slightly more skewed towards one (prolate) as redshift increases.
We also note that for halos with masses $\gtrsim 10^{12.3}h^{-1}M_{\odot}$ it appears 
to be a good approximation to model the halos as prolate, so the shapes can be adequately 
described by the single minor-to-major axis ratio. We will make use of this approximation 
to simplify some measurements of the halo shape correlations in section~\ref{sec:alignments}.

\begin{figure*}[htpb]
 \centerline{
 	 \includegraphics[scale=0.75]{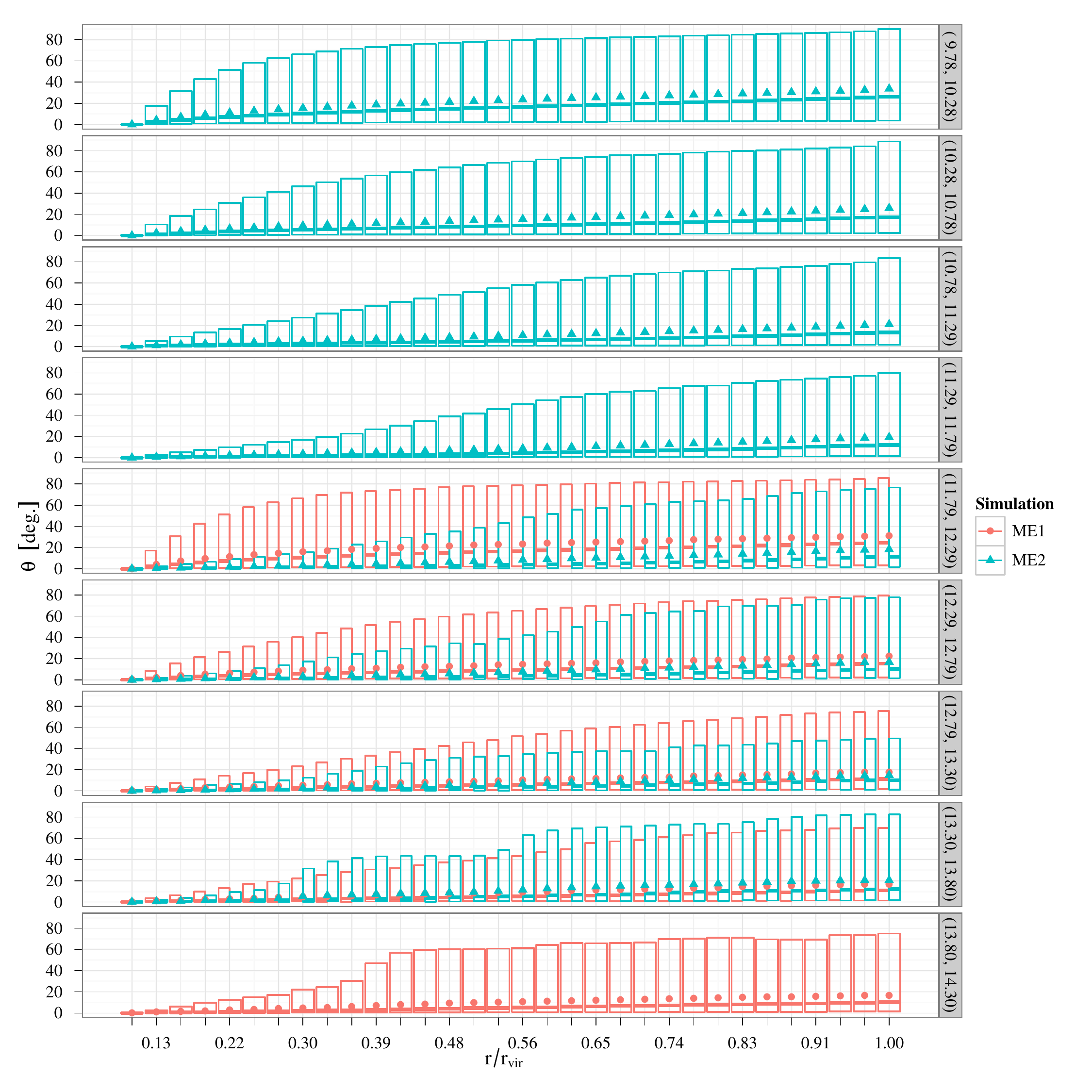}
 }
 \caption{\label{fg:axisalignments}Angle between the major axes at different radii within 
 individual halos at $z=1/2$.
  Points indicate mean angles at a given radius while the boxes show the 25-75\% central quantile range with the median of the distribution denoted by a horizontal line within 
  each box.
  Each panel shows a different mass range with labels $\log_{10}(M_{200} / h^{-1}M_{\odot})$.
  The halo radius on the abscissa is normalized by $r_{\rm vir}\equiv \rvir$ (not $r_{200}$).}
\end{figure*}

In figure~\ref{fg:axisalignments} we plot the distributions of the angle between the 
halo major axis at $0.1\rvir$ and the major axis at 32 logarithmically spaced 
larger radii out to $\rvir$.  The 
central lines of the boxes in figure~\ref{fg:axisalignments} denote the median of the angles 
for all halos in each simulation while the box upper and lower edges denote the first and 
third quartiles of the angle distributions.  
It is interesting to see that the angle distributions for the two simulations are in good agreement 
in the largest mass bin (bottom panel) but are in larger disagreement as the halo mass decreases.  
We believe this is an artifact of the limited mass resolution in the Millennium-1 simulation wherein
some substructures are not fully resolved by the SubFind algorithm and are included as part of the 
main halo, leading to larger scatter in the halo shapes at different radii.  In the higher resolution 
Millennium-2, more substructures are resolved and removed by SubFind, so that the parent halo 
has less twisting of the major axes as the radius increases.
This is an important effect to note when modeling the alignments of BCG galaxies added to $N$-body 
simulations as in~\cite{2009RAA.....9...41F, 2009ApJ...694L..83O}. (We will return to this 
issue in section~\ref{sec:projcorr}.)

Because we have used the reduced inertia tensor (defined in equation~\ref{eq:redinertiatensor})
to compute the halo shapes, we expect that our measurements of the misalignment angles of halo 
major axes at different radii will be skewed more towards zero than if the un-reduced 
inertia tensor,
\begin{equation}
	I^{{\rm un-red}}_{ij} \equiv \sum_{n=1}^{\nparthalo} x_{n,i}x_{n,j},
\end{equation}
were used to measure the shapes. This is because the reduced inertia tensor puts more 
weight on particles at small radii and 
is therefore more sensitive to the cumulative mass distribution in the halo. The choice of inertia 
tensor definition should therefore be kept in mind when comparing our results in 
figure~\ref{fg:axisalignments} with other works. 

The wide distributions of angles in figure~\ref{fg:axisalignments} is also remarkable.  
Because we have mapped all angles onto the interval $(0,90]$ degrees, the distributions of angles 
are highly skewed towards zero (indicated by the differences between means and medians).  
However, for radii $\gtrsim0.5\rvir$ the distributions have large tails showing that $\sim$25\% of halos 
have outer major axes nearly perpendicular to the major axes at $0.1\rvir$. 
To check that errors in the measurement of the halo orientation at $0.1\rvir$ have not skewed 
the distributions in figure~\ref{fg:axisalignments}, we recomputed the alignment angles relative to the 
axes at $\rvir$ and confirmed that the relative alignment angles at different radii are robust to the 
reference radius.
The wide distributions of alignment angles in figure~\ref{fg:axisalignments} illustrate 
the necessity of simulations that can accurately resolve the inner regions of 
halos when modeling galaxy intrinsic alignments for analyzing weak lensing surveys.
It is also possible that further investigations into these axis alignment distributions could show that 
the Milky Way is not unusual in having a distribution of satellites in a plane perpendicular to 
its disk~\cite{2009MNRAS.394.2223M}. 

\section{Halo alignments}
\label{sec:alignments}
To describe the alignments of the principal axes of different halos we measure the halo-halo 
and halo-mass 
3D correlation functions binned in both the separation distance and the angles between either the halo major 
axes or the halo major axis and the separation vector to a tracer of the mass density (which we 
will call the ``alignment angles'').
Because the alignment effect is weak compared to the spatial clustering of the halos, we 
present most of our measurements in terms of the quantity,
\begin{equation}\label{eq:excessalignment}
	\excor(r,\theta)\equiv\frac{1+\xi(r,\theta)}{1+\xi(r)},
\end{equation}
which gives the excess probability to find a halo or mass overdensity within a distance $[r,r+dr]$ from 
a given halo and with the alignment angle within
$[\theta, \theta+d\theta]$. 
Note that,
\begin{equation}
  \xi(r) \equiv \int_{0}^{\pi} \sin\theta\, d\theta\, \xi(r, \theta) ,
\end{equation}
so that the halo-halo or halo-mass correlation function $\xi(r)$ is the angle average of the
alignment correlation function $\xi(r, \theta)$.
When reporting our measurements we map all measured $\theta$ values onto the interval $(0, \pi/2)$ with 
the assumption that $\xi(r,\theta) = \xi(r, \pi-\theta)$ due to the symmetry of the triaxial halo ellipsoids.
We also ignore any potential dependence of the halo alignment correlation function on the 
orientations of the minor and intermediate axes.

Another motivation for measuring the ratio defined in eq.~(\ref{eq:excessalignment}) 
comes from \cite{2011JCAP...05..010B} who recently showed that within the linear alignment model that assumes 
halo orientations align with the gradients of the large-scale gravitational potential, 
\begin{equation}
  \excor(r, \theta_p) = 1+\sqrt{\frac{\pi}{2\left<\gamma_{+}\gamma_{+}\right>}}
  \cos(2\theta_p)
  \frac{\left<\delta\gamma_{+}\right>(r)}{1+\xi(r)},
\end{equation}
where $\theta_p$ is the angle between the apparent major axis projected on the sky and the 
projected separation vector to the mass overdensity $\delta$, and 
$\gamma_{+}=\left(1-s_p^2\right) / \left(1 + s_p^2\right)\cos2\theta_p$ with $s_p$ the ratio of the projected 
minor and major axes.
While this is defined in terms of the projected halo shape and orientation it gives us a reference
for interpreting the radial and $\theta$ dependence of $\excor$ as defined in 
eq.~(\ref{eq:excessalignment}) and is directly related to eq.~(\ref{eq:excessalignment}) when the 
halo pairs are in the plane of the sky.

\begin{figure*}[htpb]
  \centerline{
  \includegraphics[scale=0.45]{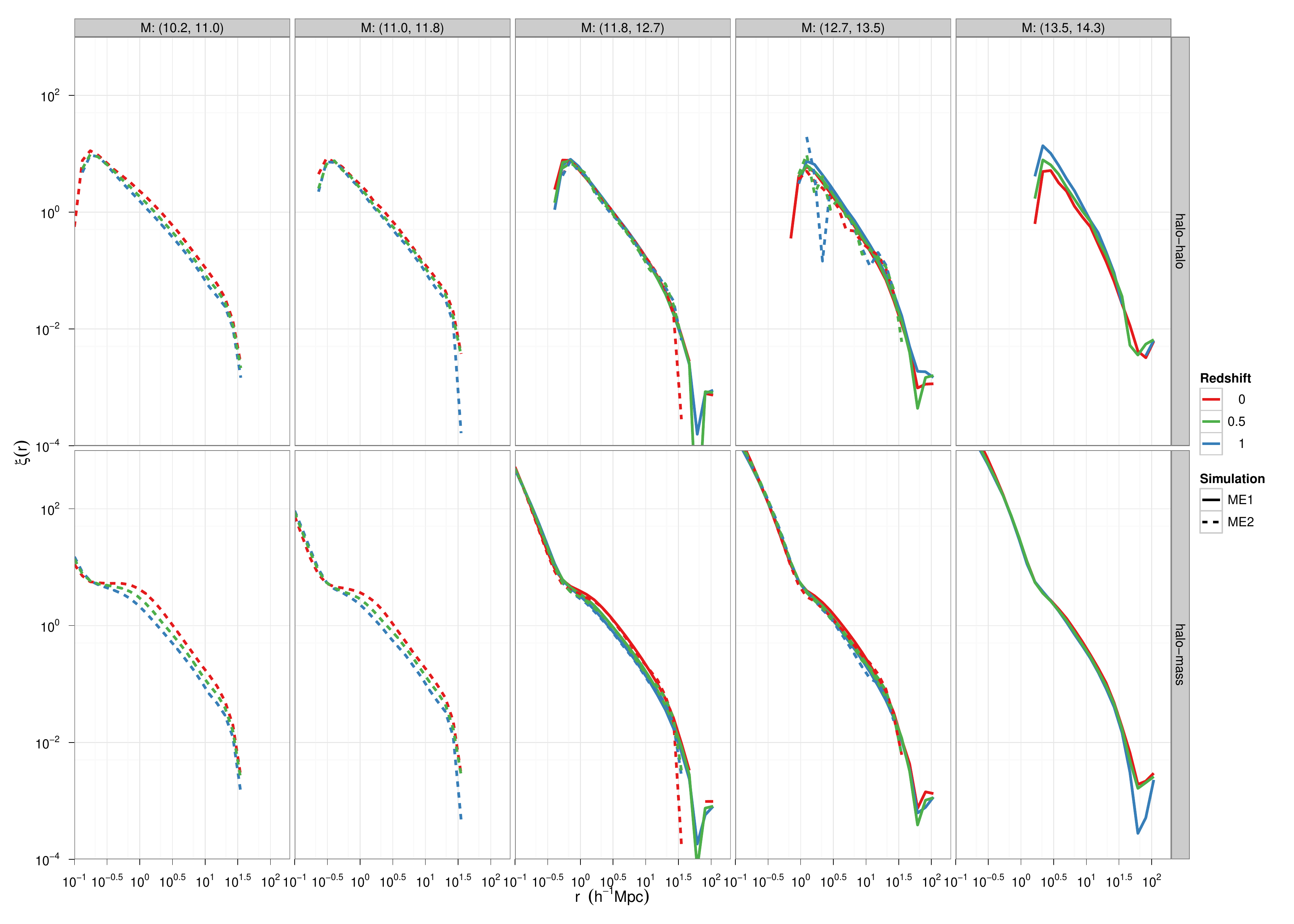}
  }
  \caption{\label{fg:corrfcns} Halo-halo and halo-mass correlation functions in 
  halo mass bins (units are $M_{\odot}/h$) at redshifts 0, 0.5, and 1.
  The points where the halo-halo correlation functions turn over at small separations 
  clearly the show the halo exclusion scales.  
  As halo mass increases the large-scale correlations become more similar for all $z$ because of the
  increasing halo bias with redshift.
}
\end{figure*}

We show measurements of the halo-halo and halo-mass position correlation functions in mass bins, 
$\xi\left(r; M_{1}, M_{2}=M_{1}\right)$, in figure~\ref{fg:corrfcns}~\cite[see also][]{2008MNRAS.388....2H}.  
The correlation functions are in good agreement between the two 
simulations in all overlapping mass bins from the minimum measured radii out to 
$r\sim 20 h^{-1}$Mpc where the correlation functions in the Millennium-2 simulation begin to 
show artifacts due the imposition of the integral constraint in the 100~$h^{-1}$Mpc simulation 
volume.
In both rows of panels the separation of the correlation functions at different redshifts 
is clearly visible in the lowest mass bin due to the linear growth.  For larger halo mass bins 
the increasing halo bias with redshift compensates for the decreasing growth and the correlations 
have similar amplitudes at all three measured redshifts. For the highest mass bin in the 
halo-halo correlation function (top-row, right panel), the effects of the nonlinear halo bias 
can be seen at small $r$ where the amplitudes of the correlation functions are 
increasing with increasing redshift.
At small-radii in the top row of panels in figure~\ref{fg:corrfcns}, the correlation functions 
turn towards zero due to halo exclusion (so the turn-over scale increases with increasing halo mass).
In the bottom panels of figure~\ref{fg:corrfcns} the small-scale correlation instead show the 
correlations of the halos with the mass density interior to the halo (with the scale again 
set by the mass-dependent halo radii).  
We will use the correlation functions in figure~\ref{fg:corrfcns} as references for 
assessing the effects of halo alignments in the next section.

\subsection{Angle-binned correlation functions}
\label{sec:anglebinnedcorrelations}
\begin{figure}[htpb]
  \centerline{
  \includegraphics[scale=0.4]{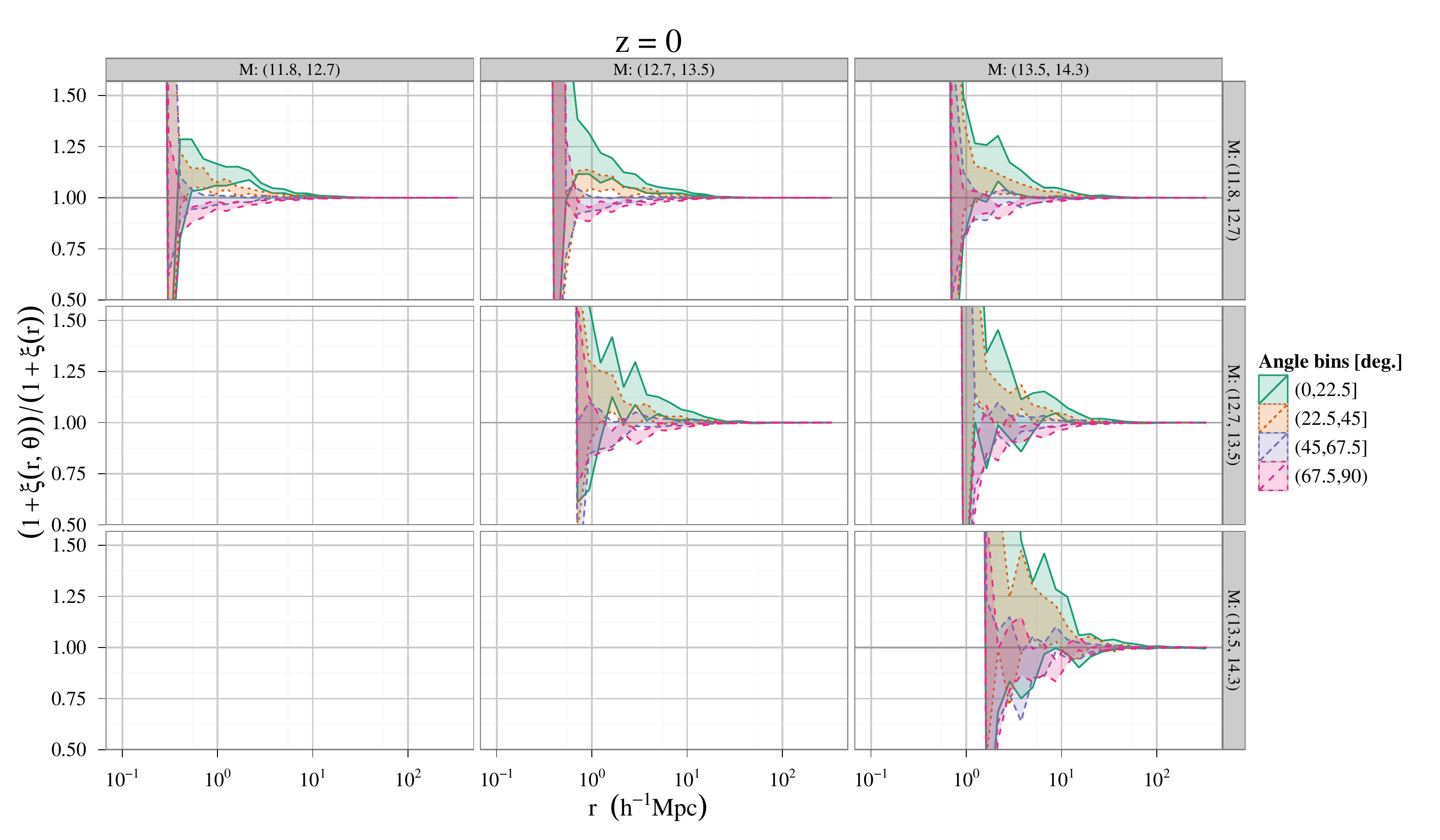}
  }
  \centerline{
  \includegraphics[scale=0.4]{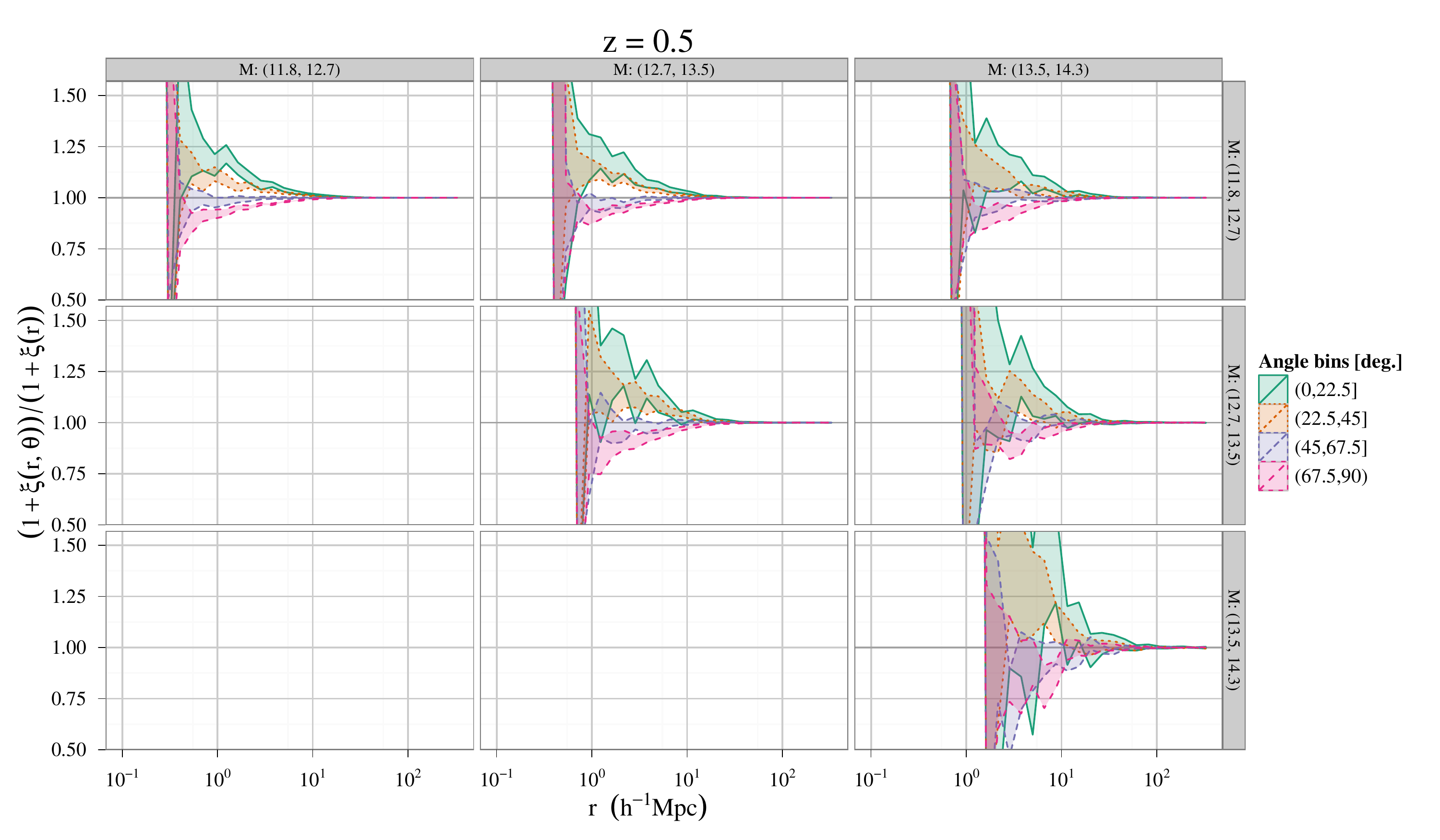}
  }
   \centerline{
  \includegraphics[scale=0.4]{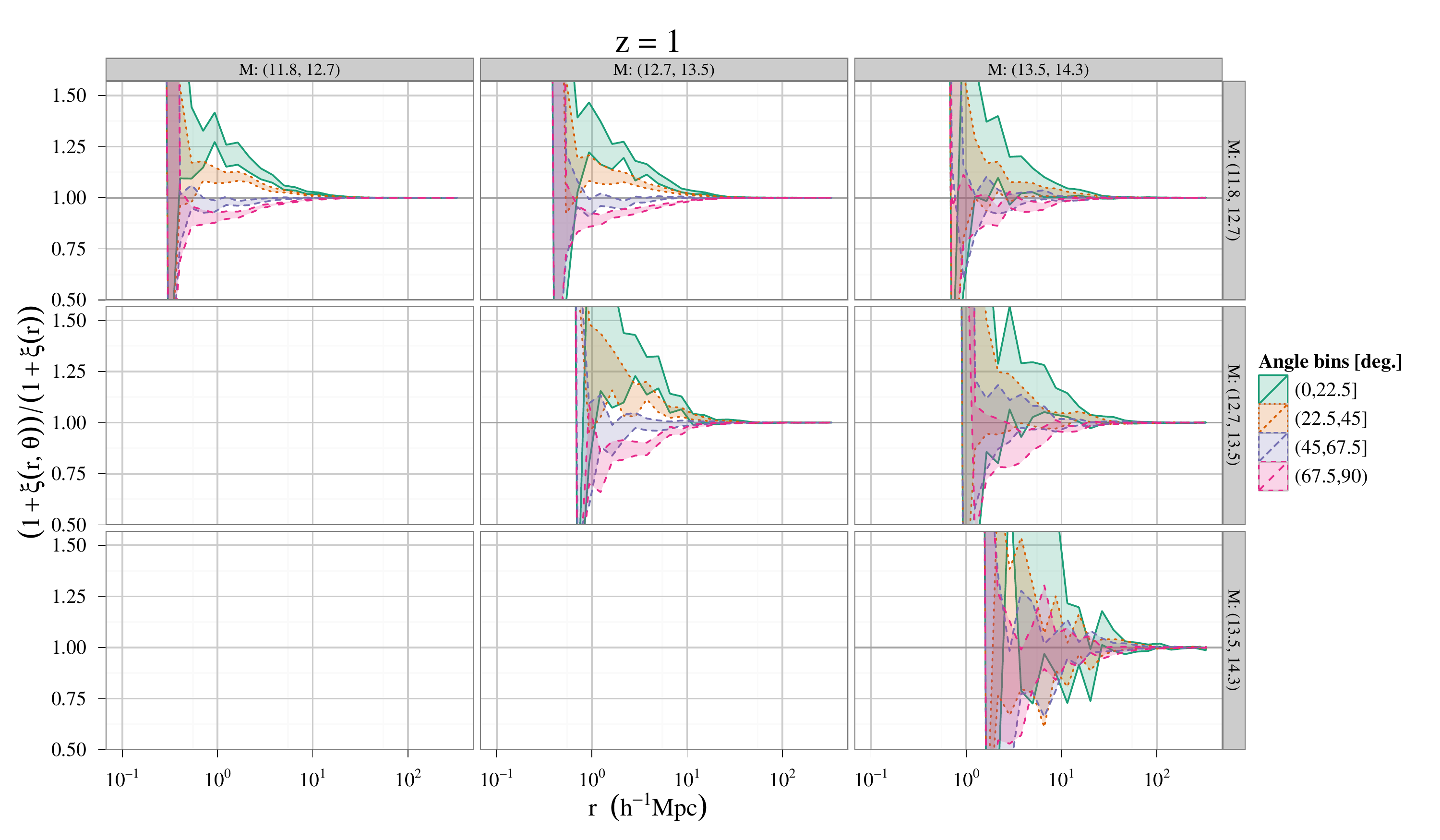}
  }
  \caption{\label{fg:thetacorrouterouter}Halo-halo $\excor\left(r,\theta; M_{1}, M_{2}\right)$ 
  versus $r$ for bins in mass, $\log_{10}(M_1)$ and $\log_{10}(M_2)$, and the angle, $\theta$, between the 
  major axes of two halo shapes at $r=\rvir$ 
  as measured in the Millennium-1 simulation at $z=0,0.5,1$ from top to bottom.
  (The Millennium-2 measurements have larger uncertainties and are consistent with 
  no alignments.)
  The shaded bands denote the 95\% confidence intervals determined from 100 fixed block
  bootstrap samples.
}
\end{figure}

In figure~\ref{fg:thetacorrouterouter} 
we show measurements of $\excor\left(r,\cos\theta;M_{1},M_{2}\right)$, 
with $\theta$ the angle between 
the major axes of two halos in a pair and where the halo shape is measured at the $\rvir$.
The error bars are estimated using fixed block bootstrap samples of the simulation volumes 
following the ``marked-point bootstrap'' method of~\cite{2008ApJ...681..726L}.  
In the marked-point bootstrap algorithm each halo is assigned an array of ``mark'' 
values with each array entry equal to the number of halos (or mass tracers) in a given radius and angle bin 
relative to the first halo (so the mark array has length equal to the number of radius bins times 
the number of angle bins). 
We then divide each simulation volume into 64 equal-volume sub-cubes with fixed spatial locations 
and resample the sub-cubes 100 times with replacement to generate 100 realizations of the 
set of mark arrays in each simulation. For each set of resampled marks, we compute the 
correlation functions by summing the selected mark arrays for each halo, 
We then compute the ratio $\excor(r,\cos\theta)$ for each 
resampling and determine errors by sorting the resulting values in each radius and angle bin to  
extract the 95\% confidence intervals from the ordered arrays.

In figure~\ref{fg:thetacorrouterouter}, 
we have plotted only the Millennium-1 measurements in the three highest mass bins 
because the Millennium-2 measurements are noisy in these mass bins and there is no 
significant alignment signal in the two lower mass bins.
However, there is a significant alignment signal that increases with halo mass with $\excor\sim 1.1$ at 
$r\sim 1$~$h^{-1}$Mpc for the intermediate mass bins and the angle bin with $0 \le \theta \le 22$~degrees at z=0.   
The excess halo alignment correlations also increase with redshift giving an excess correlation 
of 1.25 at $r\sim 1$~$h^{-1}$Mpc  and $\theta < 22$~degrees in the middle mass bin at z=1.
In the CDM model we expect that the inner shapes of halos are in place at high redshift. As the evolution 
of the halos proceeds, the filaments feeding mass onto the halos can re-form in new orientations causing 
shifts in the orientations of the halo shapes and decreasing the correlations with surrounding halo orientations.
At high redshift there is also a greater range of length scales in the mass density perturbations collapsing 
at the same time, which also would support larger halo orientation correlations to larger halo 
separations.

\begin{figure*}[htpb]
  \centerline{
  \includegraphics[scale=0.48]{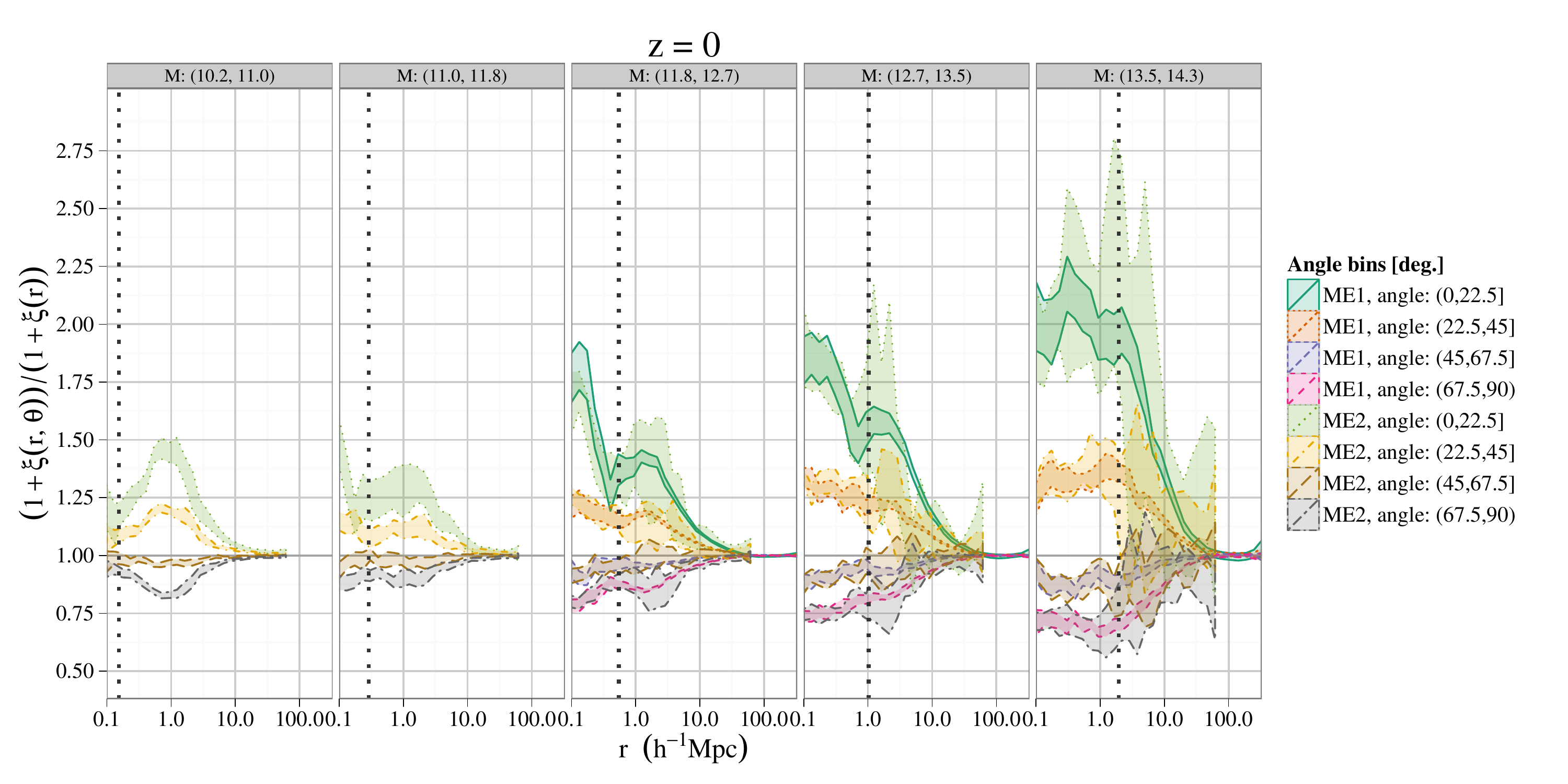}
  }
  \centerline{
  \includegraphics[scale=0.48]{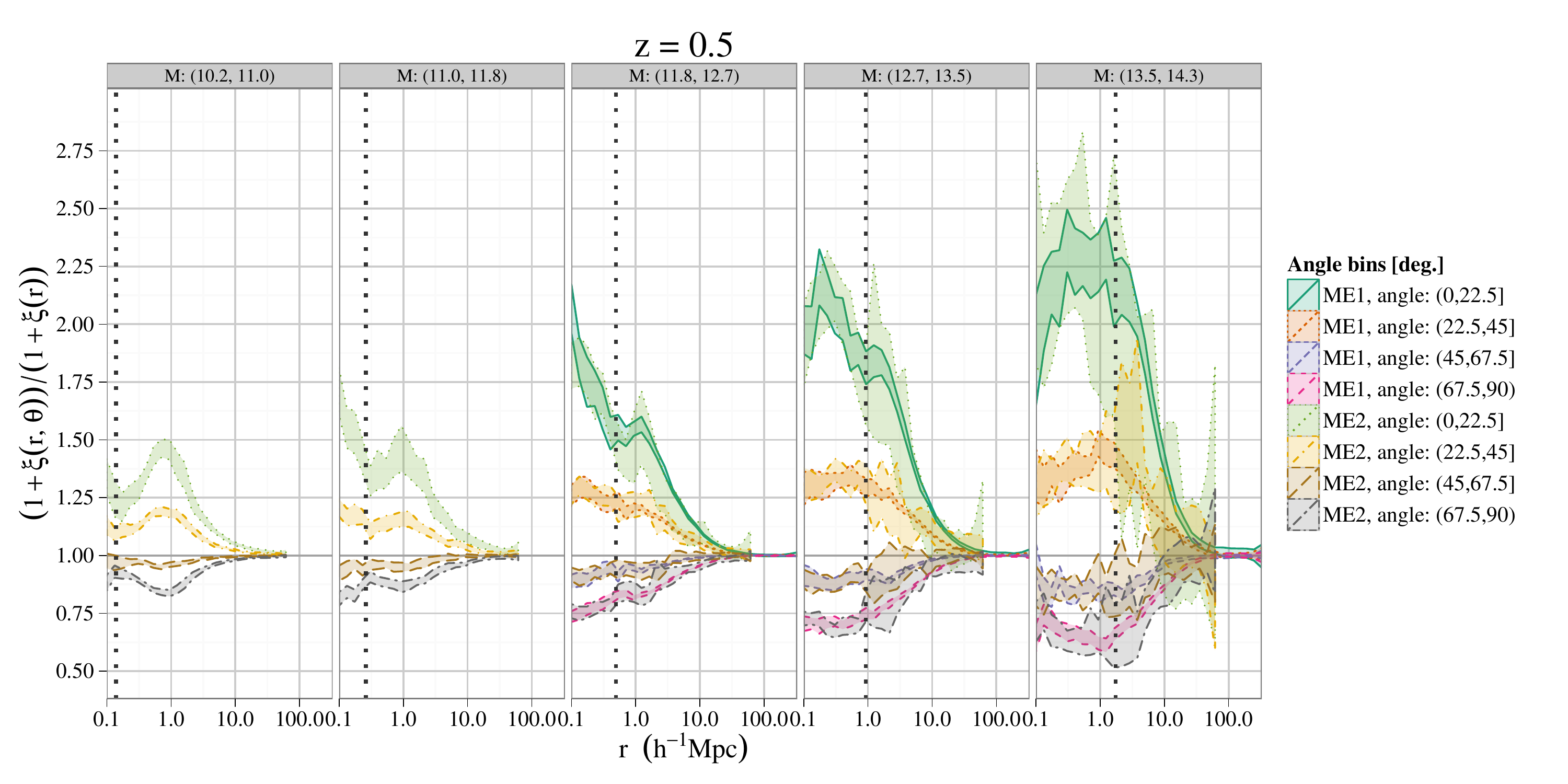}
  }
   \centerline{
  \includegraphics[scale=0.48]{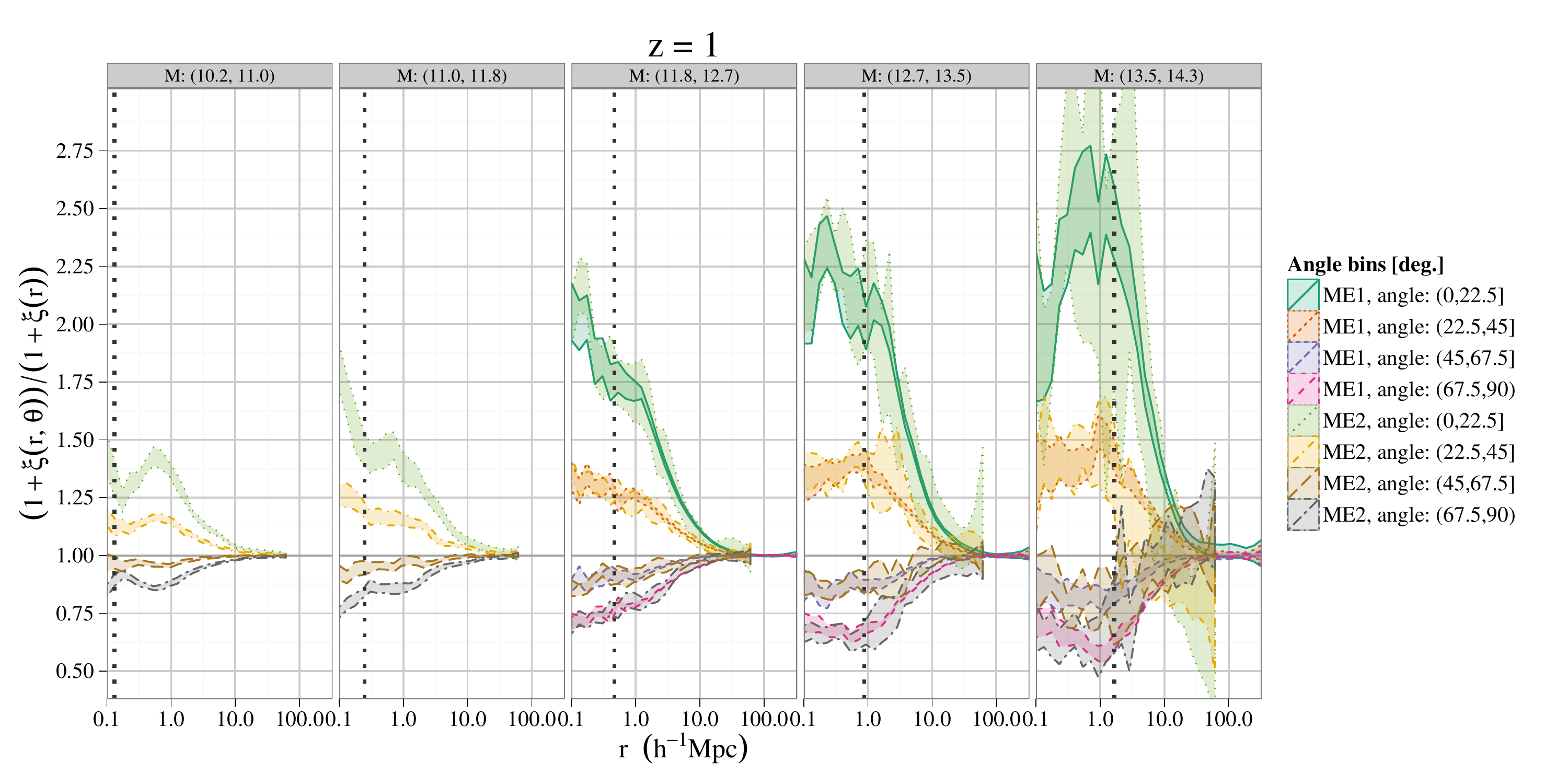}
  }
  \caption{\label{fg:thetacorroutersep}Halo-mass $\excor\left(r,\theta; M_{1}\right)$ 
  versus $r$ for bins in mass, $\log_{10}(M_1)$, and the angle, $\theta$, between the major axes of one halo 
  (measured at $\rvir$) and the position vector to a mass tracer.
  The shaded bands denote the 95\% confidence intervals determined from 100 fixed block
  bootstrap samples.
  The dotted vertical lines show the maximum $\rvir$ value in each mass bin.
}
\end{figure*}

We show the halo-mass cross-correlation functions where $\theta$ is now the angle between the major axis 
of a halo and the vector connecting the halo center to a mass tracer particle in 
figure~\ref{fg:thetacorroutersep}. To compute the cross-correlations we selected a random subset of 
$10^6$ mass tracer particles in each simulation.  
There are very strong alignment signals for the halo-mass cross-correlation in all mass bins 
and both simulations are in good agreement.
If there is any stochasticity between the alignment of halos and their surrounding mass distribution, 
then we would naturally expect the halo-mass excess alignment correlations to be much larger than the 
halo-halo excess alignment correlations.  This is because in the latter case the random component of the 
halo alignments is included twice (once for each halo).
Comparing the relative amplitudes in figure~\ref{fg:thetacorrouterouter} and 
figure~\ref{fg:thetacorroutersep}
would indicate a wide distribution in the angles between the halo major axes and the surrounding 
mass overdensities.

It is also interesting to note in figure~\ref{fg:thetacorroutersep} that the alignment correlations 
are significant out to separations of several tens of megaparsecs for all measured 
halo masses, which is consistent with observations of galaxy intrinsic 
alignments~\citep{2006MNRAS.367..611M, 2007MNRAS.381.1197H}.  
This is somewhat surprising however as the effects of the finite simulation box size are known 
to depress the position-position auto correlation functions on scales $\sim 0.1$ times the box size 
because the correlation functions must satisfy the integral constraint with the simulation volume.

The excess correlations in figure~\ref{fg:thetacorroutersep} are also consistent with 
self-similar evolution at fixed $M / M_{*}(z)$. Using the $M_{*}(z)$ values from 
table~\ref{tab:mstar} the lowest mass bin at $z=1$, the second lowest mass 
bin at $z=0.5$ and the middle mass bin at $z=0$ should represent roughly similar 
ratios of $M/M_{*}(z)$. Indeed we can see in figure~\ref{fg:thetacorroutersep} 
that the peaks in the excess correlations at, e.g.,  
$r\sim1$~$h^{-1}$Mpc have consistent amplitudes across the three redshift measurements.

\begin{figure*}[htpb]
  \centerline{
  \includegraphics[scale=0.41]{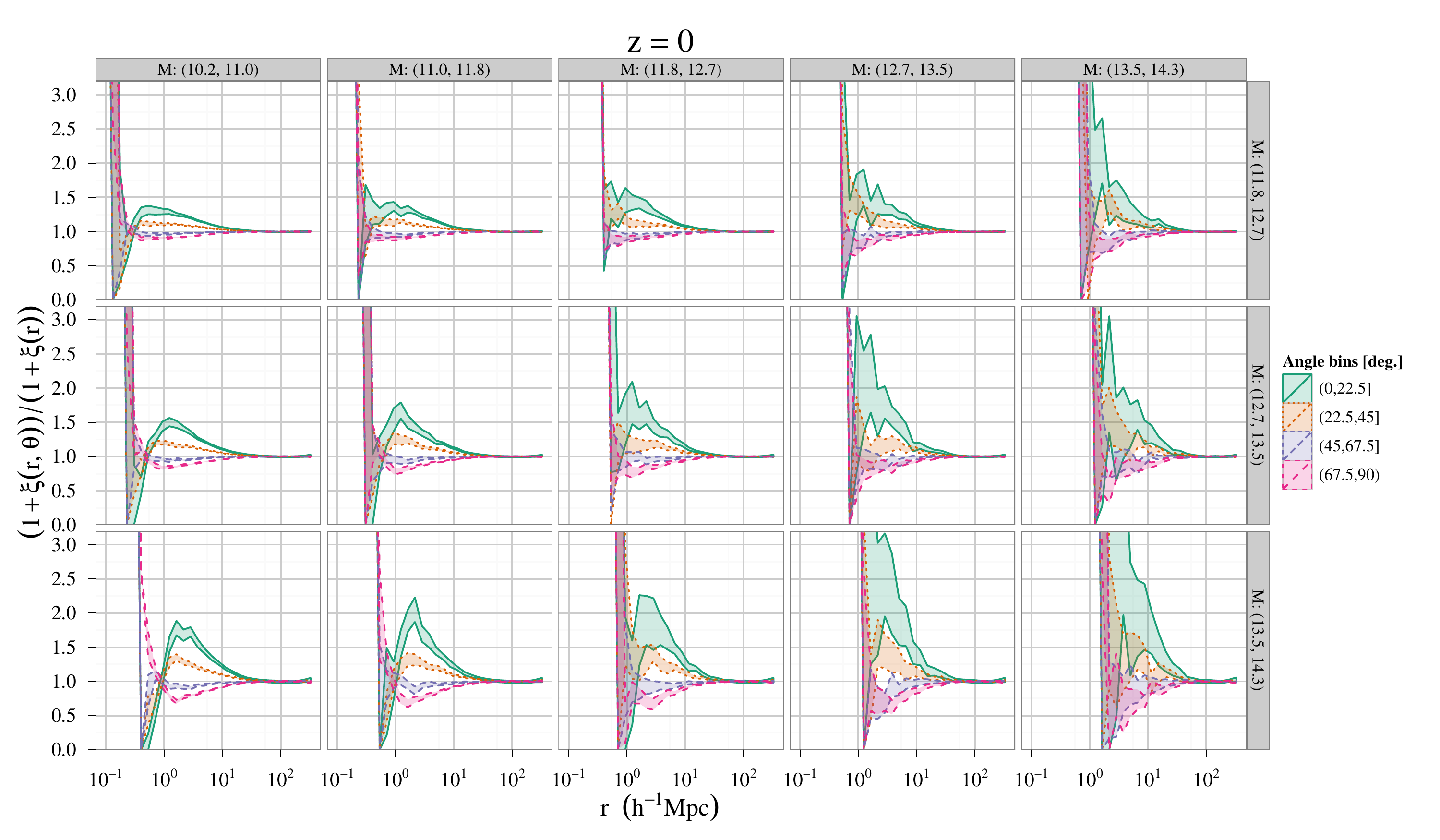}
  }
  \centerline{
  \includegraphics[scale=0.41]{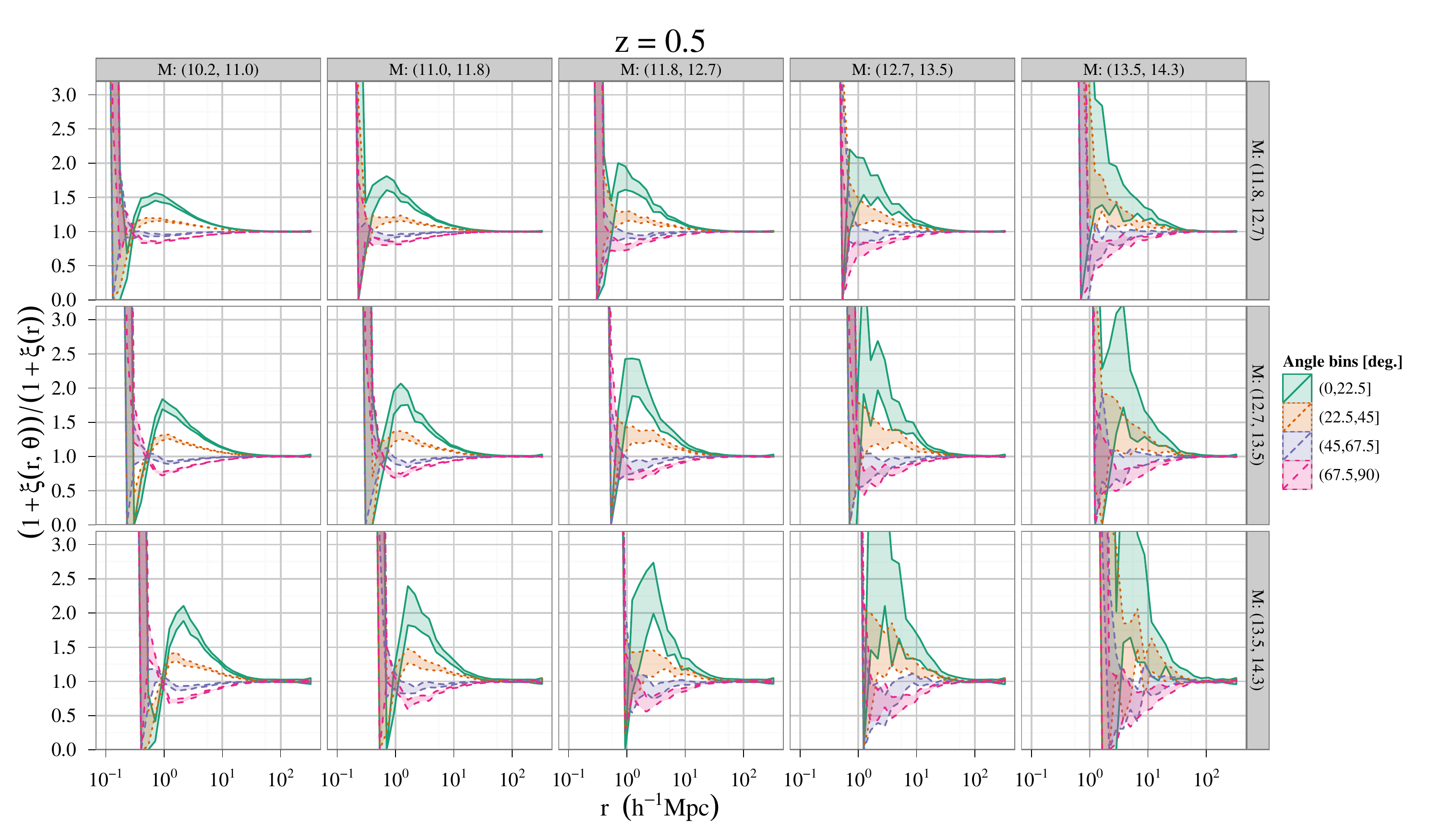}
  }
   \centerline{
  \includegraphics[scale=0.41]{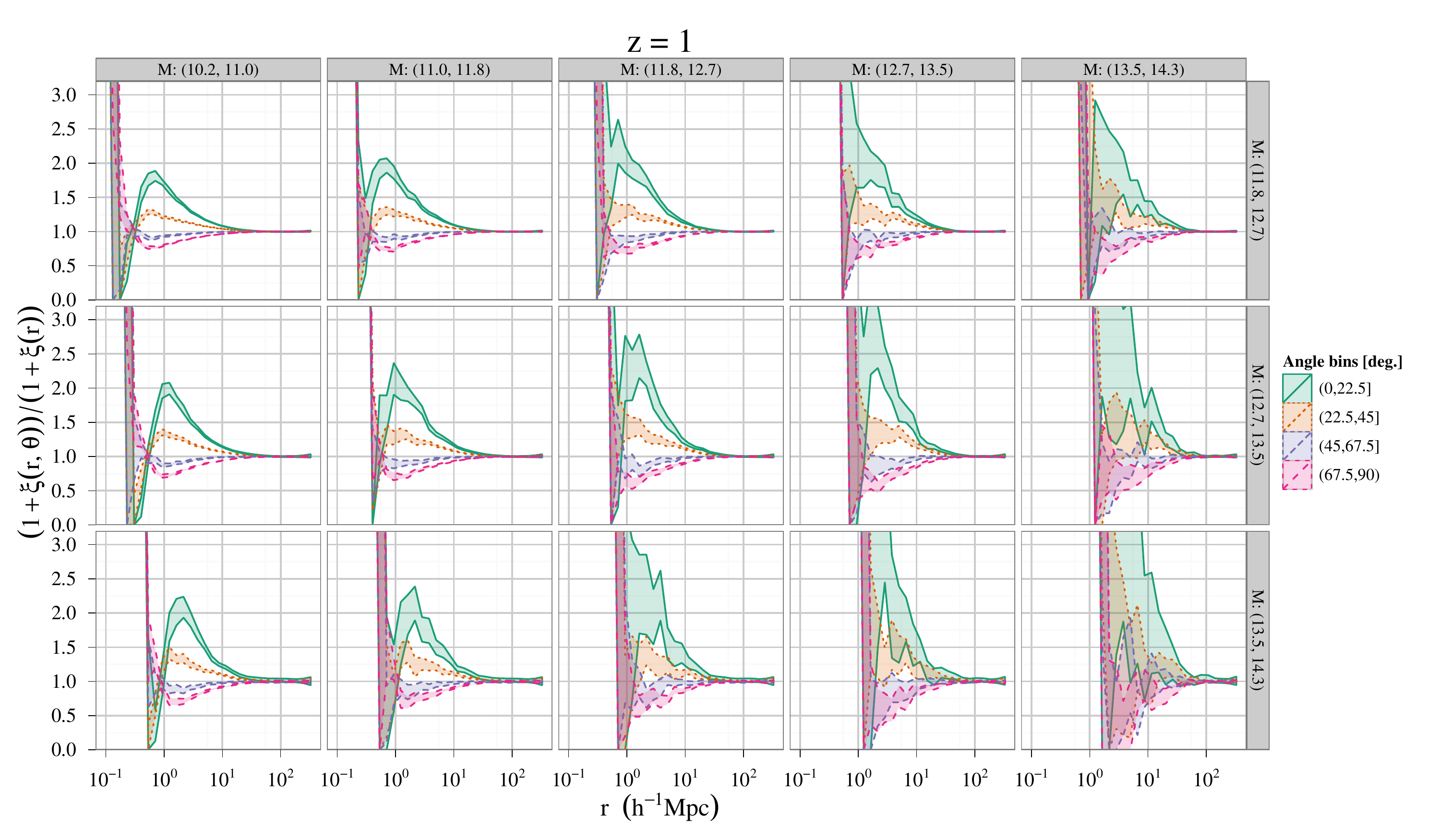}
  }
  \caption{\label{fg:thetacorroutershapehalo} $\excor\left(r,\theta; M_{1}, M_{2}\right)$ where 
  $\theta$ is the angle between the major axis of one halo (measured at $\rvir$) 
  and the separation vector to another halo. 
  The rows of panels show the mass bins of the halos with the shape measurements.
  The columns of panels show the mass bins of the halos used as mass density tracers.  
  The three plots show results for redshifts $z=0,0.5,1$ from top to bottom.}
\end{figure*}

To better understand the peaks in the excess correlations around $r=1$~$h^{-1}$Mpc in 
figure~\ref{fg:thetacorroutersep} we show similar correlation functions 
(for Millennium-1 only) in 
figure~\ref{fg:thetacorroutershapehalo} where the mass tracers are now SubFind-0 halos rather 
than mass particles as in figure~\ref{fg:thetacorroutersep}.
The separate rows of panels in figure~\ref{fg:thetacorroutershapehalo} show mass bins in the 
halos whose alignments are measured while the columns of panels show mass bins in those halos 
that are used as mass tracers.
Because we are now measuring the correlations of halo alignments with halo positions in 
figure~\ref{fg:thetacorroutershapehalo} the correlation functions drop to zero at small radii 
just as in figure~\ref{fg:thetacorrouterouter}. For scales larger than the typical halo radii 
the excess correlations have similar shapes in figures~\ref{fg:thetacorroutersep} and 
\ref{fg:thetacorroutershapehalo} but the amplitudes are larger in figure~\ref{fg:thetacorroutershapehalo},
possibly because of the halo bias. 
At radii smaller than the peaks in the excess correlations in figure~\ref{fg:thetacorroutershapehalo},
the excess correlations quickly change sign (about 1) before hitting the halo exclusion radius.  
This is because the radius of the halo is larger along the halo major axis. So for radii $\lesssim$ the 
typical halo major axis lengths there is greater excess probability to find a neighboring halo  
perpendicular to the major axis.
The intermediate peaks around $r=1$~$h^{-1}$Mpc in figure~\ref{fg:thetacorroutersep} are 
therefore where the two-halo excess correlations reach a maximum before changing sign.

In appendix~\ref{sec:extracorrs} we show the alignment correlation functions 
similar to those in figures~\ref{fg:thetacorrouterouter}, 
\ref{fg:thetacorroutersep}, and \ref{fg:thetacorroutershapehalo} but with the 
halo shapes measured at $0.1\rvir$ rather than $\rvir$.  The amplitudes 
of the excess correlations are consistently smaller when the shapes are measured at 
$0.1\rvir$ as expected if there is stochastic misalignment between the inner and 
outer halo shapes.

\subsection{Axis ratio correlations}
\label{sec:axisratiocorrelations}
\begin{figure*}[htpb]
  \centerline{
  \includegraphics[scale=0.48]{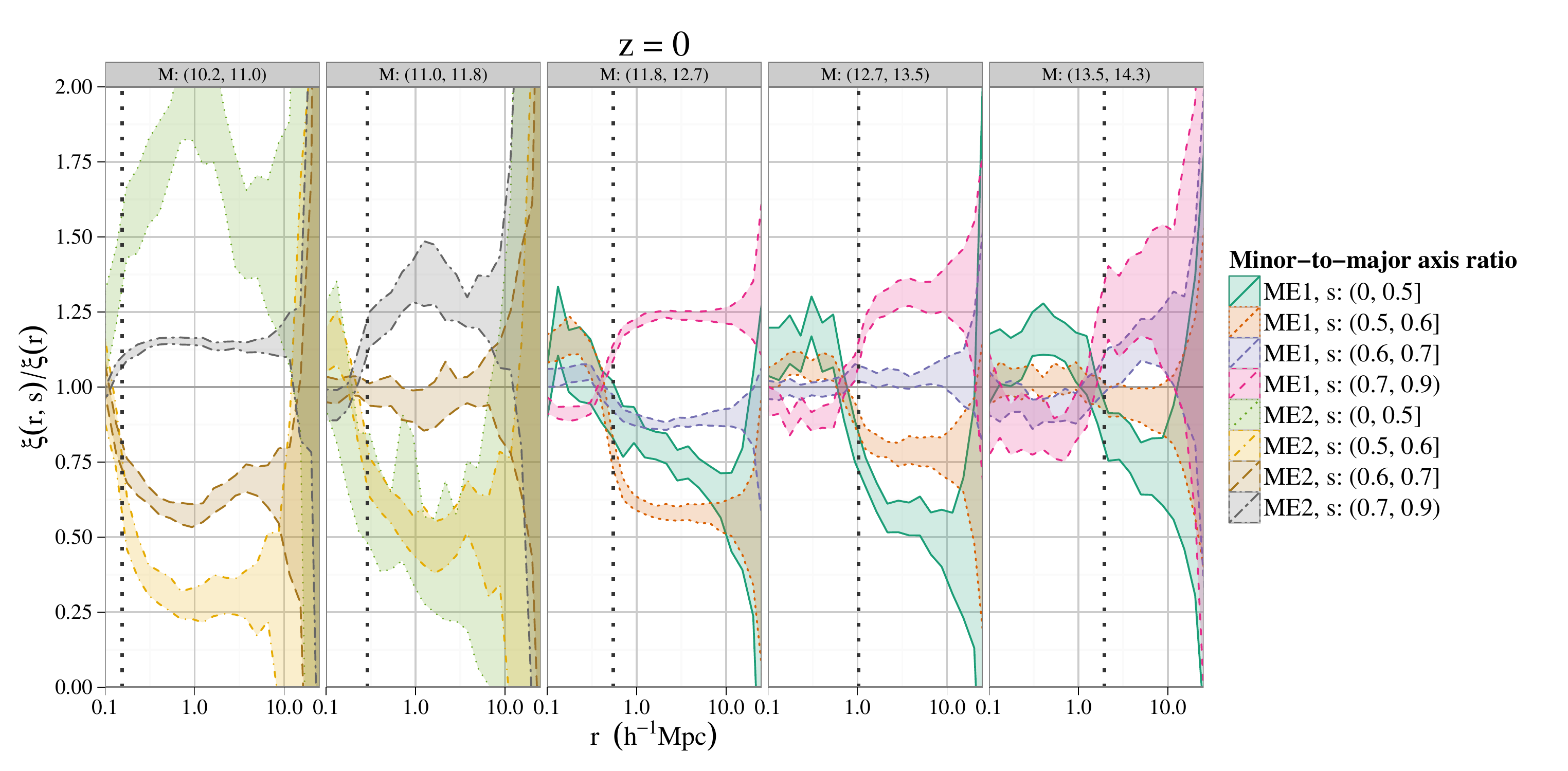}
  }
  \centerline{
  \includegraphics[scale=0.48]{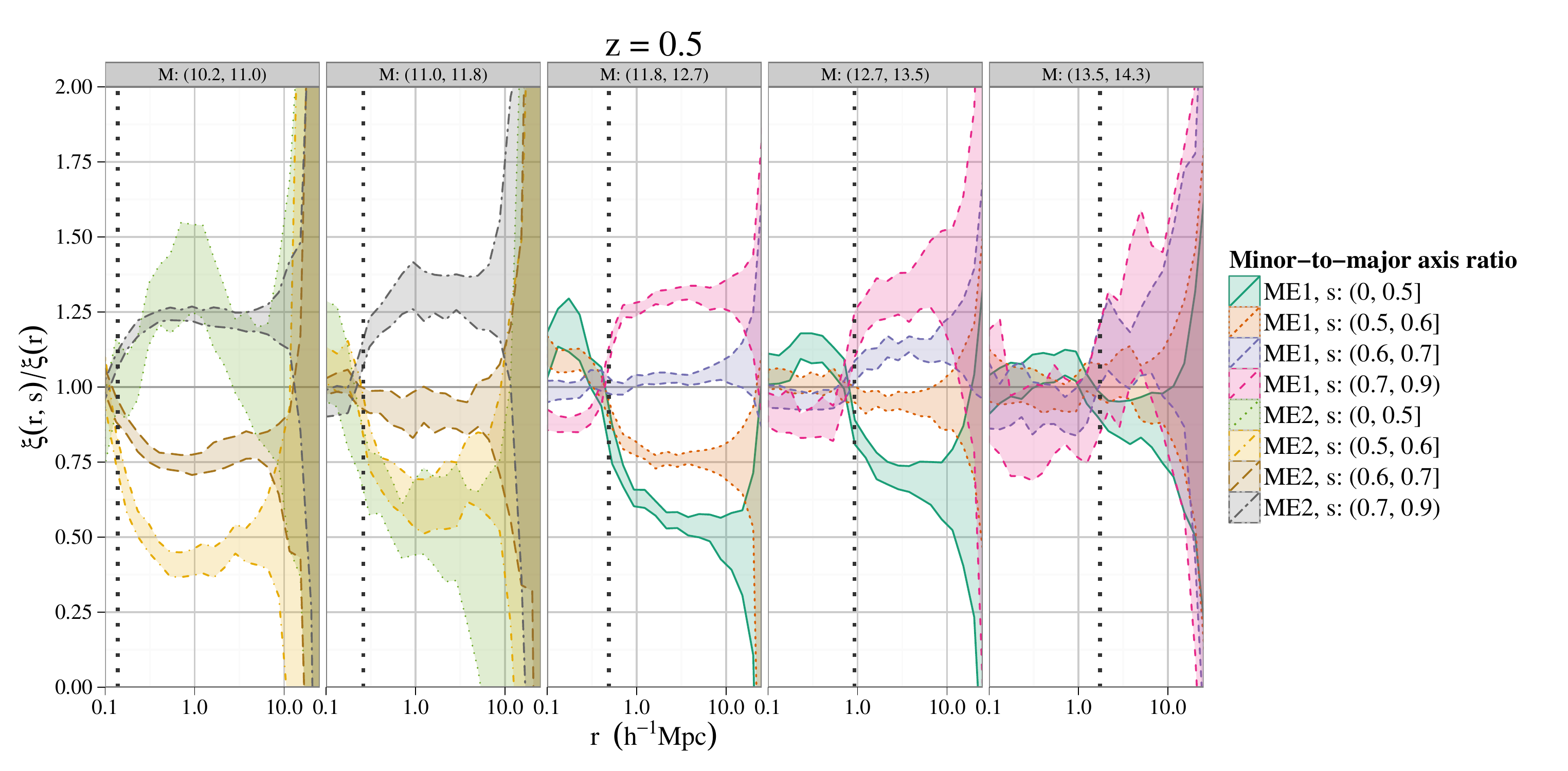}
  }
  \centerline{
  \includegraphics[scale=0.48]{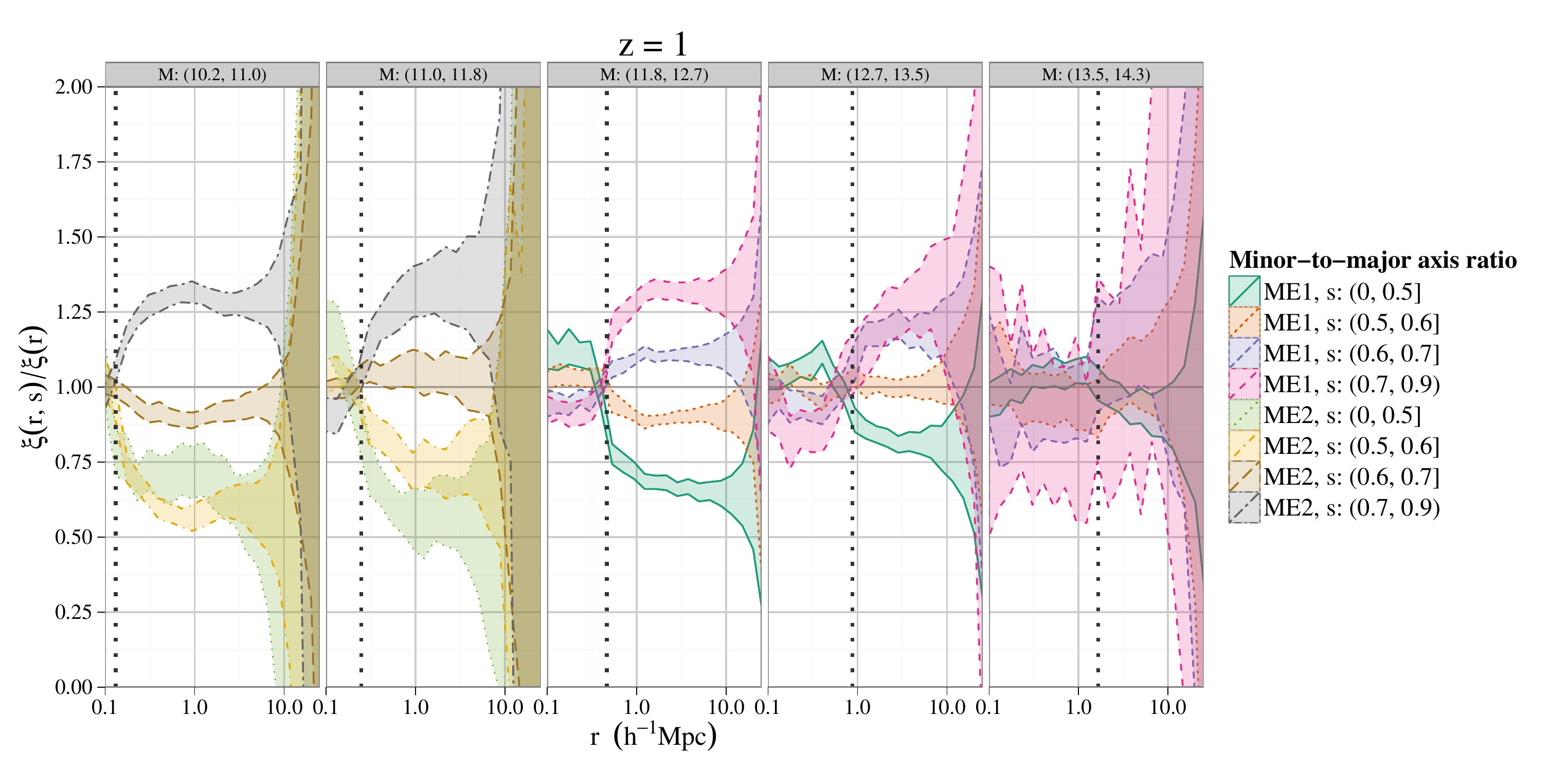}
  }
  \caption{\label{fg:axisratiocorrs}Ratio of the halo-mass correlation functions 
  binned in both separation, $r$, and minor-to-major axis ratio, $s$, to the correlation 
  binned only in $r$ at $z=0, 0.5, 1$ from top to bottom. The halo shapes are measured at $\rvir$.
  The dotted vertical lines show the maximum $\rvir$ value in each mass bin.}
\end{figure*}

Up to this point we have considered only the correlations of halo orientations. We now consider how 
the halo-mass cross-correlation functions depend on the halo shapes as parameterized by the 
minor-to-major axis ratio, $s$.
The halo-mass correlations binned in both $r$ and $s$, where $s$ is determined from the halo shapes 
at $\rvir$, are shown in figure~\ref{fg:axisratiocorrs} normalized by the halo-mass correlation 
integrated over axis ratio $s$.
The halo-mass correlations have a strong dependence on the minor-to-major axis ratio at all 
redshifts and in all but the most massive halo mass bin, where the statistics are weak at high 
redshift. At scales above the typical virial radius in each mass bin the ratio of $\xi(r,s)/\xi(r)$ 
is consistent with a constant value indicating a scale-independent excess halo bias that 
depends on the halo shape. Halos with large $s$ values, corresponding to weak asphericity, 
are more biased than the total halo population while the most elliptical halos with small $s$ 
values are less biased, as previously found by~\cite{2010ApJ...708..469F}. 
The excess halo shape bias in 
figure~\ref{fg:axisratiocorrs} appears to have only weak dependencies on halo 
mass and redshift. 
Also note that the bins with $s\in(0,0.5]$ break the 
general trend of the correlation function ratios with $s$ in a few panels in 
figure~\ref{fg:axisratiocorrs} ($z=0$ middle mass bin and the lowest mass bin 
at all $z$). 

At scales smaller than the typical virial radii in each mass bin in 
figure~\ref{fg:axisratiocorrs}, there is also a weak dependence on the axis ratio $s$, which 
is opposite that at large scales. \cite{2002ApJ...574..538J} found that more prolate halos tended 
to have lower concentrations, which would boost the one-halo correlation function in the 
direction seen at small scales in figure~\ref{fg:axisratiocorrs}. However, to limit the scope  
of the paper, we do not pursue this issue further.

In principle, the halo shapes might also be correlated with their orientations. 
To test for such correlations we recompute the angle-binned halo-mass cross-correlations from 
section~\ref{sec:anglebinnedcorrelations} with each pair weighted by the quantity,
\begin{equation}
  w \equiv \frac{1-s^2}{1+s^2},
\end{equation}
which is similar to the definition of the projected galaxy ellipticity given the 
projected axis ratio in weak lensing surveys. We compute the weighted correlation functions
rather than computing a correlation function binned in both the axis ratios and 
orientation angles to increase the signal-to-noise ratio of the measurement.

\begin{figure*}[htpb]
  \centerline{
  \includegraphics[scale=0.48]{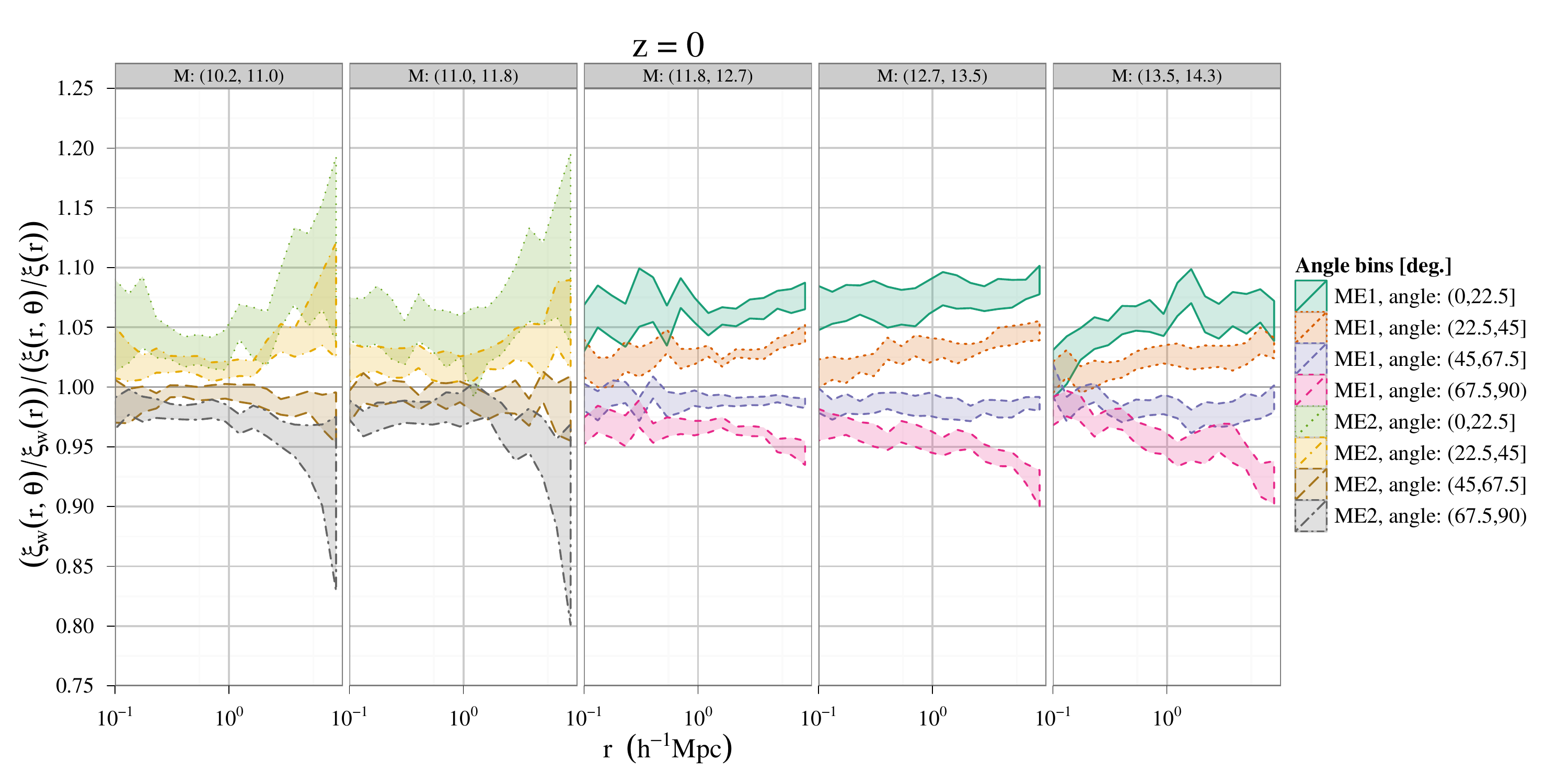}
  }
  \centerline{
  \includegraphics[scale=0.48]{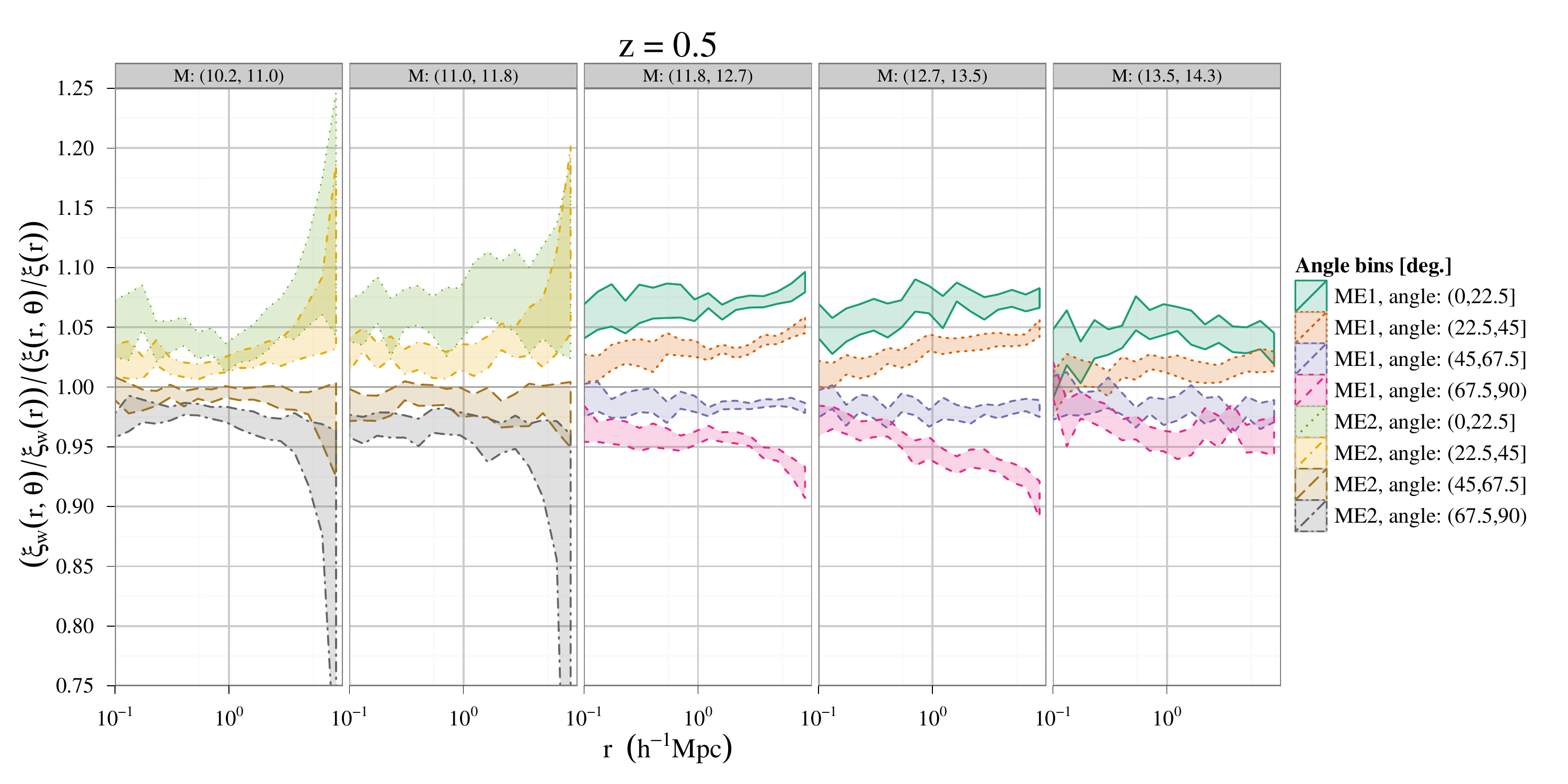}
  }
  \centerline{
  \includegraphics[scale=0.48]{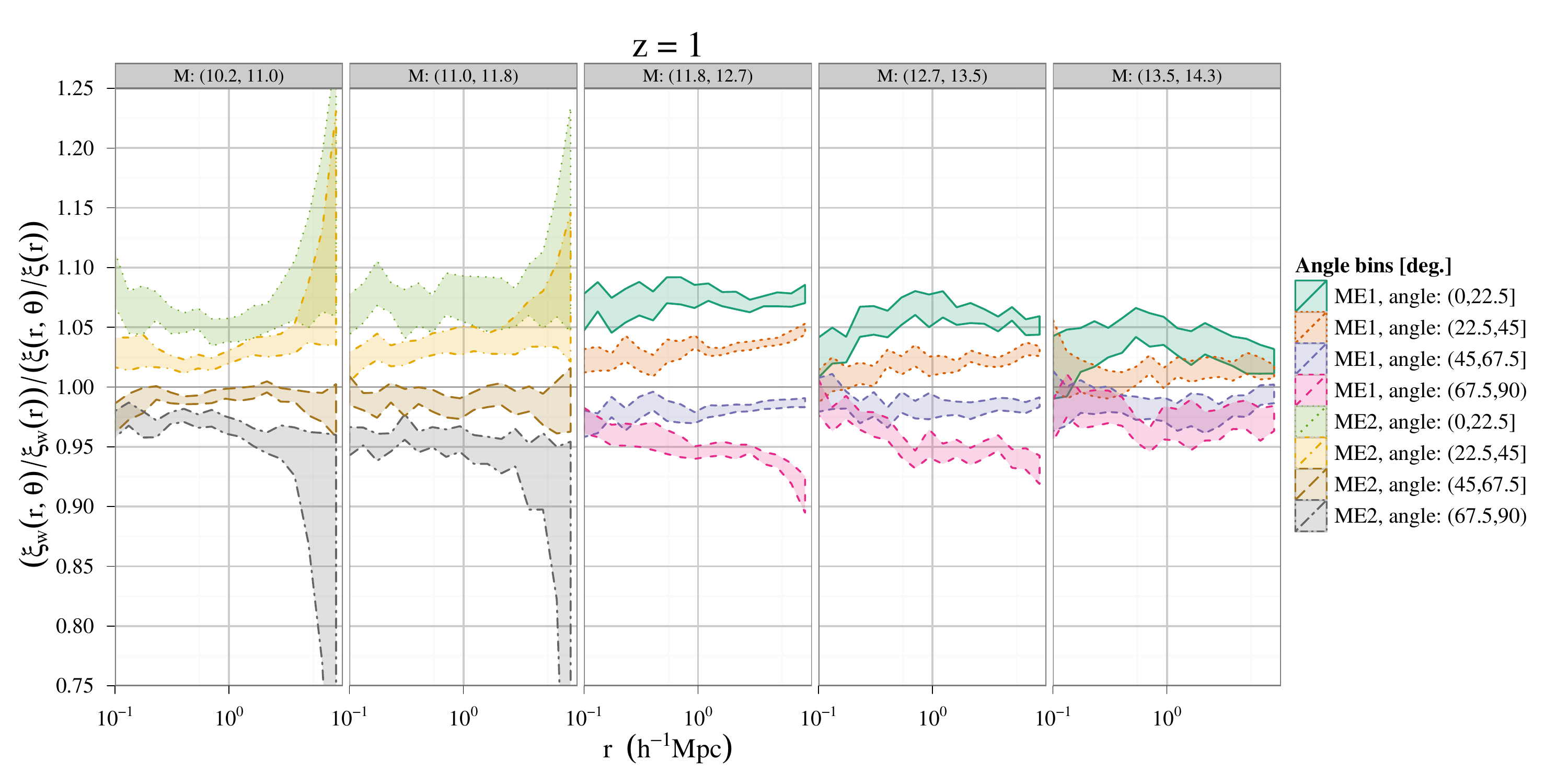}
  }
  \caption{\label{fg:markedcorrs}Ratio of the halo-mass alignment correlation functions weighted 
  by $1-s^2/1+s^2$ to the alignment correlation with no weighting at $z=0, 0.5, 1$ from top 
  to bottom. The halo shapes are measured 
  at $\rvir$.}
\end{figure*}

The resulting weighted correlation functions are shown in figure~\ref{fg:markedcorrs}.
We have first divided the weighted correlation function binned in alignment angle $\xi_{w}(r,\theta)$
by the weighted correlation function integrated over alignment angles $\xi_{w}(r)$ to isolate 
the excess alignment correlation.  We then divided by the identical ratio for the 
unweighted alignment correlations $\xi(r,\theta)/\xi(r)$ (shown in 
figure~\ref{fg:thetacorroutersep}).
If the weight $w$ was uncorrelated with the alignment angle $\theta$ then the 
``ratio of ratios'' plotted in figure~\ref{fg:markedcorrs} would be equal to one for all $r$. 
The deviation from one indicates the degree to which halos with large values of $w$
(i.e. highly aspherical halos with small $s$ values) have excess 
alignment correlations relative to the entire halo population.

We have plotted the measurements from the Millennium-2 simulation only in the two lowest mass 
bins in figure~\ref{fg:markedcorrs}
to avoid cluttering the less-noisy Millennium-1 results in the three higher mass bins.
For all mass bins the ratios are significantly different from one across the radius range from 
0.1 to 10~$h^{-1}$Mpc, although the deviations are no more than 5\%. 
Note that we truncated the range on the abscissa at 10~$h^{-1}$Mpc 
because the measurements are very noisy and consistent with no excess correlations on 
larger scales. The excess correlations are also roughly constant over the plotted 
radius range and show very little change with halo mass or redshift.  

Both the tidal stretching of halos and anisotropic accretion of mass along filaments will 
tend to introduce both asphericity and halo alignments, so it is perhaps not surprising that 
the halo axis ratios should be correlated with their orientations.  However, the 
mechanism causing the correlations of the halo alignments and shapes remains to be shown.

\subsection{Implications for measuring BCG alignments with their host halos}
\label{sec:projcorr}
Previous studies~\citep[most recently][]{2009RAA.....9...41F, 2009ApJ...694..214O, 2009ApJ...694L..83O} have concentrated on the 
projected alignment correlation functions in $N$-body simulations in order to more directly 
compare with observations. In this section we show that we can reproduce previous 
measurements of the projected alignment correlation function by changing the 
definition of the inertia tensor used to compute halo orientations rather than 
invoking a stochastic misalignment of BCGs with their host halos as in the cited 
previous studies. This pertains to our earlier point that our reduced inertia tensor 
definition is appropriate for modeling the shapes and alignments of central galaxies.

We compute $w_p(r_p,\theta_p)$ 
using an algorithm similar to that used to compute $\xi(r, \theta)$. 
To determine the projected angle $\theta_p$ between the major axis of the projected halo 
shape and the projected pair separation vector $\rv_p$, we compute the 3D 
iterative inertia tensor as before to determine the particles belonging to a 3D ellipsoid 
but then compute the eigenvalue decomposition of the 2D inertia tensor of the final 
projected mass density. As in previous sections, the halo shapes are determined from SF0 
particles only (i.e. with bound sub-structures removed) and only for halos passing 
our convergence and relaxation criteria described in Section~\ref{sec:methods}. 
From the projected alignment angles, we compute 
the anisotropic correlation function $\xi(r_p, \Pi, \theta_p)$ and integrate the 
line-of-sight separation $\Pi$ over a slab of width 80~$h^{-1}$Mpc.

\begin{figure*}
  \centerline{
    \includegraphics[scale=0.7]{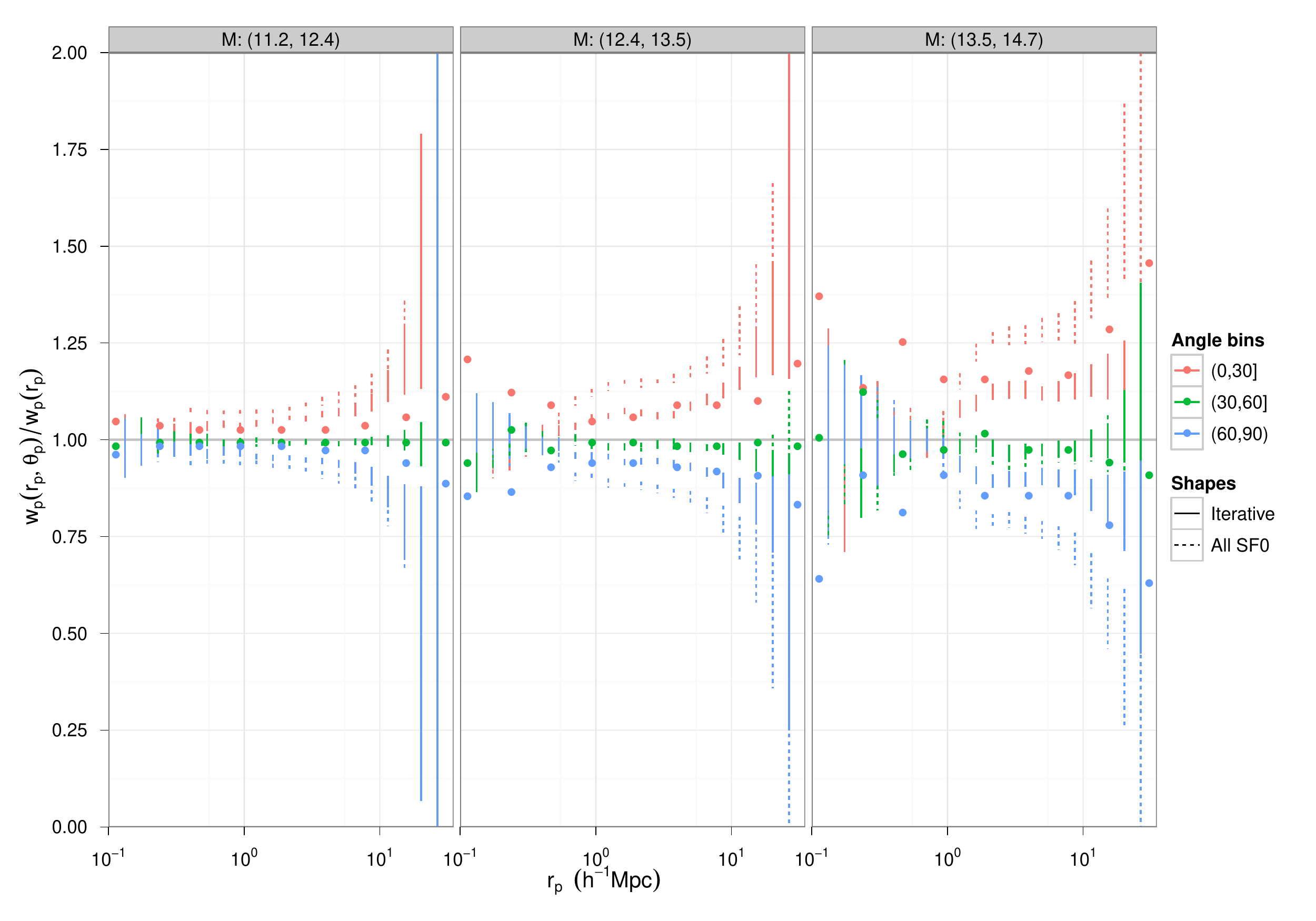}
  }
  \caption{\label{fg:faltcomparison} Projected alignment correlation function from 
  ME1 using two different 
  algorithms for determining the outer SubFind-0 halo shapes.  The ``iterative'' algorithm is 
  described in Section~\ref{sec:methods} and is used througout this paper.  The ``all SF0'' method 
  uses the {\it unweighted} inertia tensor of all SubFind-0 halo particles. 
  The points show the measurements from~\cite{2009RAA.....9...41F} using ``red'' galaxies in the SDSS (the top 
  panel in their figure~1.).
  The SDSS measurements were made in galaxy luminosity bins that are only approximately 
  related to the halo mass bins plotted here.  
  The 3 luminosity bins for the points from left to right are, $-21 < M_{0.1_r} < -20$, 
  $-22 < M_{0.1_r} < -21$, $-23 < M_{0.1_r} <-22$.}
\end{figure*}

The solid lines in figure~\ref{fg:faltcomparison} show the 
projected halo-mass cross-correlation $w_p(r_p, \theta_p)$  
binned in the angle $\theta_p$ between the 2D halo major axis measured at $\rvir$
and the projected separation vector to the mass tracers.  
The dashed lines in figure~\ref{fg:faltcomparison} show the analogous measurement when 
our calculation of the inertia tensor in the iterative determination 
of the halo shape is modified 
from equation~\ref{eq:redinertiatensor} to the unweighted inertia tensor $I_{ij}\propto\sum x_i x_j$, 
which is the inertia tensor used in \cite{2009RAA.....9...41F} to define their 
``outer'' halo shapes.
Finally, the points in figure~\ref{fg:faltcomparison} 
show the measurements of $w_p(r_p, \theta_p)$ using ``red'' galaxies in the SDSS 
by \cite{2009RAA.....9...41F} (their figure~1 upper panels). 
To plot the points in our halo mass bins we have made a crude mapping of the 
magnitude bins used in \cite{2009RAA.....9...41F} to halo mass according to the 
Halo Occupation Distribution model of \cite{2009ApJ...707..554Z} (their figure~3) 
for Luminous Red Galaxies in the SDSS catalogue. Note however that the ``red'' color 
selection in~\cite{2009RAA.....9...41F} may select non-central galaxies in the lower 
luminosity / halo mass bin.

Figure~\ref{fg:faltcomparison} shows first that changing the inertia tensor from the 
unweighted to weighted definition causes the amplitude 
of the excess projected alignment correlation function to drop by a factor of $\sim2$ in the largest 
mass bin (dashed to solid lines).  
Our improved match to the observations using the weighted inertia tensor is consistent with 
the result in~\cite{2009RAA.....9...41F} who found the best match to their observations using only the 
inner halo orientations (defined to enclose the equivalent BCG stellar mass) 
to model the BCGs because the weighted inertia tensor is also more sensitive 
to the inner halo shape than the unweighted inertia tensor.
In contrast, \cite{2009ApJ...694..214O, 2009ApJ...694L..83O} defined halo shapes using the 
unweighted inertia tensor of all halo particles and then concluded a broad distribution of 
BCG alignment angles are necessary to fit their measured shape correlations. 
This suggests that
(1) observations of $w_p(r_p, \theta_p)$ (or similar statistics) 
must be carefully compared with 
simulations (including a careful definition of ``halo shapes'') 
when drawing conclusions about halo alignments or 
alignments of BCGs with their parent halos,
(2) the most useful definition of the ``shape'' of a halo is application 
specific and the reduced inertia tensor may be more useful for modeling the shapes of central 
galaxies (whereas the ``shape'' definitions advocated in \cite{2011ApJS..197...30Z} may be more 
relevant for the shapes derived from, e.g., weak lensing studies),
(3) any misalignments between BCGs and the outer shapes of their parent halos might be 
    sufficiently modeled by the misalignments of dark matter halo shapes at different radii 
    (as shown in figure~\ref{fg:axisalignments}).
Comparing the BCG alignment with the spatial distribution of satellite galaxies in 
groups~\citep[e.g.][]{2008MNRAS.385.1511W}
can be a complimentary way to constrain BCG alignment angle distributions, which could 
potentially remove some of the ambiguity in modeling BCG orientations in simulations.

\section{Interpretation of halo alignment correlations}
\label{sec:alignmentmodels}
In this section we consider how our measured halo alignment correlations could be incorporated into the halo model~\citep[see, e.g., ][]{2002PhR...372....1C} for the dark matter two-point clustering statistics.  A triaxial halo model was previously presented by~\cite{2005MNRAS.360..203S}, but lacked a well-motivated model for the correlations of halo shapes and orientations.  Using a toy model for the triaxial halo correlations, \cite{2005MNRAS.360..203S} and \cite{2006MNRAS.365..214S} showed that including non-spherical halo shapes can lead to observationally detectable changes in the predicted matter power spectrum with even more significant changes in the matter bispectrum. Perhaps even more critically for near-term observations, the halo model can be used to predict the contamination to galaxy weak lensing correlations from intrinsic alignments of galaxy shapes~\cite{2010MNRAS.402.2127S}. However, assuming galaxy shapes are correlated with the shapes of their parent halos, the correlations between triaxial dark matter halo orientations and the surrounding mass overdensities (that would lens background galaxies) is a necessary input to any such halo model. 

Key components in the triaxial halo model are therefore the halo $n$-point correlations including dependencies on halo masses, ellipticities, and alignments.
\cite{2005MNRAS.360..203S} call these the halo ``seed'' correlations, and we adopt their nomenclature.  
Using the shorthand from \cite{2005MNRAS.360..203S} for the joint probability distribution of $N_i$ halos with positions $\xv_i$, masses $M_i$, axis ratios $\axis_i$, and orientations $\orient_i$,
\begin{eqnarray}
  &&p(1,\dots,N_i) \equiv \\
  &&p(\xv_1,\dots,\xv_{N_i}, M_1,\dots,M_{N_i}, \axis_1,\dots,\axis_{N_i},\orient_1,\dots,\orient_{N_i}), \nonumber
\end{eqnarray}
the two-point halo seed correlation function is defined by the relation,
\begin{equation}
  p(1,2) = p(1)p(2) \left(1 + \seed(1,2)\right) .
\end{equation}  
We now interpret the halo-halo alignment correlation functions 
from section~\ref{sec:anglebinnedcorrelations} 
as the seed correlation functions of \cite{2005MNRAS.360..203S}.
We may alternately choose to use the correlations measured at $\rvir$ or $0.1\rvir$ depending on the 
application.  For example, the inner halo shapes are most relevant for modeling the alignments of 
central galaxies, while the outer halo shapes would be a better choice for modeling the alignments of clusters probed by the SZ effect or x-ray surveys~\cite{2012MNRAS.421.1399B}.

The halo-mass correlations presented in section~\ref{sec:anglebinnedcorrelations} can also 
be readily interpreted as seed correlations in the halo model with the assumption that the 
integral over the masses of the second halo in each pair has already been done.
In figures~\ref{fg:thetacorroutershapehalo} and \ref{fg:thetacorrinnershapehalo} we present measurements of the halo-mass alignment correlations
when SubFind-0 halos are used to trace the mass density.  These provide the seed correlations 
for modeling the halo-mass correlation in the halo model with the mass dependence of both halos in each pair explicit.

\subsection{Multipole decomposition of alignment correlations}
To make use of our measured alignment correlation functions in the halo model, it will be helpful to decompose the dependence on the alignment angle $\theta$ and the axis ratio $s$.

We fit the angular dependence of the seed correlation functions 
with the separable model,
\begin{equation}\label{eq:alignmentfit}
  \seed_{hX}(r, \theta, s) = \xi_{hX}(r, s)\left[1 + \sum_{n\in2\mathbb{Z}_{+}}
  f_{n}(r, s)P_{n}(\cos(\theta))
  \right],
\end{equation}
where $\xi_{hX}$ denotes the halo-mass or halo-halo correlation functions,
$P_{n}$ are the Legendre polynomials of degree $n$, and 
$f_{n}(r, s)$ parameterizes any radial and shape dependence that deviates from that in $\xi_{n}(r)$.
Redshift dependence of the correlation functions and fit parameters is implicit.
Because $\int_{-1}^{1}P_{n}(x) dx=0$ the constraint 
that $\int \xi(r,\theta, s) d\cos\theta = \xi(r, s)$ is satisfied by equation~\ref{eq:alignmentfit}.
As pointed out by~~\cite{2005MNRAS.360..203S,2011JCAP...05..010B}, $\xi(r,\theta)$ also has 
the symmetry constraints $\xi(r,\theta)=\xi(r, \theta+\pi)$ and 
$\xi(r,\theta) = \xi(r,-\theta)$, which are satisfied by our restriction of $n$ to the positive 
even integers in equation~\ref{eq:alignmentfit}.

We have factored out $\xi_{h,X}(r,x)$ in 
equation~\ref{eq:alignmentfit} to provide a simple fitting form for $\excor(r,\theta)$ as 
presented in the previous section. By dividing $\seed_{hX}$ by $\xi_{hX}$ we explicitly 
remove the linear halo bias, leaving any residual stochastic or nonlinear bias to be 
absorbed in the fit parameters $f_{n}(r, s)$.

\begin{figure}[htbp]
	\centerline{
		\includegraphics[scale=0.6]{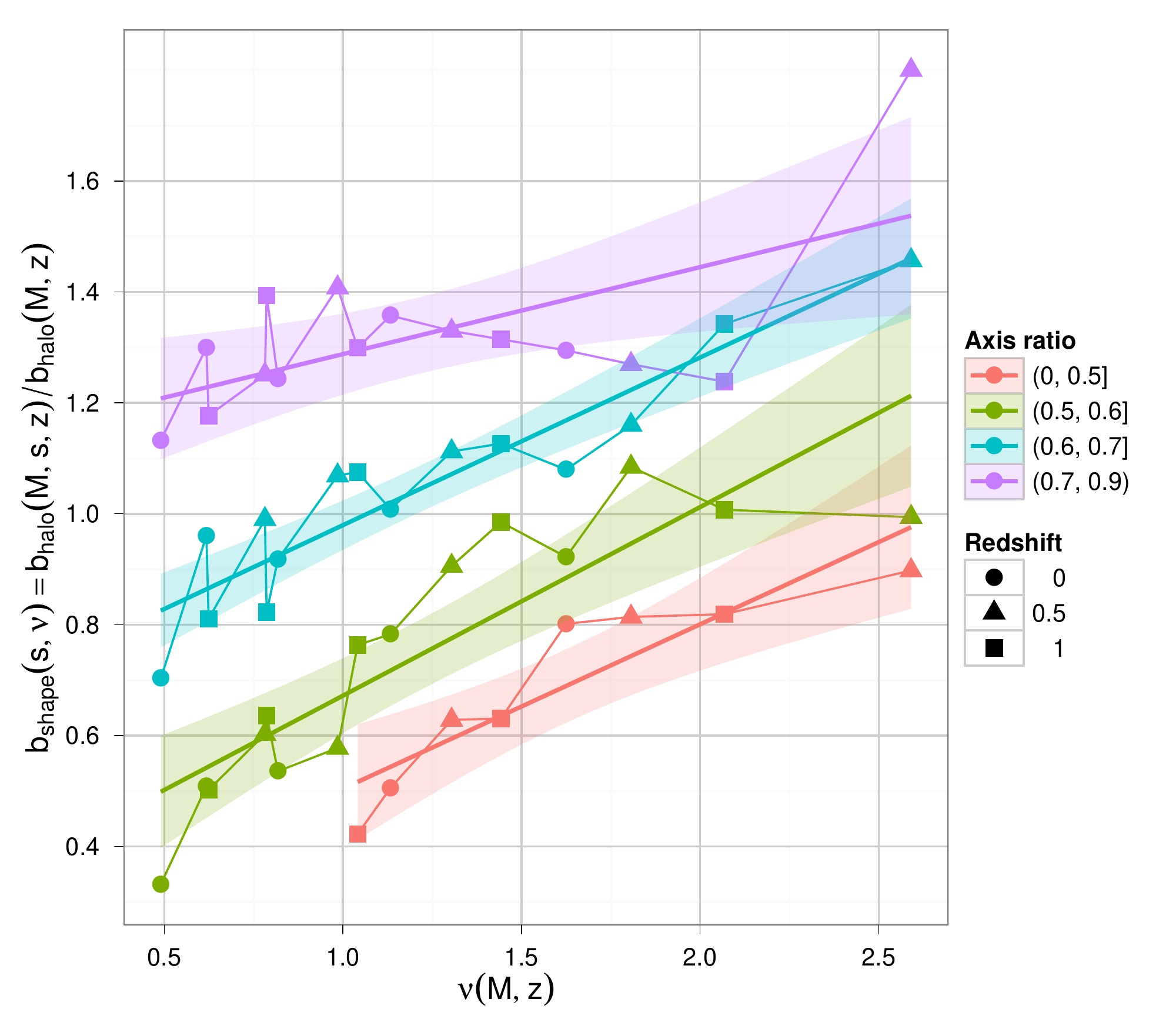}
	}
	\caption{Halo ``shape bias'' as defined in equation~\ref{eq:shapebias} as a function 
	of the peak-height threshold $\nu(M,z)\equiv\delta_{\rm SC}(z)/\sigma(M,z)$ in bins 
	in the minor-to-major axis ratio $s$. The colors show bins in the axis ratio while the 
	point shapes denote the different redshifts where the halo-mass correlation functions were measured.
	The thick solid lines show linear regressions in each axis ratio bin while the shaded bands 
	show standard errors on the regressions. We have excised the low-$\nu$ values in the lowest 
	bin in axis ratio (bottom red line and points) as measurements of these values are particularly noisy 
	(see the top left panels in figure~\ref{fg:axisratiocorrs}) }
	\label{fg:shapebias}
\end{figure}

Because $\xi_{h\delta}(r,s) / \xi_{h\delta}(r)$ at large scales in figure~\ref{fg:axisratiocorrs} 
is roughly constant as a function of $r$
at fixed $s$ and halo mass and is also roughly constant as a function of halo mass at 
fixed $r$ and $s$, we model,
\begin{equation}\label{eq:shapebias}
	\xi_{h\delta}(r,s) = b_{h}(M)\,\baxis(s,M)\, \xi_{\delta\delta}(r),
\end{equation}
where $b_{h}$ is the usual linear halo bias as a function of halo mass,
$\baxis(s, M)$ is a ``shape bias'' as a function of halo minor-to-major axis ratio and halo mass
(see \cite{2006ApJ...652...71W} for a similar bias factorization),
and $\xi_{\delta\delta}$ is the usual mass autocorrelation function.
We estimate shape bias values by minimizing the mean squared error between the log of the halo-mass 
correlation functions binned in axis ratio $s$ and the log of the halo-mass correlation integrated over $s$
over the radius interval $3\rvir$ to 20~$h^{-1}$Mpc~\cite{2005MNRAS.363L..66G}.
The estimated shape bias values are shown in figure~\ref{fg:shapebias} versus the 
peak-height threshold $\nu(M,z)\equiv\frac{\delta_{\rm SC}(z)}{\sigma(M,z)}$ where 
$\delta_{\rm SC}(z)$ is the linear overdensity for spherical collapse~\cite{1996MNRAS.282..263E} and 
$\sigma(M,z)$ is the linear mass variance in top-hat spheres extrapolated to the redshift $z$.
Plotting the shape bias as a function of $\nu(M,z)$ allows us to combine the halo mass and redshift 
dependencies into a single variable and exposes the roughly linear increase in the shape bias for fixed 
$s$ at all masses and redshifts. 
The coefficients of the linear regressions shown by the thick solid lines in figure~\ref{fg:shapebias} 
are given in table~\ref{tb:shapebiascoefs} where $\baxis(s, \nu)\equiv b_0(s) + b_{1}(s)\nu$.
\begin{table}
\begin{center}
\caption{\label{tb:shapebiascoefs}Linear regression coefficients for $\baxis(s, \nu)$}
\begin{tabular}{lcc}
\hline
Axis ratio bin & $b_0$ & $b_1$ \\
\hline
(0, 0.5] & 0.21 & 0.30 \\
(0.5, 0.6] & 0.33 & 0.34 \\
(0.6, 0.7] & 0.68 & 0.30 \\
(0.7, 0.9) & 1.13 & 0.16 \\
\hline
\end{tabular}
\end{center}
\end{table}
Note that $\baxis(s,M)$ must obey the integral constraint (for given $M$ and $z$),
\begin{equation}\label{eq:shapebiasnorm}
	\int_0^1 \baxis(s,M)\, p(s,M)\, ds = 1,
\end{equation}
which must be imposed when using the values in table~\ref{tb:shapebiascoefs} in numerical 
models such as the halo model.

Integrating equation~\ref{eq:alignmentfit} over axis ratio, $s$, gives a model for the angle-binned 
alignment correlation functions presented in section~\ref{sec:alignments},
\begin{equation}\label{eq:aligncorrmodel}
	\seed_{hX}(r,\theta) \equiv \int \seed_{hX}(r,\theta,s)\, p(s)\, ds 
	= \xi_{hX}(r)\left[1 + 
	\sum_{n\in2\mathbb{Z}_{+}} \tilde{f}_{n}(r)P_{n}(\cos(\theta))\right],
\end{equation}
where,
\begin{equation}\label{eq:ftilde}
	\tilde{f}_{n}(r) \equiv \frac{\int \xi_{hX}(r,x)f_{n}(r,s)\, p(s)\,ds}{\xi_{hX}(r)}.
\end{equation}
Fitting the excess correlations in figure~\ref{fg:thetacorroutersep} for $\tilde{f}_{n}(r)$ with $n=2,4$ at each $r$ yields small positive values of $\tilde{f}_{4}(r)$ that are consistent with zero at the 2.5-$\sigma$ level for all halo mass bins and redshifts. Because we used only four bins in $\cos\theta$ when measuring the alignment correlations we should expect to have a weak detection of higher-order angular dependencies. Below, we choose to focus only on the quadrupole term $\tilde{f}_{2}(r)$.

Finally, the $s$ dependence of $f_{n}(r,s)$ can be extracted using the weighted alignment correlation
functions shown in figure~\ref{fg:markedcorrs}. The weighted correlation functions with 
weights, $w$, are related to the binned correlation functions by,
\begin{align}\label{eq:xiw}
	\xi_{w}(r,\theta) &= \frac{\int \seed_{hX}(r,\theta,s)\, w(s)\,p(w(s))\, 
	\left|dw/ds\right|\,ds}
	{\int w\, p(w)\, dw}
	\notag\\
	&\equiv \xi_{w}(r)\left[
	1 + \sum_{n\in2\mathbb{Z}_{+}} f_{w,n}(r)P_{n}(\cos(\theta))\right].
\end{align}

\begin{figure}[htpb]
	\centerline{
		\includegraphics[scale=0.48]{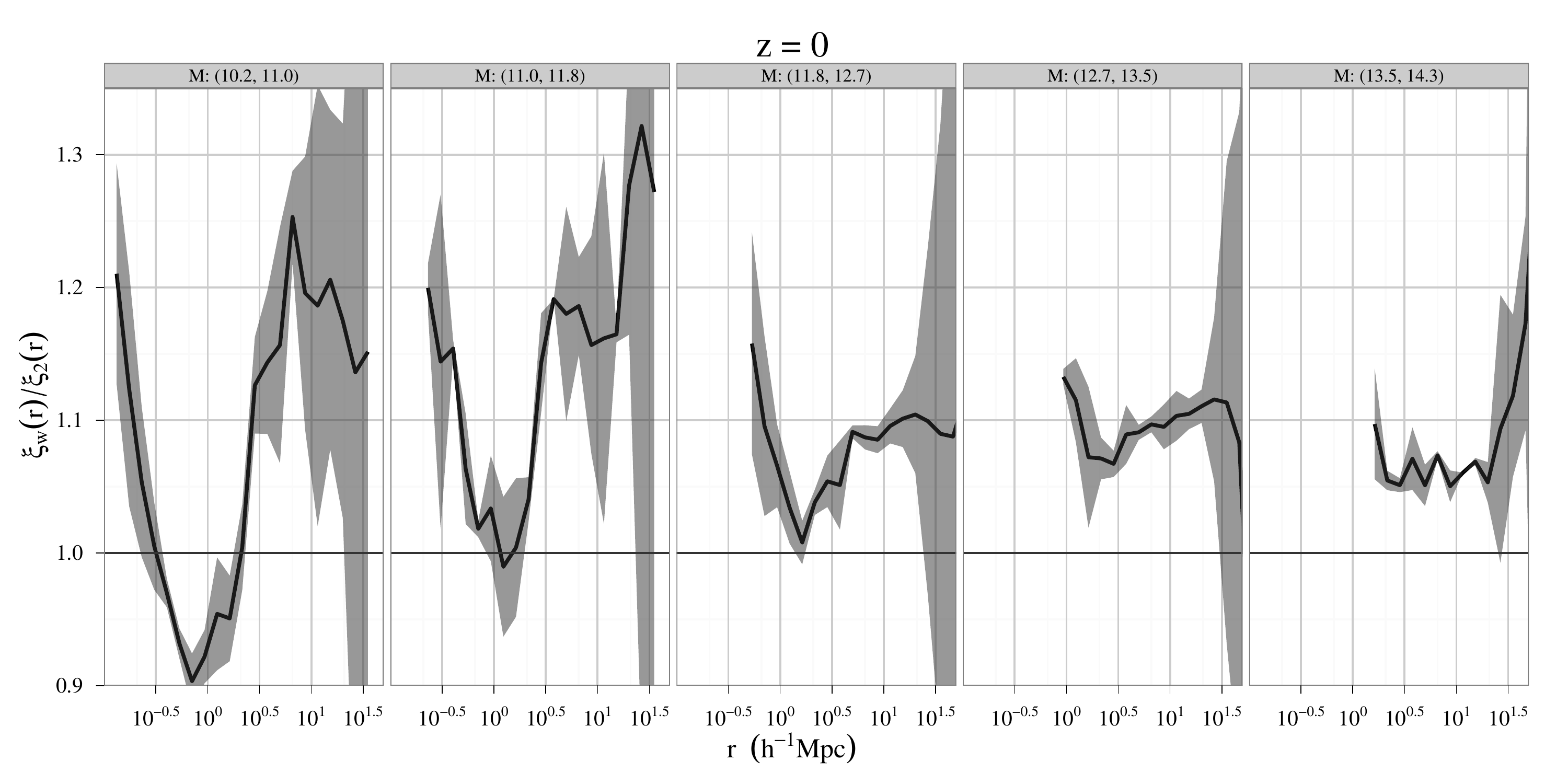}
	}
	\centerline{
		\includegraphics[scale=0.48]{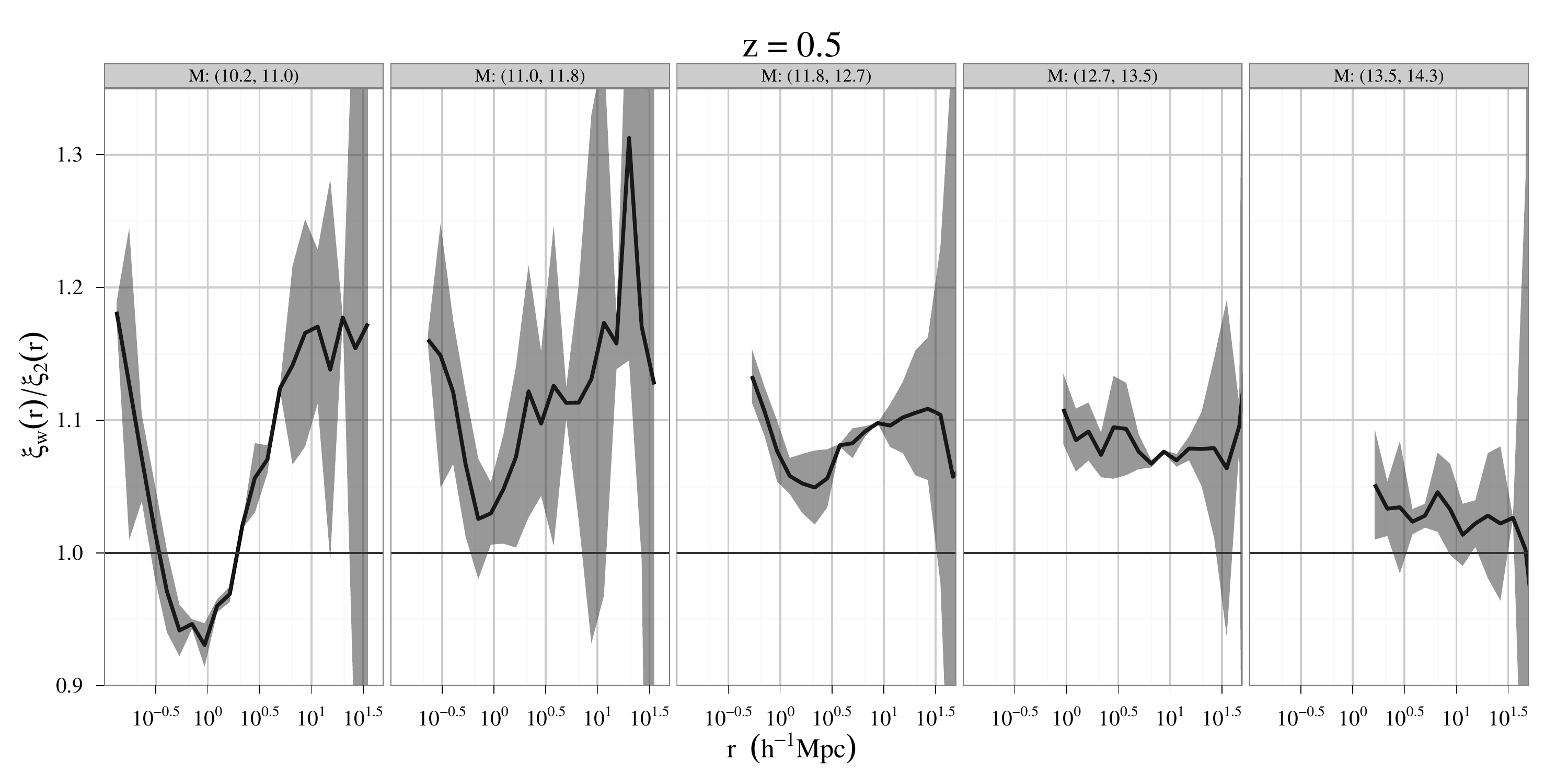}
	}
	\centerline{
		\includegraphics[scale=0.48]{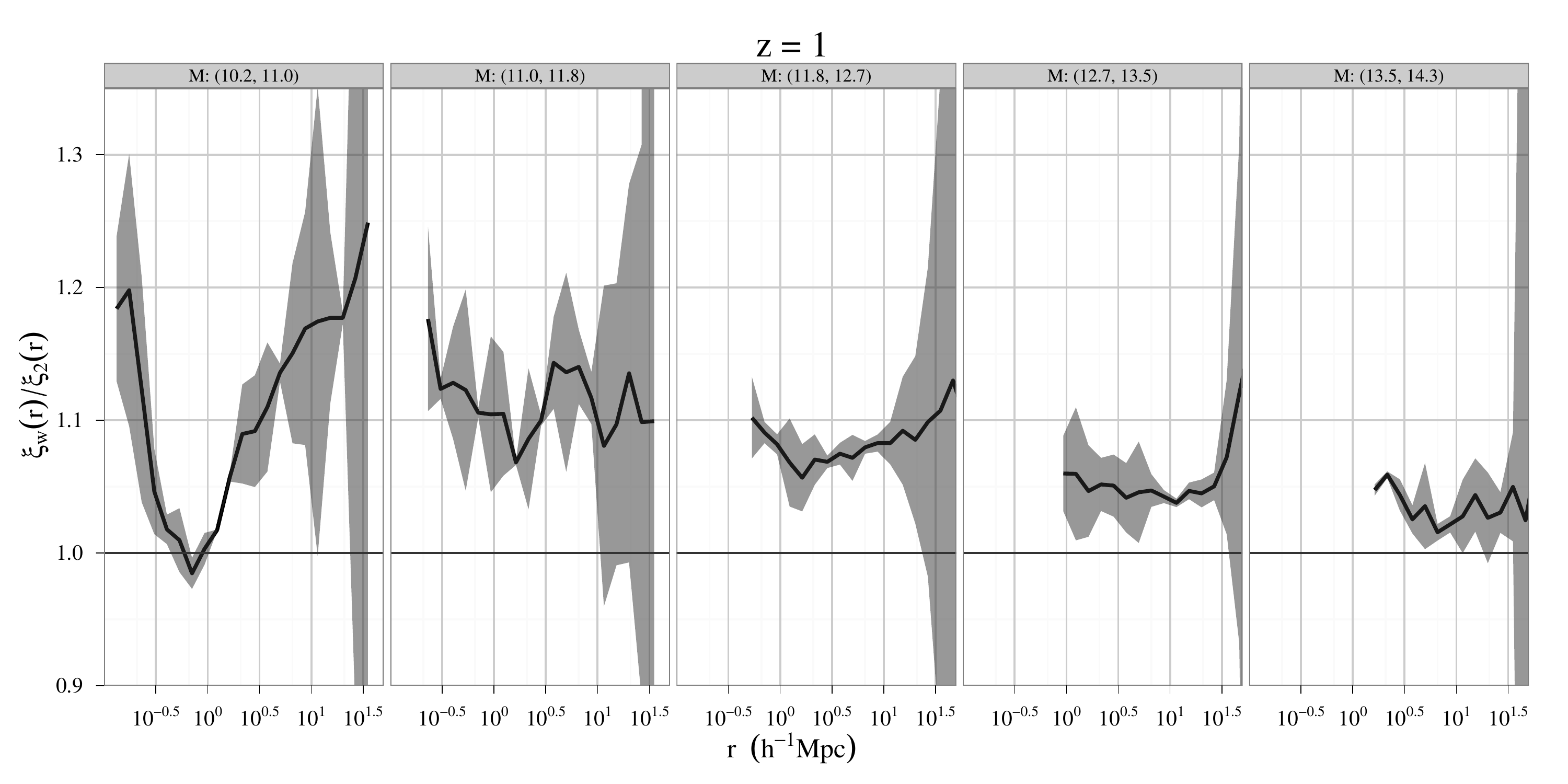}
	}
	\caption{\label{fg:f2ratios}Ratios of weighted and unweighted quadrupole moments of the
	halo mass correlations.}
\end{figure}

Equations~\ref{eq:ftilde} and \ref{eq:xiw} can be rewritten with the aid 
of equation~\ref{eq:shapebias} to yield two integral equations for the shape-dependence 
of the functions $f_{n}(r,s)$,
\begin{align}
	\int_{0}^{1} f_{n}(r,s)\, \baxis(s)p(s)ds &= \tilde{f}_{n}(r) 
	\frac{\xi_{hX}(r)}{b_h\xi_{\delta\delta}(r)} \equiv F^n_{1}(r)
	\label{eq:F1n}
	\\
	\int_{0}^{1} f_{n}(r,s)\, \frac{w(s)}{\left<w\right>}
	\baxis(s)p(w(s))\left|\frac{dw}{ds}\right|ds &= f_{w,n}(r) 
	\frac{\xi_{w}(r)}{b_h\xi_{\delta\delta}(r)} \equiv F^n_{2}(r).
	\label{eq:F2n}
\end{align}
The functions of $r$ on the right-hand sides of equations~\ref{eq:F1n} and \ref{eq:F2n}
are the multipole moments of $\xi(r,\theta)$ and $\xi_{w}(r,\theta)$ normalized by the 
linear halo-mass correlation function.
From equation~\ref{eq:shapebiasnorm} we see that equation~\ref{eq:F1n} is satisified 
with $f_n(r,s) = F_1^n(r)$. However, if $f_n(r,s)$ is indeed independent of $s$ then 
equation~\ref{eq:F2n} implies,
\begin{equation}
	\int_0^1 \frac{w(s)}{\left<w\right>}
	\baxis(s)p(w(s))\left|\frac{dw}{ds}\right|ds = \frac{F^n_2(r)}{F^n_1(r)} = \text{constant}
	\qquad \forall n,r\,| M.
\end{equation}
We plot the ratio $F_2^n(r, M)/F_1^n(r, M)$ for $n=2$ 
measured from the halo-mass cross-correlation functions 
in figure~\ref{fg:f2ratios}. Note the two lowest mass bins show measurements in Millennium-2 while 
the three higher mass bins show Millennium-1 measurements.
The ratio is roughly consistent with a constant value in each mass bin and redshift for scales 
larger than a few $h^{-1}$Mpc (note that measurements on scales larger than $\sim40h^{-1}$Mpc 
are very noisy due to low statistics and are omitted from the plot). 
However, the ratios in figure~\ref{fg:f2ratios} show large variations on scales between 
$r_{200}$ and $\sim 3 h^{-1}$Mpc at $z=0$, but become flatter as $z$ increases in all but the 
least massive bin. 
It would appear consistent with our measurements then 
to neglect the $s$ dependence of $f_{2}(r,s)$ for $r\gtrsim10 h^{-1}$Mpc at all $z$.
At $z=1$ and $M > 10^{11}\,h^{-1}M_{\odot}$ the $s$ depedendence of the quadrupole can 
be neglected for all $r > r_{200}$.

We note that the deviations of the ratios in 
figure~\ref{fg:markedcorrs} from unity does not necessarily require $f_n(r,s)$ to depend 
on halo shape, but could in principle be described by the shape dependence of the 
halo bias as described by the term $\baxis(s, M, z)$. 
This can be seen explicitly by 
noting that the left-hand sides of equations~\ref{eq:F1n} and \ref{eq:F2n} would 
both integrate to $f_n(r)$ if $f_n(r,s)=f_n(r)$ and $\baxis(s)$ were independent of 
$s$. But the correlation functions on the right-hand sides of these equations would 
also be equal in this case giving equal Legendre expansion coefficients 
$f_{w,n}(r)$ and $\tilde{f}_n(r)$ and therefore equal ratios in figure~\ref{fg:markedcorrs}.
However, because of the results in figure~\ref{fg:f2ratios}, we are led to conclude that 
both the shape dependence of the halo bias and the correlations of halo shapes and orientations 
could contribute to the behaviour in figure~\ref{fg:markedcorrs}.

\begin{figure}[htpb]
  \centerline{
    \includegraphics[scale=0.55]{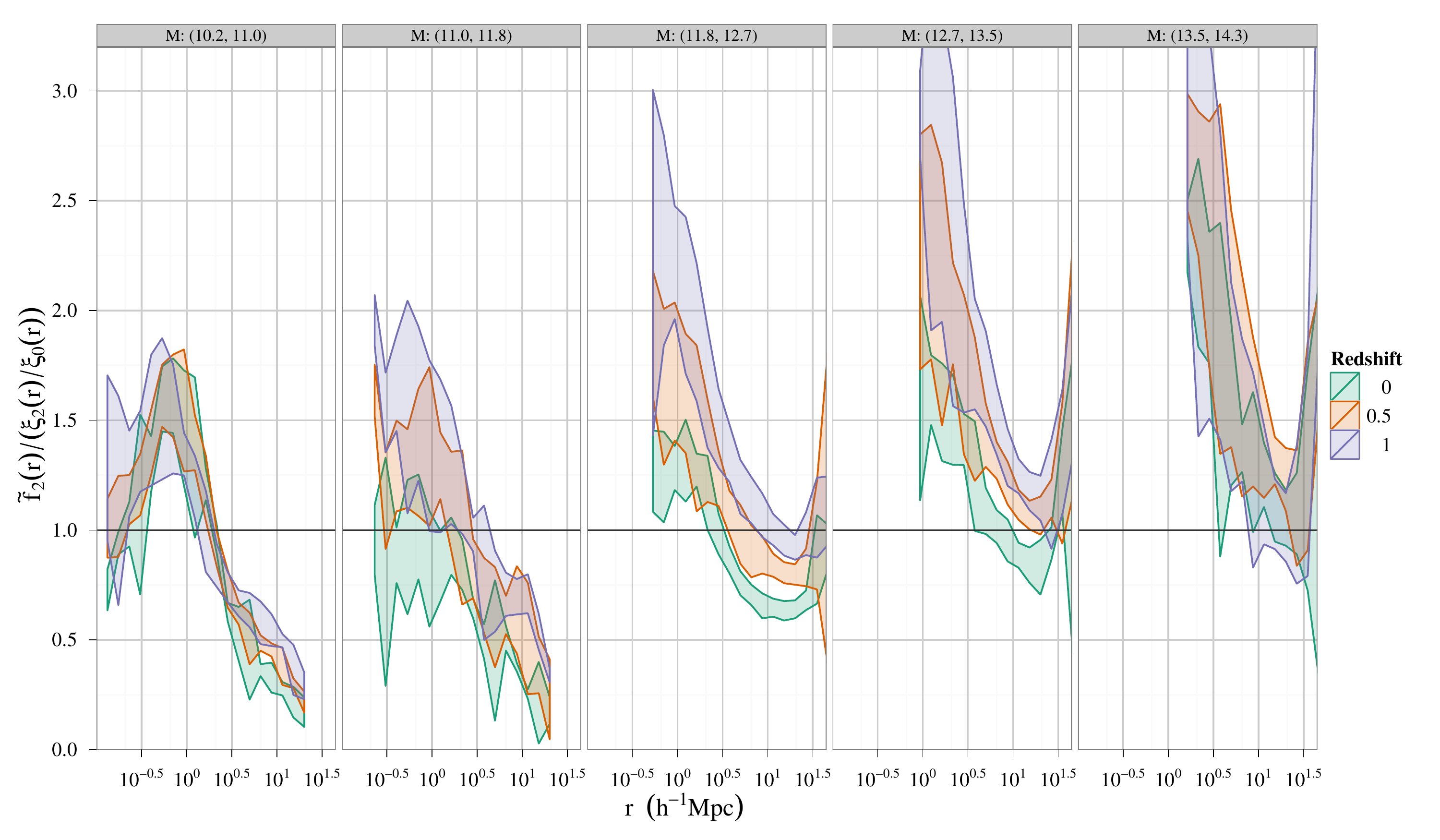}
  }
  \caption{\label{fg:f2fits}Estimates of the quadrupole moment, $\tilde{f}_2(r)$, 
  of the halo-mass alignment correlation function, $\xi_{h\delta}(r,\theta)$ at $r > r_{200}$. 
  The estimates have been 
  normalized by the ratio of the linear mass correlation function quadrupole and monopole moments.
  If the halo alignments were sourced entirely by the quadrupole moment of the linear mass density field, then these ratios would be constant functions of $r$. The two lowest mass panels (left) 
  show results from the Millennium-2 simulation while the three higher mass panels show the 
  Millennium-1 results.}
\end{figure}

It is also interesting to compare the fitted quadrupole coeficients to the 
quadrupole moment of the linear mass power spectrum. 
These should be proportional under the assumption of the linear alignment (LA) 
model~\cite{Heavens:2000p1896, 2001MNRAS.320L...7C, 2000ApJ...545..561C, 2002MNRAS.332..788M, Hirata:2004p17, 2011JCAP...05..010B} that the alignments of 
halo orientations follow the gradients of the large-scale (linear) gravitational potential.
That is, because the LA model is based on the linear, Gaussian, mass density perturbations, 
the linear mass power spectrum is the only source for the halo-mass multipole moments.
The amplitude of the LA model quadrupole is not determined by the theory, but deviations from the 
predicted scale-dependence should indicate where the LA model breaks down.
It is expected that the LA model should be a poor approximation on scales of 
a few Mpc where it is known that filamentary structures dominant the shape of the gravitational potential 
around halos~\cite{2007ApJ...655L...5A, VeraCiro:2011nb}.
In figure~\ref{fg:f2fits} we show the 
ratio of $\tilde{f}_{2}(r)\xi_{h\delta}(r) / \xi_{2,\delta}(r)$, where,
\begin{equation}\label{eq:corrmultipoles}
  \xi_{n,\delta}(r) \equiv \int_{2\pi/L_{\rm box}}^{10} 
  \frac{k^2dk}{2\pi^2} P_{\delta}^{\rm lin}(k) j_{n}(kr),
\end{equation}
are the multipole moments of the linear mass power spectrum (which is largely insensitive 
the upper integration limit), and $j_n$ are the spherical Bessel functions 
of degree $n$. Note the lower integration limit in eq.~\ref{eq:corrmultipoles}
is set by the simulation box size, $L_{\rm box}$, which skews the large-scale shape of the 
multipoles compared to those in an infinite volume.
Each panel of figure~\ref{fg:f2fits} is truncated at the lower range by $r_{200}$ corresponding 
to the most massive halo in each panel. The quadrupole moment ratios have steep negative slopes 
for all masses over the range $r_{200} < r < 40h^{-1}$Mpc 
indicating significant differences from the LA model prediction 
(the Millennium-2 
results in the two lowest mass panels are truncated at $1/4$ of the box size, or 25$h^{-1}$Mpc).
From this we can conclude that the alignments of halos with the surrounding mass overdensities 
over these length scales 
is sourced significantly by higher-order correlations in the nonlinear mass density field.

\subsection{Weak lensing cluster mass bias from correlated line-of-sight structures}
\label{sec:wlmassbias}
Galaxy cluster masses estimated from weak lensing are known to be contaminated by the lensing from 
line-of-sight structures external to the cluster because the lensing kernel is much larger than the 
size of a cluster~\cite{2001ApJ...547..560M,Marian:2009wi}.
The line-of-sight lensing contamination is largest when the line-of-sight structures are spatially 
correlated with the cluster (so that there is on average more or less mass along the cluster line-of-sight 
than along a random line-of-sight). \cite{Marian:2009wi} showed that the contribution to the 
projected mass density 
in the vicinity of a cluster of mass $M$ and orientation $\orient$
from external line-of-sight structures within a slab of width $L$ centered on the 
cluster is given by,
\begin{equation}\label{eq:surfacemass}
  \left<\Sigma_{\rm ext}(\rperp| M, \orient)\right> = 2\rho_m
  \int_{\sqrt{\rvir^2(M)-\rperp^2}}^{\frac{L}{2}} d\rpara
  \left(1 + \xi(\rv, \theta | M, \orient)\right)
\end{equation}
where $\xi$ is the halo alignment cross-correlation function as shown in 
figure~\ref{fg:thetacorroutersep}.
An estimate for the total halo mass can be obtained by integrating the projected 
surface mass density over an aperture defined by $W(\rperp)$.  The contribution 
to the halo mass estimate from external mass along the line-of-sight is then,
\begin{equation}
  \left<M_{\rm ext}(M)\right> = \int d\rperp\, W(\rperp)\, 
  \left<\Sigma_{\rm ext}(\rperp| M, \orient)\right>.
\end{equation}
Inserting eq.~(\ref{eq:aligncorrmodel}) into eq.~(\ref{eq:surfacemass}), the contributions to the 
external mass are,
\begin{equation}
  \left<M_{\rm ext}(M, \vartheta)\right> = 
  \left<M_{\rm uni}(M)\right> + 
  \left<M_{\rm corr}(M)\right> +
  \left<M_{\rm align}(M, \vartheta)\right>,
\end{equation}
where,
\begin{equation}
  \left<M_{\rm uni}(M)\right> \equiv 2\rho_m\int d\rperp\, W(\rperp)\, 
  \int_{\sqrt{\rvir^2(M)-\rperp^2}}^{\frac{L}{2}} d\rpara,
\end{equation}
and
\begin{equation}
  \left<M_{\rm corr}(M)\right> \equiv 2\rho_m\int d\rperp\, W(\rperp)\, 
  \int_{\sqrt{\rvir^2(M)-\rperp^2}}^{\frac{L}{2}} d\rpara\, \xi_{h\delta}(\rv),
\end{equation}
were considered by \cite{Marian:2009wi}. The new term we present here is,
\begin{equation}
  \left<M_{\rm align}(M, \vartheta)\right> \equiv 2\rho_m\int d\rperp\, W(\rperp)\, 
  \int_{\sqrt{\rvir^2(M)-\rperp^2}}^{\frac{L}{2}} d\rpara\,
  \xi_{h\delta}(r)\, \tilde{f}_2(r) P_{2}(\cos\theta),
\end{equation}
and we have introduced the explicit dependence on the angle between the halo major axis 
and the line-of-sight, $\vartheta$.  Therefore, $\left<M_{\rm corr}(M)\right>$
is a measure of the weak lensing signal from correlated mass along the line-of-sight to a 
cluster that is isotropically distributed about the cluster.  
The new term $\left<M_{\rm align}(M, \vartheta)\right>$ on the other hand, is a measure of extra 
mass that is correlated with the cluster orientation.

\begin{figure}[htpb]
  \centerline{
    \includegraphics[scale=0.42]{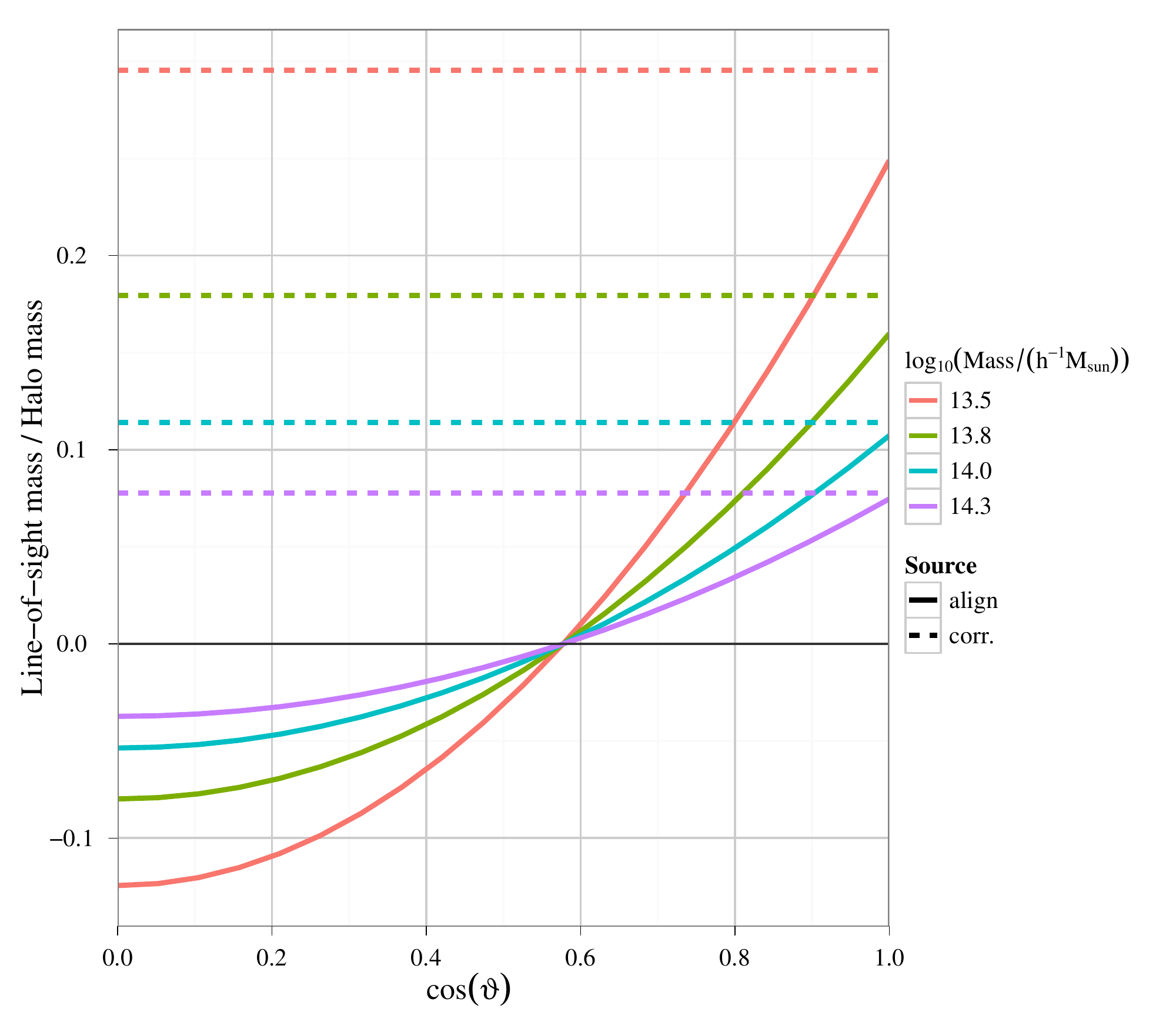}
    \includegraphics[scale=0.42]{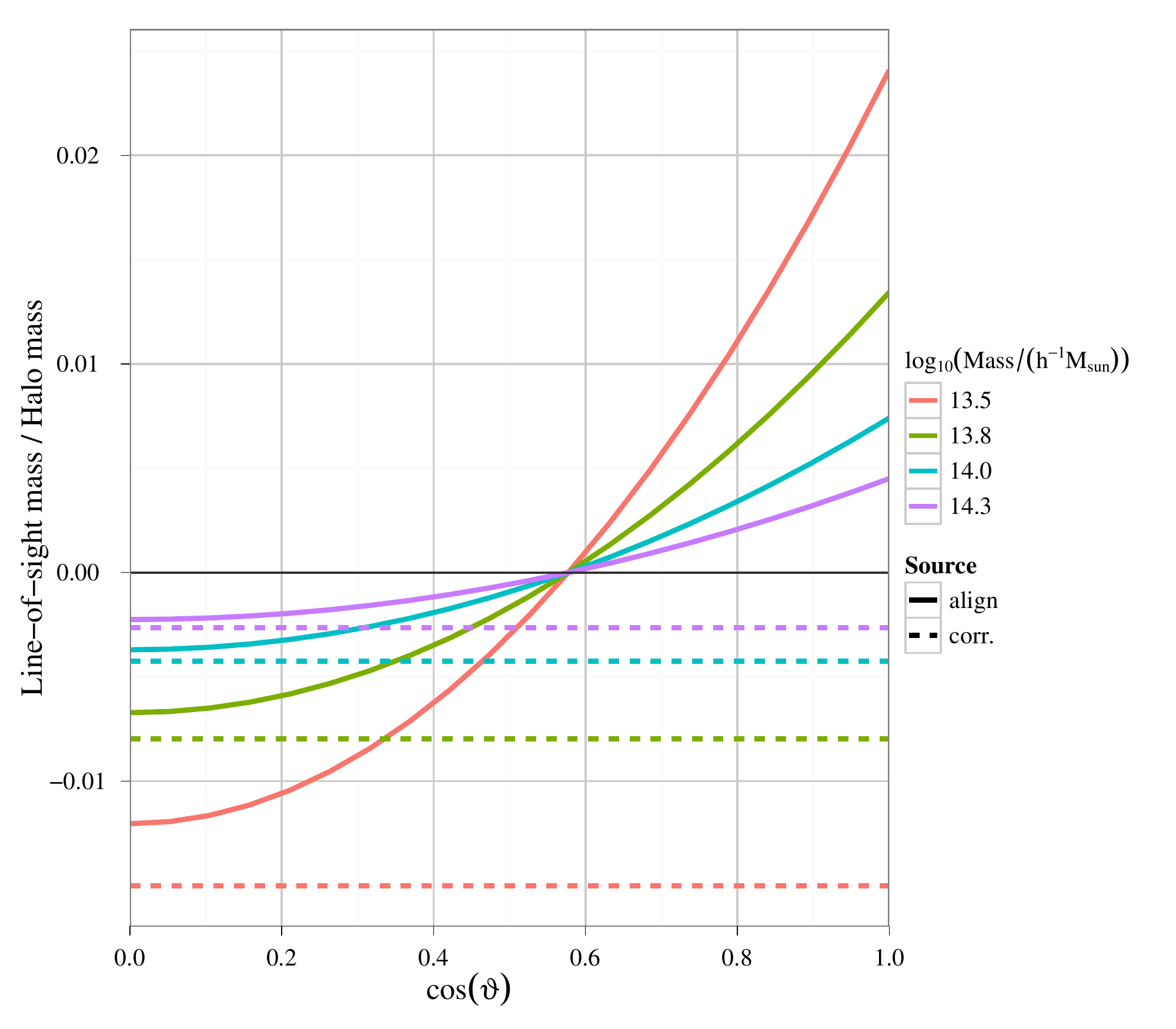}    
  }
  \caption{\label{fg:wlmassbias}Bias in the weak-lensing inferred halo mass from correlated structures 
  along the line of sight as a function of the cosine of the angle of the halo major axis 
  with respect to the line-of-sight. The left panel assumes a top-hat filter is used to integrate the projected 
  surface mass density to estimate the total halo mass.  The right panel assumes a compensated 
  filter that is chosen to maximize the signal-to-noise ratio of the halo mass estimate 
  for a spherically symmetric NFW halo.}
\end{figure}

We compare $\left<M_{\rm corr}(M)\right>$ and $\left<M_{\rm align}(M, \vartheta)\right>$ 
with the total halo mass in figure~\ref{fg:wlmassbias} for two choices of $W(\rperp)$ 
presented in \cite{Marian:2009wi}. The left panel shows the fraction of the mass estimates 
from line-of-sight structures when integrating the projected surface mass density over 
a top-hat aperture with radius equal to $r_{200}$.  When $\vartheta\sim0$, 
$\left<M_{\rm align}(M, \vartheta)\right> \approx \left<M_{\rm corr}(M)\right>$ so that 
for triaxial clusters oriented along the line-of-sight the contamination from external 
line-of-sight mass is roughly twice as large as previous estimates. 
For $\vartheta \sim \pi/2$, there is a decrement of mass along the line-of-sight 
so the total mass bias will be decreased relative to the isotropic case.
In the right panel of figure~\ref{fg:wlmassbias} a compensated filter from 
\cite{Marian:2009wi} is used that is chosen to maximize the signal-to-noise ratio of the halo 
mass estimate when the halo has a spherically symmetric NFW density profile.
Both $\left<M_{\rm corr}(M)\right>$ and $\left<M_{\rm align}(M, \vartheta)\right>$ are 
reduced to negligible levels for current cluster weak lensing studies.  However, the 
compensated filter requires finding a good guess for the true halo mass and concentration.
Any errors in this guess will result in increased mass bias terms. 

As we mentioned in Section~\ref{sec:shapes}, the predicted bias in cluster 
weak lensing mass estimates could change with a different choice of inertia tensor 
and halo constituent particles used to measure the halo-mass alignment correlation function.
These choices will depend on the cluster selection method and the estimator for 
cluster ellipticity when applying our predictions to interpret observations. For example, 
our results as presented may be useful if the BCG shape and orientation is used as a proxy 
for the cluster orientation while a different shape estimator may be more useful to compare 
with x-ray or Sunyaev-Zel'dovich morphologies. We leave the determination of shape 
estimators for these applications to future work.

\section{Conclusions}
\label{sec:conclusions}
We have measured the triaxial shapes and the alignment correlations of dark matter halos over wide ranges in halo radius (0.1$\rvir$ -- $\rvir$) and halo mass ($10^{10}$ -- $2\times10^{14}\,h^{-1}M_{\odot}$) by combining measurements in two similuations with different volumes and mass resolutions. While we largely find good agreement between the two simulations in the mass ranges where they overlap, there are some important discrepancies. First, near the lower mass limit for the Millennium simulation (with particle mass $8.6\times 10^{8}\, h^{-1}M_{\odot}$) the distributions of axis ratios are slightly skewed towards more spherical halos than in the Millennium-2 simulation (with particle mass $6.885\times 10^{6}\, h^{-1}M_{\odot}$). Second, the distributions of angles between the halo major axes at small radii and large radii are narrower and closer to zero in the Millennium-2 simulation indicating less misalignment between halo shapes at different radii. 

We draw several conclusions based on the distributions of halo shapes and the agreement or discrepancies between our two simulations:
\begin{itemize}
	\item The mass resolution of the $N$-body simulation and the method for computing the inertia tensor of halos can significantly alter the alignments of halo axes as a function of halo radius as well as the correlation of halo orientations with the surrounding mass distribution. Because we are using halo particles that are identified to be graviationally bound but not part of any bound substructures, we have speculated that more substructures are resolved and removed in the higher-resolution simulation. In general, we would expect unresolved substructures to both make the inferred halo shapes more spherical and to randomize the orientations of the halo shapes at different radii, in agreement with our results. 
	Constraints on the alignments of BCGs with their host halos should account for these issues.
	\item The mean angles between the halo major axes at small and large radii are typically $\sim20^{\circ}$, which is significantly less than other claims in the literature~\cite{2008ApJ...675..146F,2009ApJ...694..214O}. However, there is very large scatter in the distributions of alignment angles so that $\sim25$\% of halos of all measured masses have nearly perpendicular major axes at small and large radii. This is a somewhat surprising result, indicating that even if we assume central galaxies trace the inner shape of their parent halos, we should frequently expect large misalignments with tracers of the outer halo shape such as the satellite distribution or gravitational lensing measurements. The large misalignments of inner and outer halo shapes therefore confound constraints on the dynamical accretion of satellite galaxies based on their spatial distributions.
	\item Halo shapes become less spherical with increasing halo mass and redshift and decreasing spherical halo radius, confirming the results of previous simulation studies\cite{2002ApJ...574..538J,2006MNRAS.367.1781A}, but extending to larger mass and radius ranges. However, simulations including 
    baryons~\cite{2004ApJ...611L..73K} indicate the 
    axis ratios can increase by as much as 0.4 at small fractions of the virial radius (0.1$r_{180}$), 
    with systematic increases in the axis ratio for all radii. This remains an important issue to consider 
    when confronting our measured halo shape distributions with observations.
    As shown in figure~\ref{fg:triaxparameter}, halos of all masses become systematically more prolate for decreasing spherical halo radius (equivalent to constant enclosed mass density in our measurements). For halos with virial masses $\lesssim 10^{11}h^{-1}M_{\odot}$ the shapes at the virial radius tend to be oblate while remaining strongly prolate in the interior of the halo. For larger halo masses the halo shapes are predominantly prolate at all radii, indicating that for many applications high-mass halo shapes can be reasonably modeled with only the value of the minor-to-major axis ratio.
\end{itemize}

To study the correlations in both the shapes and orientations of halos with the large-scale structure, we computed several types of correlation functions binning in the minor-to-major axis ratios of the halos or the angles between the major axes of two halos or the major axis of one halo and the separation vector to a mass density tracer.
The conclusions we draw directly from these measured correlations include:
\begin{itemize}
	\item The excess halo alignment correlations are strongly increasing functions of halo mass, in agreement with the paradigm that high-mass halos are both younger than lower-mass halos and more biased (i.e. are located at the connecting points between filaments).
	\item The excess halo-mass alignment correlations are significant to several tens of megaparsecs, again in agreement with measurements of shear-mass correlations~\cite{2007MNRAS.381.1197H} and group and cluster 
    alignment correlations~\cite{2011arXiv1109.6020S, 2011MNRAS.414.2029P}.
	\item The excess alignment correlations are both much larger and more significant in the halo-mass correlation functions (figure~\ref{fg:thetacorroutersep}) than in the halo-halo correlation functions (figure~\ref{fg:thetacorrouterouter}). We speculate that there is a large stochastic component to the alignments of halo major axes with the surrounding mass distribution, which serves to further decrease the amplitude of the halo-halo alignment correlations because the alignments of both halos in each correlated pair must be averaged over.
	\item Halos with small minor-to-major axis ratios are less biased (i.e. have smaller correlation function amplitudes) than the average halo population at fixed mass and redshift. This is consistent with previously established relations between the age and bias of halos~\cite{2005MNRAS.363L..66G,2006MNRAS.367.1039H,2006ApJ...652...71W,2007MNRAS.377L...5G} if we understand that more spherical halos typically formed earlier (at fixed halo mass and redshift).
	\item Our measured halo-mass alignment correlation function can be adequately described by only a quadrupolar angular dependence, although we do detect a statistically significant nonzero coefficient of the fourth order multipole for higher mass halos. We are limited in this measurement by our use of only four angular bins in measuring the alignment correlations. 
	\item The alignment quadrupole has a shape that is significantly different from the quadrupole of the linear matter correlation function (see figure~\ref{fg:f2fits}), which is the predicted shape in the linear alignment model~\cite{2011JCAP...05..010B}.
\end{itemize}

As one direct application of our measured halo-mass alignment correlation functions, we updated forecasts for the contamination to galaxy cluster mass estimates from weak gravitational lensing observations due to correlated (but not gravitationally bound) structures along the line-of-sight to the cluster. Previous studies such as~\cite{Marian:2009wi} showed that the halo-mass correlation function implies average lensing contamination for clusters that can be $\sim10-20$\% of the cluster mass when using a top-hat aperture to estimate the mass. We introduced a new contamination term that is correlated with the orientation of the prolate cluster with respect to the line-of-sight and that can give additional mass biases similar in magnitude to those estimated in~\cite{Marian:2009wi}. 
It has already been shown that the bias in cluster mass estimates from line-of-sight contamination is significant for constraining cosmological parameters from cluster counts~\cite{2011PhRvD..84j3506E} and our new bias term will only increase the error in inferred parameters. However, if the orientation of the cluster with respect to the line-of-sight can be independently observationally determined~(e.g.~\cite{2012MNRAS.419.2646S}), then it may be possible to reduce the scatter in weak lensing mass estimates by subtracting the mean bias from correlated structures as predicted by the halo-mass alignment correlations we have measured. It is also possible, but remains to be investigated, whether the mass correlated with cluster orientations is a significant bias for strong lensing measurements in clusters~(see e.g. \cite{2012MNRAS.420L..18H}).

Our alignment correlation function measurements will also be useful for building halo models for intrinsic alignments of galaxies as contamination in cosmic shear surveys~\citep{2010MNRAS.402.2127S, Kirk:2010wa}. The so-called ``intrinsic-intrinsic'' (II) contamination for cosmic shear is best modeled using our halo-halo alignment correlations based on the inner halo shapes as shown in figure~\ref{fg:thetacorrinnerinner}. Note that such an II model would be built by first populating central galaxies with 3D shapes that match the 3D shapes of their parent halos and then performing the line-of-sight projection to predict the 2D cosmic shear contamination. This would further reduce the amplitude of the alignment signal, so we expect any predicted II models to be quite small on scales larger than a few megaparsecs. The so-called ``galaxy-intrinsic'' (GI) cosmic shear contamination term is likewise best modeled with the halo-mass alignment correlations shown in figure~\ref{fg:thetacorrinnersep} or \ref{fg:thetacorrinnershapehalo} and modeled in equation~\ref{eq:aligncorrmodel}. 

Because our measured alignment correlations have halo mass dependence that is roughly self-similar (see the discussion in section~\ref{sec:anglebinnedcorrelations}), and because we do not expect the scaled distributions of halo shapes to be strong functions of cosmology~\cite{2006MNRAS.367.1781A}, it may become possible to include the cosmology dependence in models of galaxy intrinsic alignments simply through the dependence of the alignment correlations on the characteristic mass scale $M_{*}(z)$ and the matter power spectrum. However, one should keep in mind that halo occupation statistics and galaxy properties influencing their observational selection probably depend on a number of halo properties that are related by the as yet un-modeled ``assembly bias''~\cite{2005MNRAS.363L..66G, 2006ApJ...652...71W, 2007MNRAS.377L...5G, 2010ApJ...708..469F, 2011arXiv1110.6174L}, which 
will confuse any cosmological dependencies in the intrinsic alignments in cosimc shear surveys.

Because the halo shape statistics at small halo radii (where the galaxy alignments are often modeled) are slow to converge with mass resolution, galaxy intrinsic alignment models may be more accurate when using a halo model calibrated from high-resolution simulations than when populating halos with galaxies in a large-volume simulation (which will generally have lower mass resolution due to computational constraints). We will pursue this approach in a forthcoming publication.

Finally, the alignment between galaxies and dark matter is an additional complication when comparing our results with observations~\citep{Hahn:2010p2884,Deason:2011hp}. Both the effects of gas on the shapes of the dark matter halos and the relative orientations of galaxy shapes and their parent halos remain promising avenues for future research.

\section*{Acknowledgments}
We thank Phil Bett, Jonathan Blazek, and Andreas Faltenbacher for 
helpful comments on the first draft of this paper.
The calculations for this paper were performed on the ICC
Cosmology Machine, which is part of the DiRAC Facility jointly funded by STFC,
the Large Facilities Capital Fund of BIS, and Durham University.
Part of this work performed under the auspices of the 
U.S. Department of Energy by Lawrence Livermore National Laboratory under Contract DE-AC52-07NA27344.

\bibliographystyle{JHEP}
\bibliography{library}

\appendix
\section{Halo counts}
\label{sec:halocounts}
We present counts of the numbers of halos with shape measurements in both our simulations 
in tables~\ref{tab:ME1counts} (for Millennium-1) and \ref{tab:ME2counts} (for Millennium-2).
The three columns in each table show counts for the three simulation snapshots considered 
in this paper. The rows in the tables show the numbers of halos when different cuts 
are applied to the halo shape catalogues. The first row, labeled ``No cuts,'' gives the 
total numbers of halos with shape measurements irrespective of any cuts, except that the 
Millennium-2 halos with virial masses less than $9.3\times10^{9}~h^{-1}M_{\odot}$ were not 
considered.
The second rows in each table, labeled ``In shape mass bins,'' give the halo counts in all 
the halo mass bins depicted by the vertical dotted lines in figure~\ref{fg:massfunctions}.
The third rows, labeled ``Quality cuts in shape mass bins,'' are the remaining counts in the 
mass bins after halos are discarded that had unconverged centers or virial masses.
The fourth rows, labeled ``Substructure cuts in shape mass bins,'' give the halo counts when 
an halos with more than 10\% mass in substructures are discarded, in addition to the previously 
applied cuts.
The fifth rows, labeled ``Axis ratio cuts in shape mass bins,'' give the remaining halo counts 
when an additional cut is applied that the minor-to-major axis ratio $s \le 0.9$.
The sixth rows, labeled ``In corr. mass bins,'' give the counts in the mass bins 
used for measuring the halo correlation functions as depicted by the vertical dashed lines 
in figure~\ref{fg:massfunctions}.
The final seventh rows, labeled ``All cuts in corr. mass bins,'' give the counts in 
the correlation function mass bins when all the previously listed cuts are applied.
\begin{table}
\begin{center}
\caption{\label{tab:ME1counts}Halo counts for Millennium-1}
\begin{tabular}{lccc}
\hline 
Description & $z=0$ & $z=0.5$ & $z = 1$ \\
\hline
No cuts                              & 20,350,503 & 20,441,331 & 20,444,688 \\
In shape mass bins                   & 16,217,712 & 17,078,569 & 16,507,354 \\
Quality cuts in shape mass bins      & 15,622,670 & 15,930,109 & 14,815,635 \\
Substructure cuts in shape mass bins & 12,372,629 & 12,124,709 & 10,979,741 \\
Axis ratio cuts in shape mass bins   & 12,349,801 & 12,115,975 & 10,976,309 \\
In corr. mass bins                   & 15,014,590 & 15,846,000 & 15,173,960 \\
All cuts in corr. mass bins   & 11,439,534 & 11,285,980 & 10,148,373 \\
\hline
\end{tabular}
\end{center}
\end{table}
\begin{table}
\begin{center}
\caption{\label{tab:ME2counts}Halo counts for Millennium-2}
\begin{tabular}{lccc}
\hline 
Description & $z=0$ & $z=0.5$ & $z = 1$ \\
\hline
No cuts                              & 11,073,314 & 15,921,384 & 12,200,000 \\
In shape mass bins                   & 8,386,084 & 9,430,983 & 9,559,984 \\
Quality cuts in shape mass bins      & 8,307,617 & 9,254,432 & 9,266,159 \\
Substructure cuts in shape mass bins & 7,147,623 & 7,626,275 & 7,313,926 \\
Axis ratio cuts in shape mass bins   & 7,075,515 & 7,588,251 & 7,295,991 \\
In corr. mass bins                   & 3,692,048 & 4,156,360 & 4,146,161 \\
All cuts in corr. mass bins   & 3,067,434 & 3,287,135 & 3,103,102 \\
\hline
\end{tabular}
\end{center}
\end{table}

\section{Halo alignment correlation supplementary plots}
\label{sec:extracorrs}
\begin{figure*}[htpb]
  \centerline{
  \includegraphics[scale=0.4]{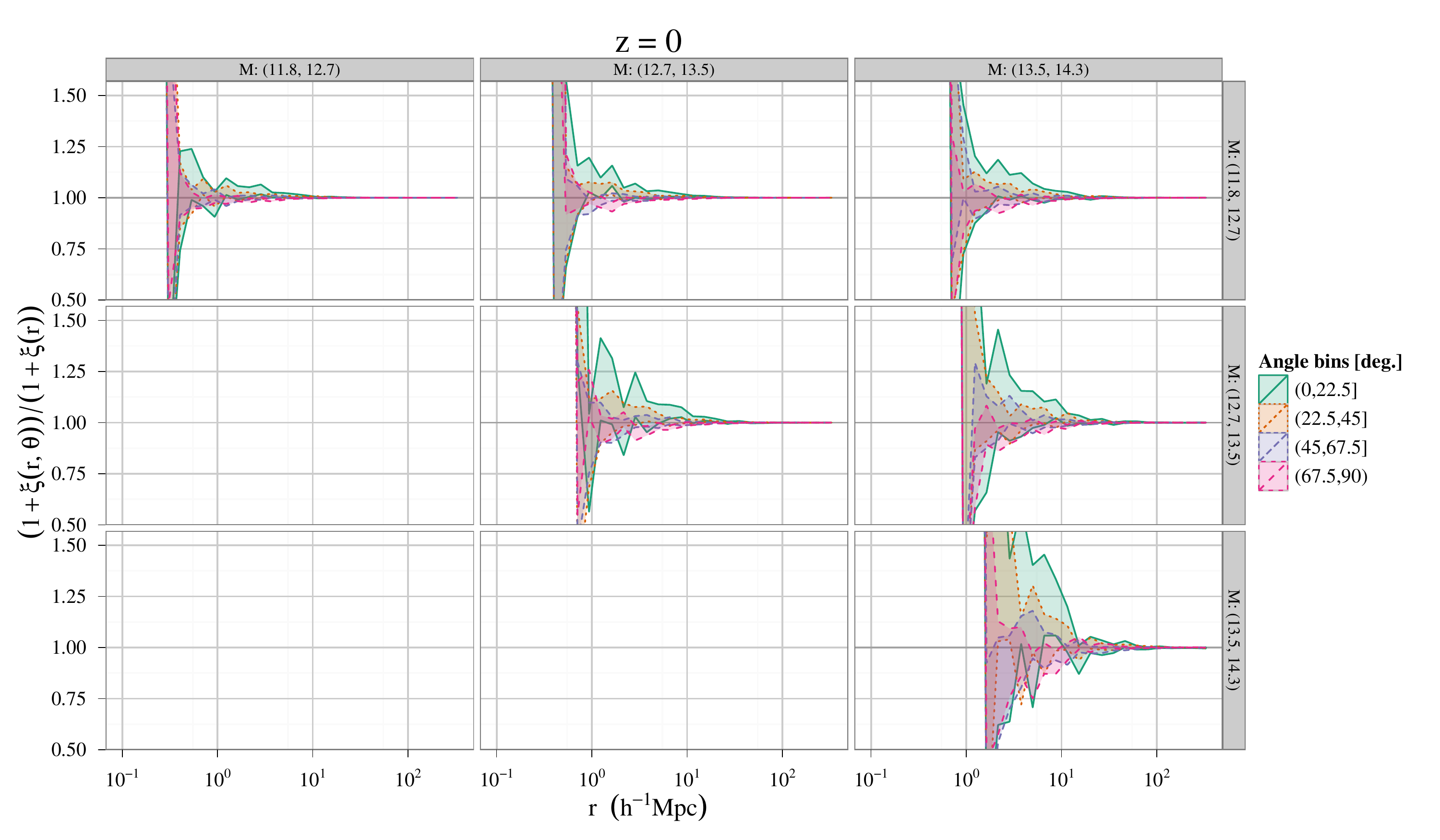}
  }
  \centerline{
  \includegraphics[scale=0.4]{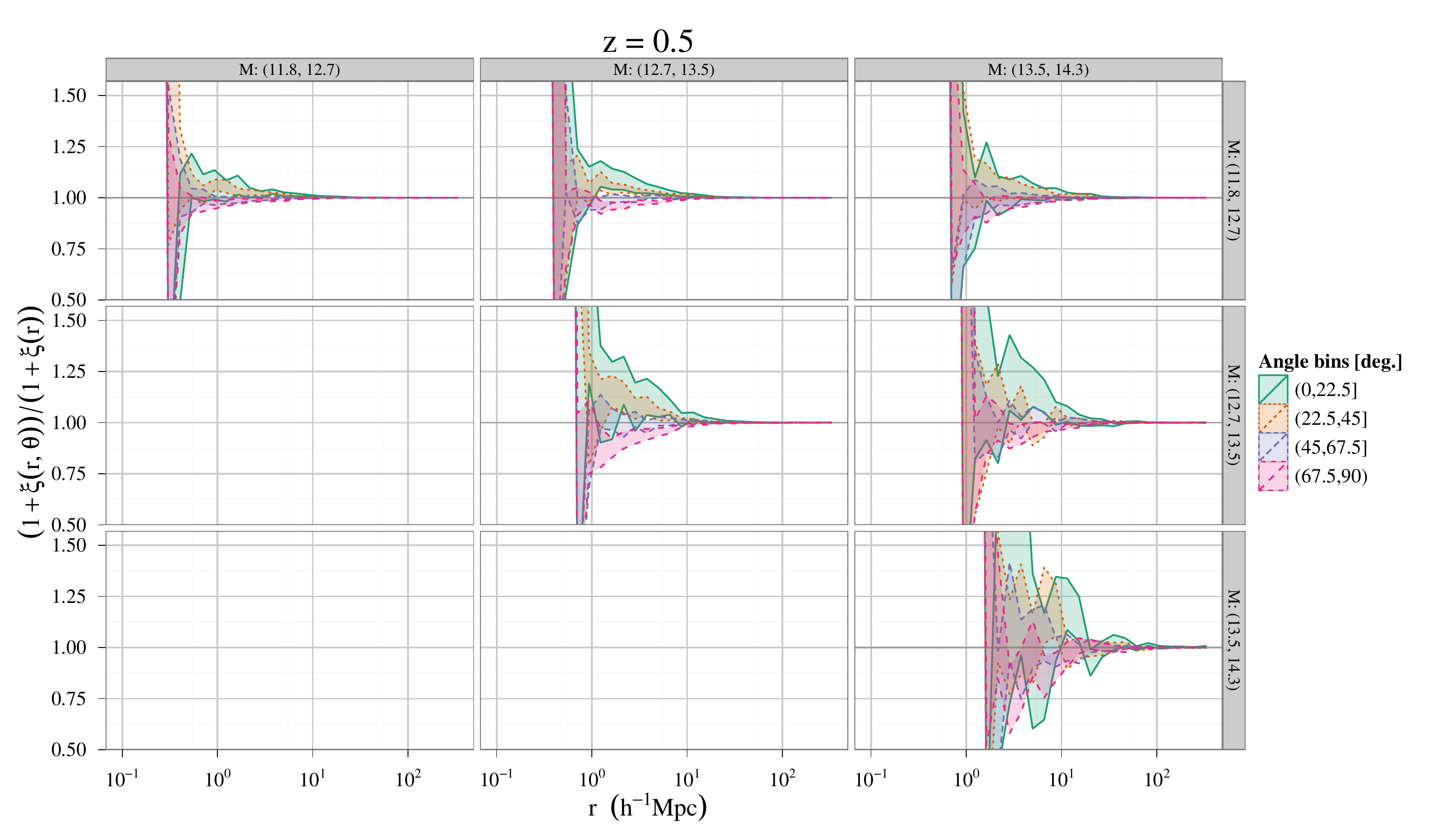}
  }
   \centerline{
  \includegraphics[scale=0.4]{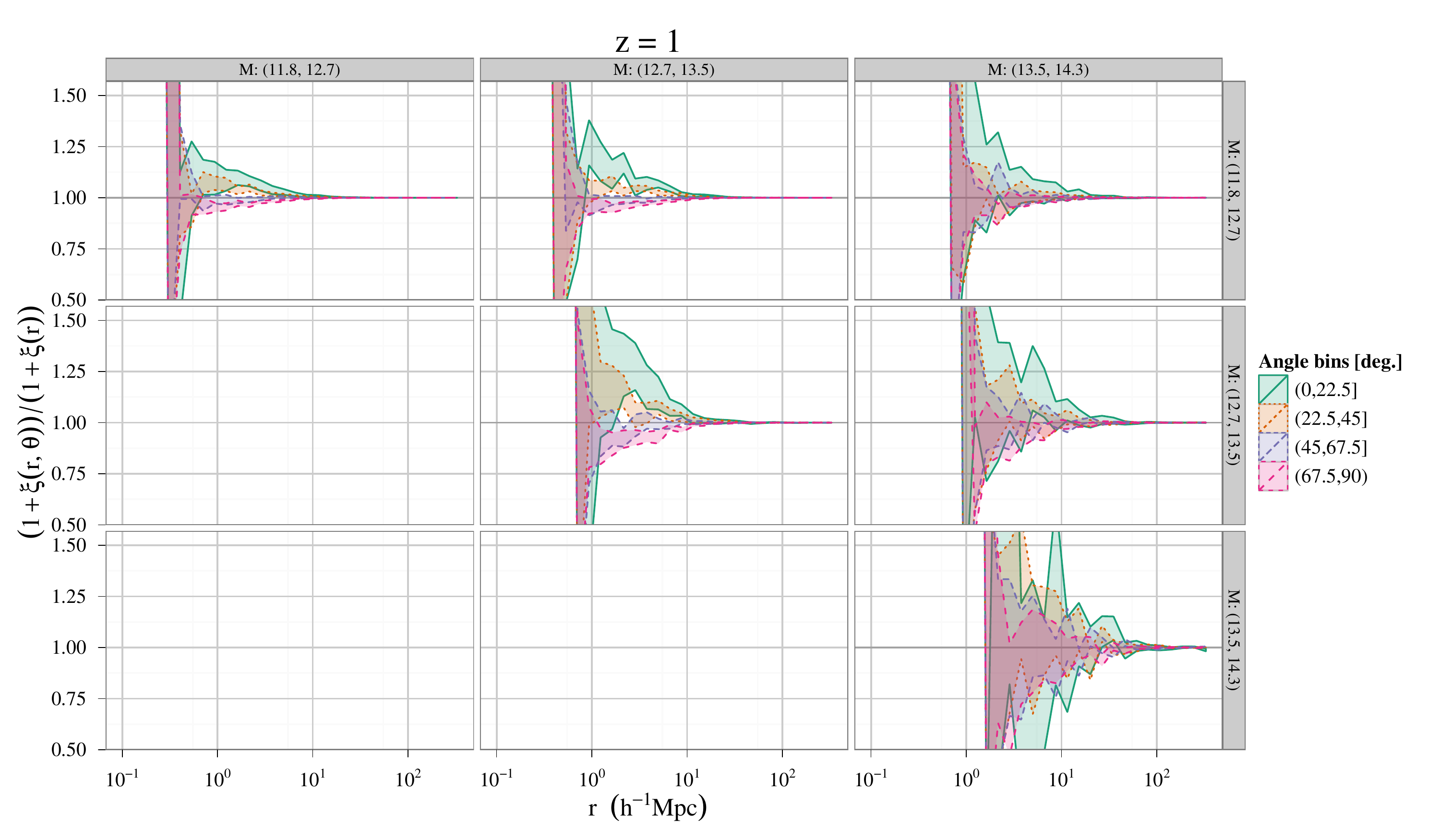}
  }
  \caption{\label{fg:thetacorrinnerinner} 
  $\excor\left(r,\theta; M_{1}, M_{2}\right)$
  as in figure~\ref{fg:thetacorrouterouter} except with shapes measured at $r=0.1\rvir$.
}
\end{figure*}

In figure~\ref{fg:thetacorrinnerinner} 
we show the halo shape-shape correlations similar to figure~\ref{fg:thetacorrouterouter} but with 
shapes measured at $0.1\rvir$. There is no detectable alignment of inner halo shapes at $z=0$, 
and only a moderate alignment signal at $z=1$.  This indicates that the misalignment of the inner 
and outer halo shapes is not strongly correlated with the intra-halo alignments.

\begin{figure*}[htpb]
  \centerline{
  \includegraphics[scale=0.48]{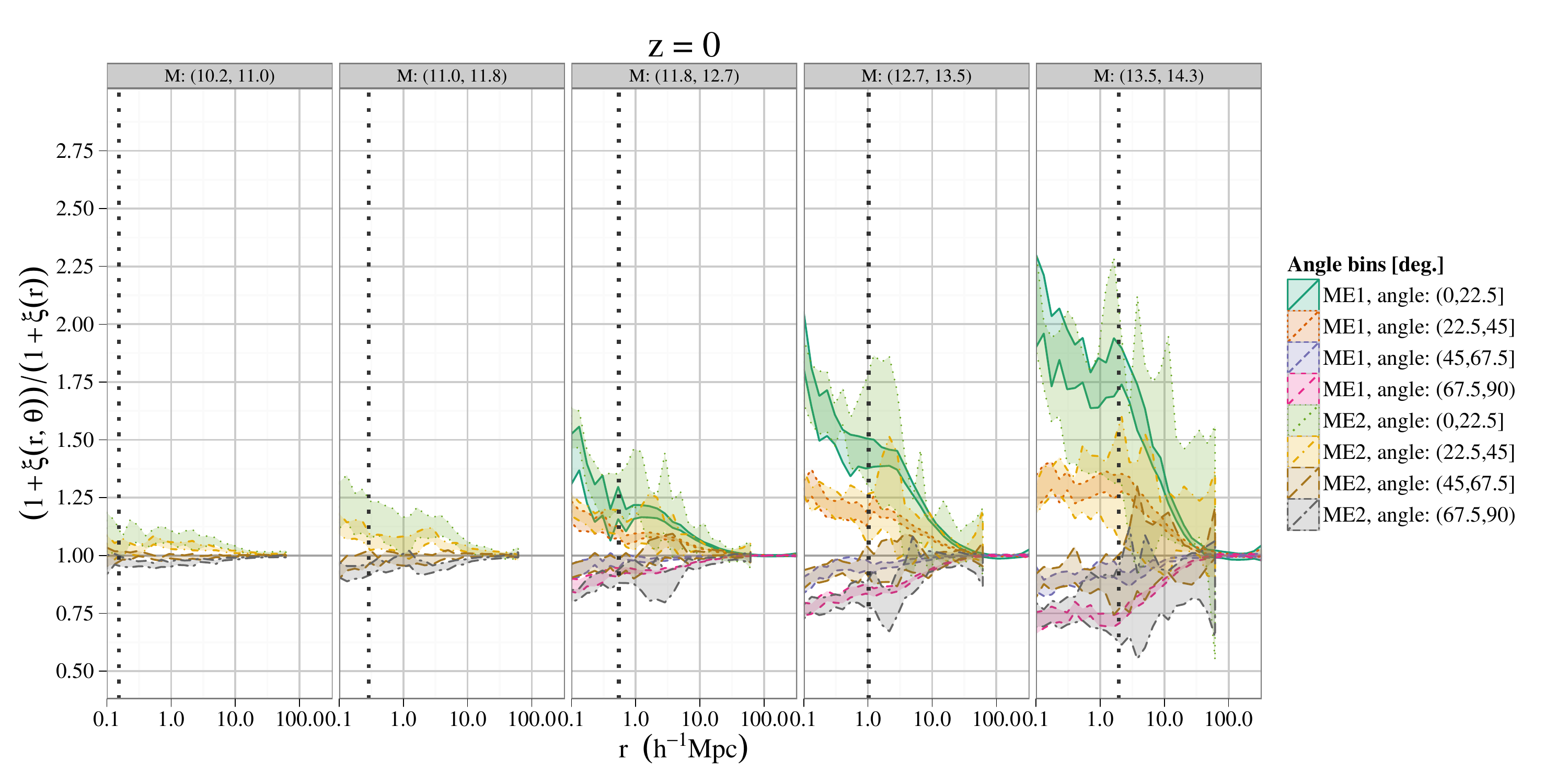}
  }
  \centerline{
  \includegraphics[scale=0.48]{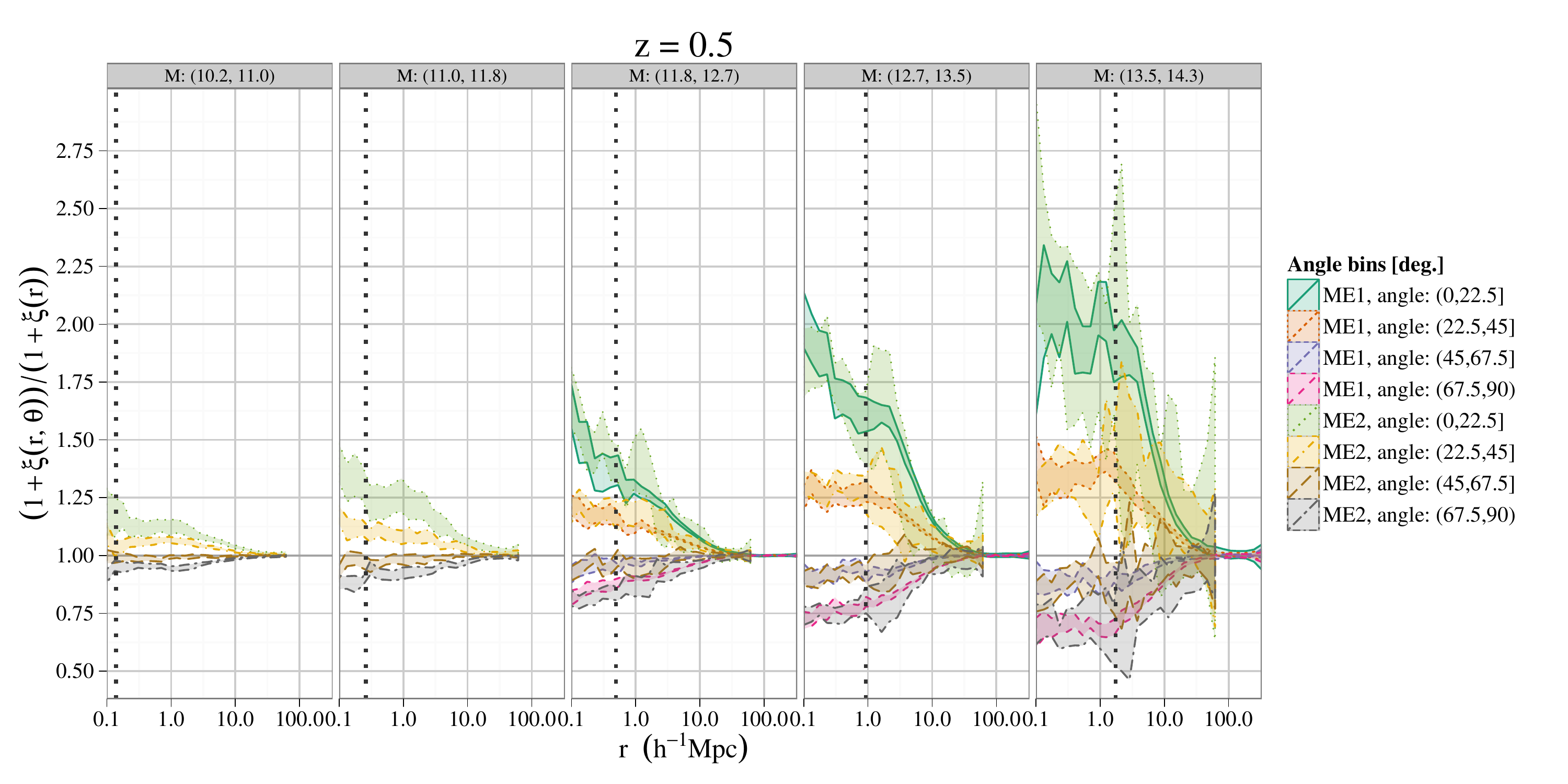}
  }
   \centerline{
  \includegraphics[scale=0.48]{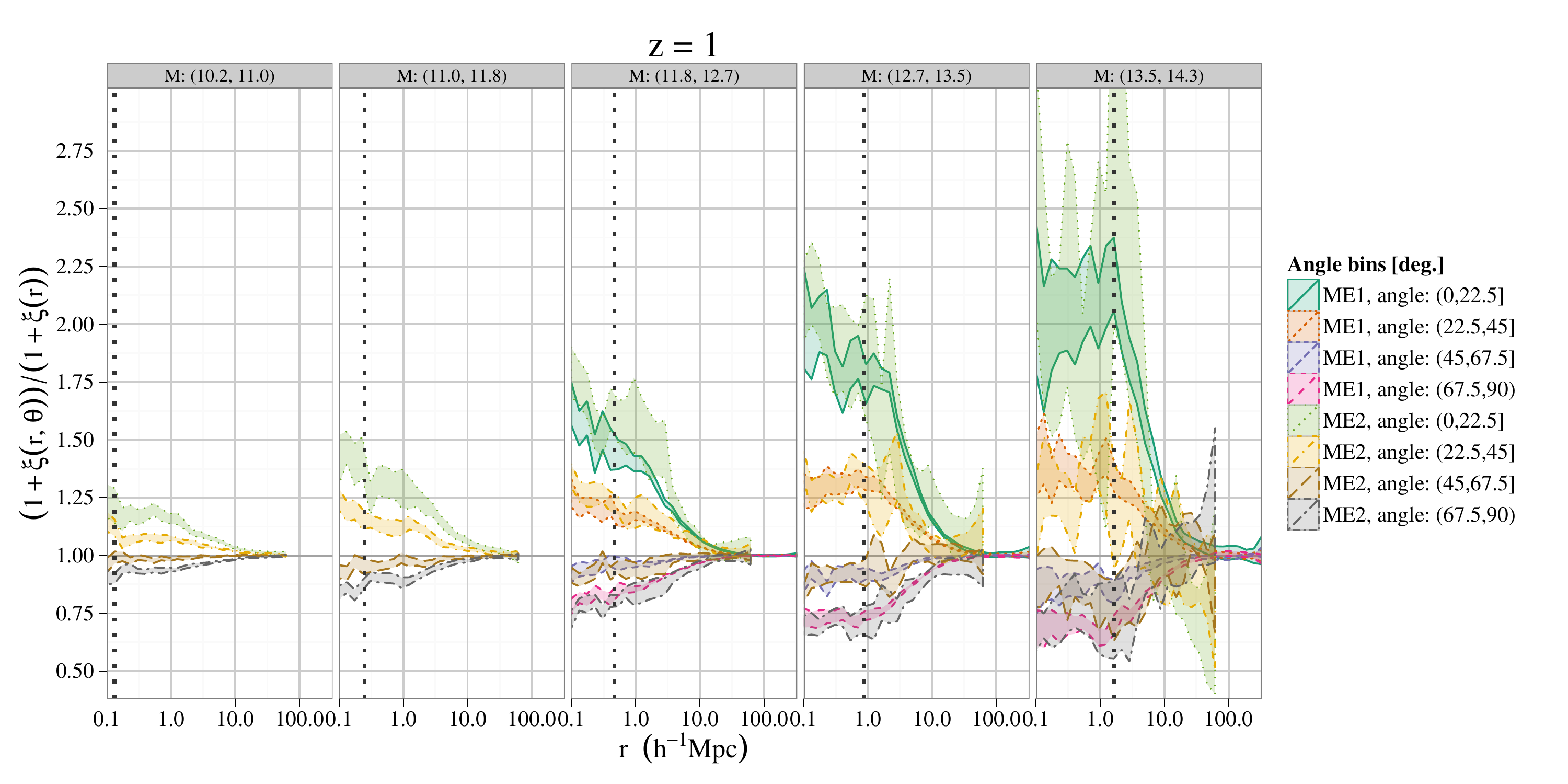}
  }
  \caption{\label{fg:thetacorrinnersep}$\excor\left(r,\theta; M_{1}\right)$ 
  with $\theta$ the angle between the major axes of one halo 
  (measured at $0.1\rvir$) and the position vector to a mass tracer.
  The shaded bands denote the 95\% confidence intervals determined from 100 fixed block
  bootstrap samples.
}
\end{figure*}

The halo-mass excess alignment correlation functions using the halo shapes at $0.1\rvir$ are 
shown in figure~\ref{fg:thetacorrinnersep}. Contrary to the halo-halo alignment correlations with the 
inner halo shapes, there is still a large alignment signal for the halo-mass correlations. 
The peak amplitudes of the excess alignments are reduced by about 20\% relative to the halo-mass 
alignments shown in figure~\ref{fg:thetacorroutersep} using the halo shapes at $\rvir$.
The shapes of the excess correlations are also changed relative to figure~\ref{fg:thetacorroutersep}, 
with a reduced ``bump'' on scales just larger than the typical virial radii in each mass bin.

\begin{figure*}[htpb]
  \centerline{
  \includegraphics[scale=0.41]{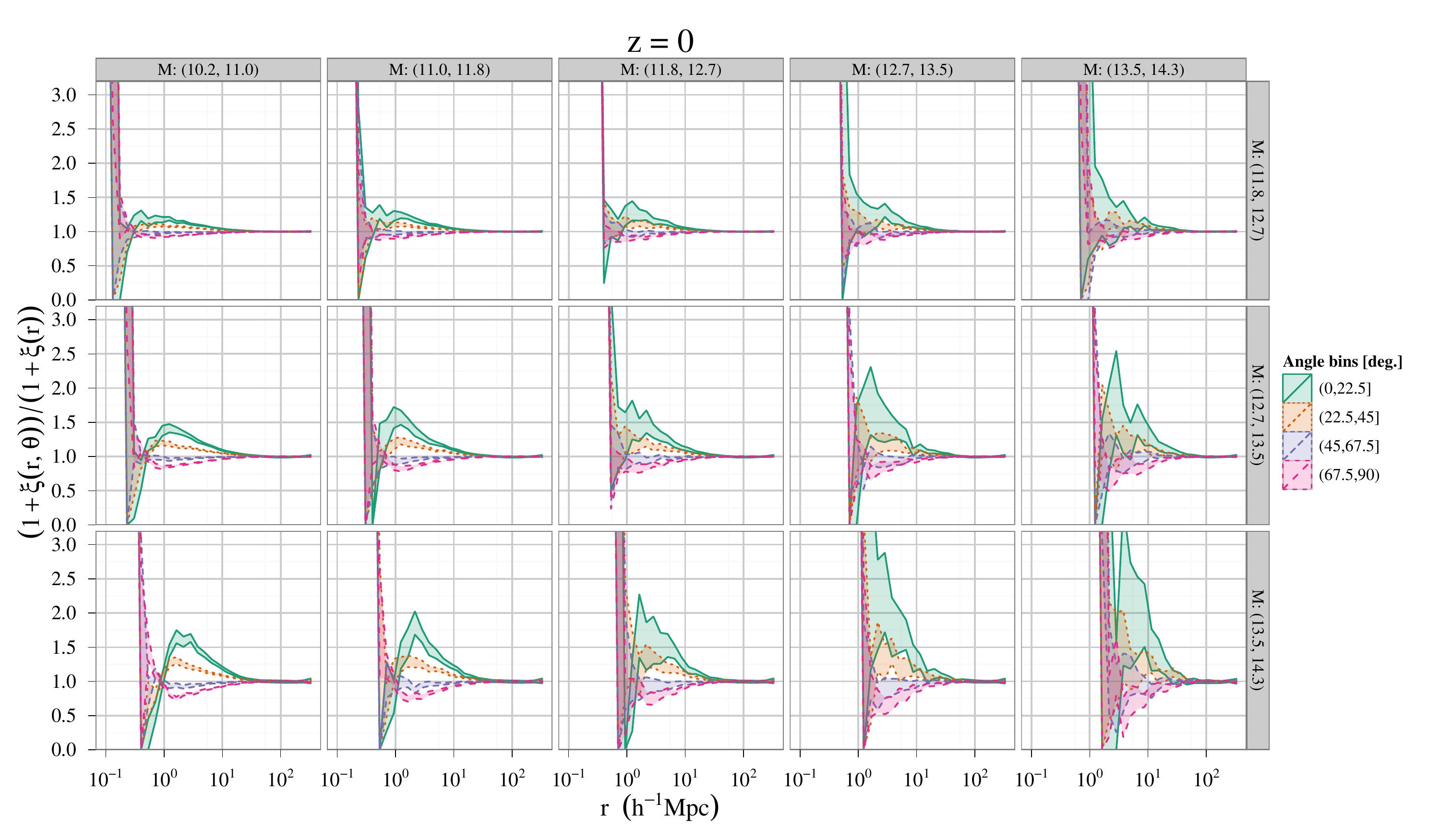}
  }
  \centerline{
  \includegraphics[scale=0.41]{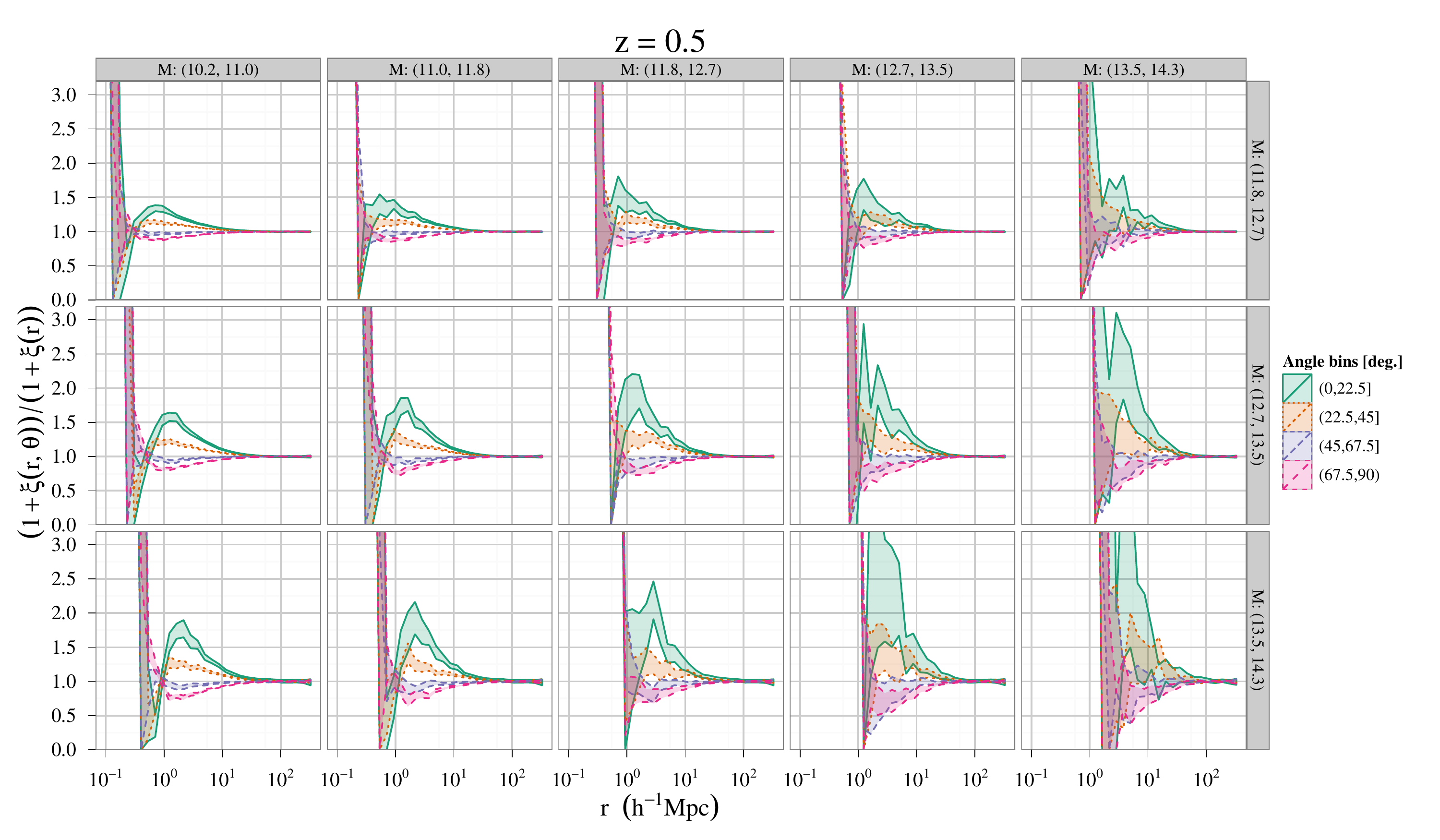}
  }
   \centerline{
  \includegraphics[scale=0.41]{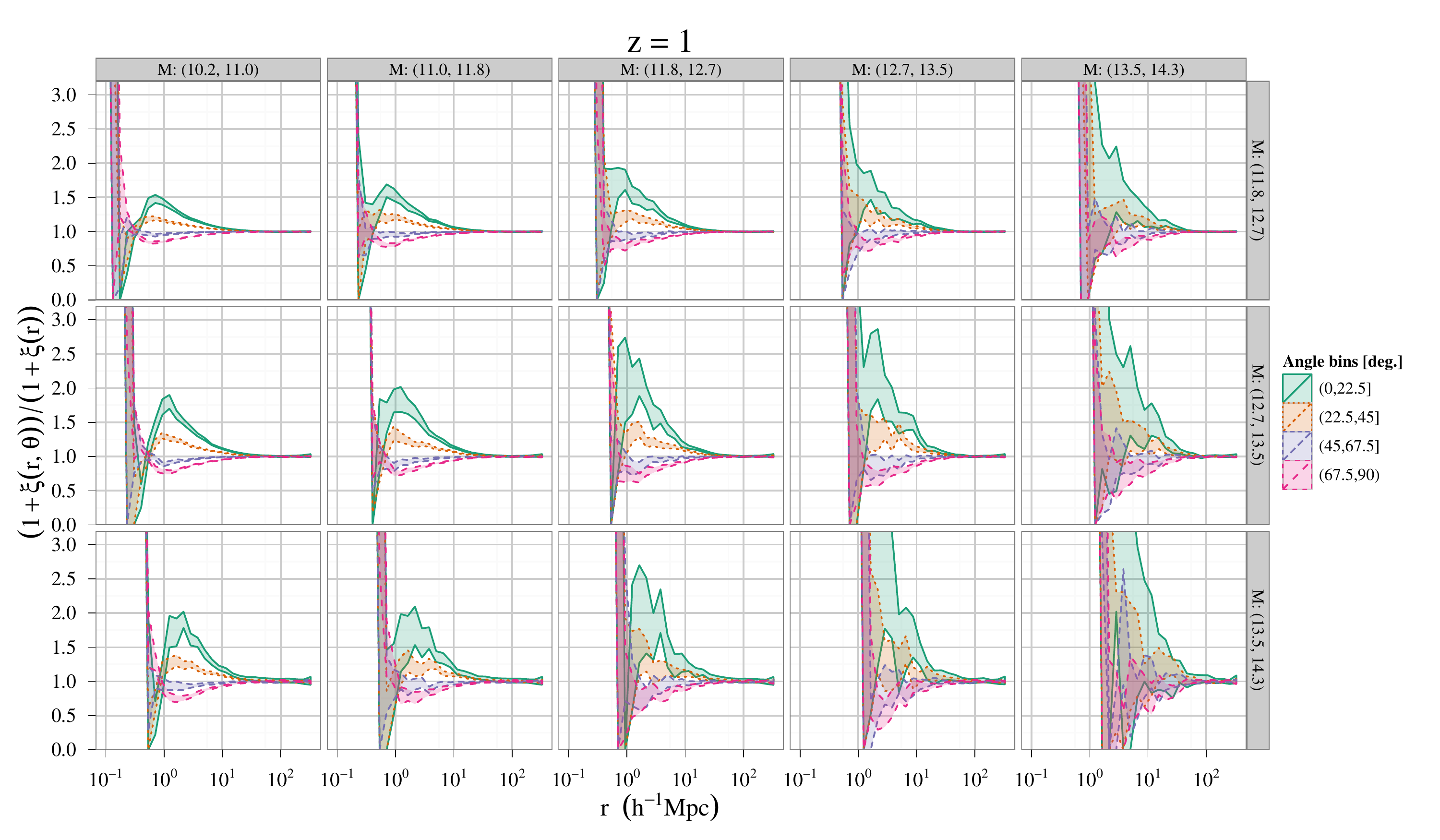}
  }
  \caption{\label{fg:thetacorrinnershapehalo} $\excor\left(r,\theta; M_{1}, M_{2}\right)$ where 
  $\theta$ is the angle between the major axis of one halo (measured at $0.1\rvir$) 
  and the separation vector to another halo. 
  The rows of panels show the mass bins of the halos with the shape measurements.
  The columns of panels show the mass bins of the halos used as mass density tracers.  
  The three plots show results for redshifts $z=0,0.5,1$ from top to bottom.}
\end{figure*}

Finally, in figure~\ref{fg:thetacorrinnershapehalo} we show the halo-mass alignment correlations where
SubFind-0 halos are used as mass density tracers and the shapes of halos are measured at $0.1\rvir$.
Again, the amplitudes of the peak alignment correlations are reduced by $\sim20$\% relative to 
those in figure~\ref{fg:thetacorroutershapehalo} based on the halo shapes at $\rvir$.
We expect the alignment correlations shown in figures~\ref{fg:thetacorrinnersep} and 
\ref{fg:thetacorrinnershapehalo} to be most useful for modeling the alignments of 
central galaxy shapes, which should more closely trace the inner rather than the outer shapes 
of their parent halos.

\label{lastpage}
\end{document}